\documentclass{aastex61} 

\def\lae{\mathrel{\raise .4ex\hbox{\rlap{$<$}\lower 1.2ex\hbox{$\sim$}}}}
\def\gae{\mathrel{\raise .4ex\hbox{\rlap{$>$}\lower 1.2ex\hbox{$\sim$}}}}

\accepted{February 12, 2018}
\submitjournal{ApJS}
\shorttitle{X-rays from Quasar Jets}
\shortauthors{Marshall et al.}

\begin{document}

\title{An X-ray Imaging Survey of Quasar Jets -- The Complete Survey}

\correspondingauthor{Herman L. Marshall}
\email{hermanm@space.mit.edu}

\author{H. L. Marshall}
\affiliation{Kavli Institute for Astrophysics and Space Research,
 Massachusetts Institute of Technology, 77 Massachusetts Ave.,
 Cambridge, MA 02139, USA}

\author{J.M. Gelbord}
\affiliation{Kavli Institute for Astrophysics and Space Research,
 Massachusetts Institute of Technology, 77 Massachusetts Ave.,
 Cambridge, MA 02139, USA}
\affiliation{Spectral Sciences Inc., 4 Fourth Ave., Burlington, MA, 01803, USA}
\author{D.M. Worrall}
\affiliation{HH Wills Physics Laboratory, University of Bristol, Tyndall Ave.,
Bristol BS8 1TL, UK}
\author{M. Birkinshaw}
\affiliation{HH Wills Physics Laboratory, University of Bristol, Tyndall Ave.,
Bristol BS8 1TL, UK}
\author{D.A. Schwartz}
\affiliation{Harvard-Smithsonian Center for Astrophysics,
 60 Garden St., Cambridge, MA 02138, USA}
\author{D.L. Jauncey}
\affiliation{CSIRO Australia Telescope National Facility,
PO Box 76, Epping, NSW 2121, Australia}
\affiliation{Research School of Astronomy and Astrophysics, Australian
National University, Canberra, ACT, 2611, Australia}
\author{G. Griffiths}
\affiliation{HH Wills Physics Laboratory, University of Bristol, Tyndall Ave.,
Bristol BS8 1TL, UK}
\author{D.W. Murphy}
\affiliation{Jet Propulsion Laboratory, 4800 Oak Grove Dr.,
Pasadena, CA 91109, USA}
\author{J.E.J. Lovell}
\affiliation{School of Mathematics and Physics, University of
 Tasmania, Hobart, TAS 7001, Australia}
\author{E. S. Perlman}
\affiliation{Dept. of Physics and Space Sciences,
 Florida Institute of Technology, 150 W. University Blvd., Melbourne, FL, 32901, USA}
\author{L. Godfrey}
\affiliation{ASTRON, the Netherlands Institute for Radio Astronomy,
Postbus 2, 7990 AA, Dwingeloo, The Netherlands}

\begin{abstract}

We {present}
Chandra X-ray imaging
of a flux-limited sample
of flat spectrum radio-emitting quasars with jet-like
structure. 
X-rays are detected from 59\% of 56 jets.
No counterjets were detected.
The core spectra are fitted by power law
spectra with photon index $\Gamma_x$
whose distribution is consistent
with a normal distribution with mean $1.61^{+0.04}_{-0.05}$
and dispersion $0.15^{+0.04}_{-0.03}$.
We show that the
distribution of $\alpha_{rx}$, the spectral index between the
X-ray and radio band jet fluxes, fits a
Gaussian with mean 0.974$\pm$0.012 and dispersion 0.077 $\pm$ 0.008.
We test the model in which kpc-scale X-rays result from inverse
Compton scattering of cosmic microwave background photons
off the jet's relativistic electrons (the IC-CMB model).
In the IC-CMB model, a quantity $Q$ computed
from observed fluxes and the
apparent size of the emission region
depends on redshift as $(1+z)^{3+\alpha}$.  We fit $Q \propto (1+z)^{a}$, finding
$a = 0.88 \pm 0.90$ and reject at 99.5\% confidence
the hypothesis that the average
$\alpha_{rx}$ depends on redshift in the manner
expected in the IC-CMB model.
This conclusion is mitigated by lack of detailed
knowledge of the emission region geometry, which requires deeper or higher resolution X-ray
observations.
Furthermore, if the IC-CMB model is valid for
X-ray emission from kpc-scale jets, then the
jets must decelerate on average: bulk Lorentz factors
should drop from about 15 to 2-3 between pc and kpc scales.
Our results compound the problems that the IC-CMB
model has in explaining the X-ray emission of kpc-scale
jets.

\end{abstract}


\section{Introduction}
 \label{sec:intro}
 
The pc-scale jets of powerful quasars are highly relativistic, with
bulk Lorentz factors ($\Gamma = (1-\beta^2)^{-1/2}$) of 10-30
\citep{2007ApJ...658..232C,2009AJ....138.1874L}.
Since radio galaxies and quasars are generally double lobed, the jets
that deliver energy to the lobes, hundreds of kpc from the core, must
also be two sided.  Because many radio jets, and practically all of
those emitting X-rays, appear to be one-sided, most models of
kpc-scale jets {invoke bulk} relativistic motion, with beaming factors,
$\delta = 1/(\Gamma[1-\beta \cos \theta ] ) >\,$1, where
$\theta$ is the angle of the jet to the line of sight.
 
On kpc scales, many fundamental physical properties of quasar jets
remain uncertain, such as the proton or positron content, whether the
particle and magnetic field energy densities are near equipartition,
and whether the jets have high $\Gamma$ tens to
hundreds of kpc from the quasar core. All of these issues bear on the
flux of useful energy carried by the
jet. The jets typically carry a significant fraction of the quasar
energy budget, and therefore potentially provide information about the
fueling and rate of growth of the central black hole.
{\em Chandra}
snapshot surveys \citep{2004ApJ...608..698S,2005ApJS..156...13M,2011ApJS..193...15M} have shown
that X-rays are easily detected from most radio jets in quasars.
One-zone (single population) synchrotron and synchrotron self-Compton
(SSC) models generally fail to explain the emission, as found
in the first {\em Chandra} observation of the kpc scale
jet emanating from the quasar PKS~0637$-$752 \citep{2000ApJ...540L..69S}
and noted in many subsequent observations of individual
sources.  For a review, see \cite{2006ARA&A..44..463H} and \cite{2009A&ARv..17....1W}.

Due to the failure of single-zone synchrotron models,
the X-ray emission of {kpc-scale} quasar jets
is usually interpreted as inverse Compton emission of relativistic jet electrons off
cosmic microwave background photons (IC-CMB).
This requires that the jet emission is Doppler boosted
with large Lorentz factor $\Gamma$, and at a small angle, $\theta$, to the
line of sight \citep{2000ApJ...544L..23T,2001MNRAS.321L...1C}.  The IC-CMB emission is
brighter than self-Compton emission
because the CMB energy density is enhanced by a
factor $\Gamma^2$ in the jet rest frame.
The model was used to explain the discovery
observations of PKS~0637$-$752 and was subsequently invoked often to explain bright
X-ray knots in individual sources as well
as for jet surveys \citep{2004ApJ...608..698S,2005ApJS..156...13M,2006AN....327..227J,2011ApJS..193...15M,2011ApJ...730...92H}.  If
valid, the model can be used to compute the jet speed along the flow
to deduce bulk deceleration \citep{2004ApJ...604L..81G,2006ESASP.604..643M,2006MNRAS.366.1465H}, or to
infer that matter is entrained \citep{2006ApJ...641..732T}.

In the past 10-15 years, however, there have been
concerns that the IC-CMB model is
inadequate or even rejected in some jets \citep{2005ApJ...622..797K,2006MNRAS.366.1465H,2006ApJ...648..900J}.
One concern with the IC-CMB model is that the lifetimes of the
electrons responsible for the X-ray emission are orders of
magnitude longer than those producing the radio emission so the
X-ray structures would be expected to extend
further downstream than the radio;
just the opposite of what is regularly observed
\citep{2003A&A...403...83T,2006ApJ...640..592S}.
Of particular interest is the observation that $\gamma$-ray emission
expected in the IC-CMB
model \citep{2014ApJ...780L..27M,2015ApJ...805..154M,2017ApJ...835L..35M,2017ApJ...849...95B}
{is not detected,}
even for PKS~0637$-$752, the prototypical case for the IC-CMB model.
An alternative class of models proposes additional synchrotron components
to explain the X-rays \citep{2004ApJ...608...95S,2006ApJ...648..900J,2006MNRAS.366.1465H}.
Either model has dramatic consequences: in the IC-CMB case,
jets should show surface brightnesses that are largely independent of
redshift \citep{2002ApJ...569L..23S}; while synchrotron models require
electrons to be accelerated to Lorentz factors $\sim$ 10$^7$
over much or all of the jet, due to their short lifetimes.

In order to find good cases for detailed study, we started a large,
shallow survey using {\em Chandra} to find X-ray emission from kpc-scale
radio jets.
This paper is a continuation of \citet[][hereafter, Paper I]{2005ApJS..156...13M}
and \citet[][hereafter, Paper II]{2011ApJS..193...15M} and presents
observations of the remainder of the quasars from the original sample
of 56.
We use this sample for a population test of the IC-CMB model's
primary predictions.
Following Paper II and \cite{2011ApJ...730...92H}, we include results from VLBI observations
from the MOJAVE program\footnote{See the MOJAVE web
page: {\tt http://www.physics.purdue.edu/astro/MOJAVE/} and
\citet{2009AJ....138.1874L}.}
that indicate the directions and speeds of relativistic jets in the
quasar cores.  If the IC-CMB model is correct, then we may test whether
the pc scale jet has changed directions or decelerated
in propagating to kpc scales.
We use a cosmology in which $H_0 = 70$ km s$^{-1}$ Mpc$^{-1}$,
$\Omega_{\rm m} = 0.3$, and $\Omega_\Lambda = 0.7$.

\section{Sample Properties}

\label{sec:sample}

Sample selection was described in Paper I.  Briefly,
56 sources were selected from 1.64 or 5 GHz VLA and ATCA imaging
surveys \citep{1993MNRAS.264..298M,lovell}.  The dominant selection criterion is on
radio core flux density -- as applied when creating the samples for
the radio imaging surveys.
The flux densities in jet-like extended
emission determine inclusion in our sample.  
Subsamples were defined in Paper I:
the ``A'' list is a complete flux-limited sample based only
on extended emission, while the
``B'' list was selected for one-sided and linear structure.
There are the same number of objects in each list
and flux-limited selection was applied first.

We reported results for the first 20
targets in Paper I, finding that 60\% of the jets could
be detected in short {\em Chandra} exposures.
In Paper II, we presented results for another 19
quasars in the sample and got the same detection rate.
Here, the observations of the remaining 17 quasars of the sample
are presented.  The radio fluxes of the extended emission in
these additional targets were somewhat lower
than for the first 39 but were observed for about the same
X-ray exposure time (5.8 ks on average).
Fourteen of the new {\em Chandra} images were obtained
as part of the completion of our survey and the other
three were taken from the {\em Chandra} archive.
For the 14 new observations, we also obtained Hubble Space Telescope
(HST) images.

As reported in Paper II,
a significant fraction of the sample is being monitored with
VLBI, mostly in the northern hemisphere.
Superluminal motions have been detected for every object in our sample that
was observed in the MOJAVE program (see table~\ref{tab:bends}).
As in Paper II, the distribution of the apparent velocities, $c \beta_{app}$,
is comparable to those of the remaining MOJAVE sources,
indicating that quasars in our sample have
{a distribution of speeds and line of sight angles that is consistent with that of} the MOJAVE program.

Five redshifts were unknown as of Paper I.
In Paper II, we reported that the redshift of PKS 1421$-$490 was 0.662.
We now include PKS 1145$-$676 in our overall analysis
with a redshift of 0.21 \citep{2009AJ....137..337S},
PKS 0144$-$522 with a redshift of 0.098
\citep{1987AJ.....94..563S}, and PKS 1302$-$82, with
a redshift of 0.87 \citep{2006AJ....131..114B}.
As noted in Paper I, PKS 1145$-$676 shows X-ray emission from a 5\arcsec\ long
region.
The redshift is still unknown for one object in the sample
for which we have an X-ray image:
PKS 1251$-$713.  We excluded this source
from sample analyses that require redshifts.

\section{Observations and Data Reduction}

The {\em Chandra} observation list
is given in Table~\ref{tab:observations}.
As in Papers I and II,
X-ray images were formed
from events in the 0.5-7.0 keV band (see Fig.~\ref{fig:images}).
The images of a few sources show readout streaks, which
do not interfere with the jets because
we selected a suitable range of observatory roll angles.

\begin{deluxetable}{lcccc}
\tablecaption{{\em Chandra} Observation Log \label{tab:observations} }
\tablehead{
\colhead{Target} & \colhead{{\em Chandra}} & \colhead{Live Time} 
	& \colhead{Date} & \colhead{Ref.\tablenotemark{a}} \\
\colhead{} & \colhead{Obs ID} & \colhead{(s)} 
	& \colhead{(UT)} & \colhead{} }
\startdata
  0106$+$013 &    9281 &  8787 & 2007-11-21  & 2\\
  0144$-$522 & 10366 &  5590 & 2009-03-26  & 1\\
  0256$+$075 & 10375 &  5604 & 2008-12-07  & 1\\
  0402$-$362 & 10374 &  5582 & 2009-03-19  & 1\\
  0508$-$220 & 10367 &  5294 & 2009-02-25  & 1\\
  0707$+$476 & 10368 &  5309 & 2009-01-20  & 1\\
  0748$+$126 & 10376 &  5610 & 2009-02-07  & 1\\
  0833$+$585 &  7870 &  3771 & 2007-01-12  & 1\\
  0859$+$470 & 10371 &  5604 & 2008-12-31  & 1\\
  0953$+$254 & 10377 &  5606 & 2009-01-20  & 1\\
  1116$+$128 & 10373 &  5579 & 2009-02-01  & 1\\
  1303$-$827 & 10365 &  5612 & 2009-11-18  & 1\\
  1502$+$106 & 10378 &  5608 & 2009-04-09  & 1\\
  1622$-$297 & 10370 &  5610 & 2009-06-17  & 1\\
  1823$+$568 & 10369 &  5578 & 2009-06-06  & 1\\
  2201$+$315 &  9283 &  9167 & 2008-10-12  & 2\\
  2230$+$114 & 10372 &  5450 & 2009-08-06  & 1\\
\enddata
\tablenotetext{a}{References refer to previous X-ray imaging
 results: 1) this paper, 2) \citet{2011ApJ...730...92H}.}
\end{deluxetable}

Table~\ref{tab:radioCont} lists the radio data used here
and radio flux contours are overlaid on the
X-ray images in Figure~\ref{fig:images}.
X-ray images were registered to radio images as in Paper I.

\begin{deluxetable}{lccrc}
\tablewidth{0pc}
\tablecaption{Radio Observations \label{tab:radioCont} }
\tablehead{
\colhead{Target} & \colhead{Instrument} & \colhead{Date}
	& \colhead{Freq.} & \colhead{$5 \times$ RMS noise} \\
\colhead{} & \colhead{} & \colhead{(UT)} 
	& \colhead{(GHz)} & \colhead{(mJy/beam)}}
\startdata
  0106$+$013 &  \mbox{\em VLA} & 2000-11-05 &   1.42 &  4.29 \\
  0144$-$522 & \mbox{\em ATCA} & 2004-05-08 &  17.73 &  0.50 \\
  0256$+$075 &  \mbox{\em VLA} & 2000-11-05 &   1.42 &  2.64 \\
  0402$-$362 & \mbox{\em ATCA} & 2002-02-01 &   8.64 &  2.52 \\
  0508$-$220 & \mbox{\em ATCA} & 2000-11-05 &   1.42 &  8.40 \\
  0707$+$476 &  \mbox{\em VLA} & 2000-11-05 &   4.86 &  1.34 \\
  0748$+$126 &  \mbox{\em VLA} & 2001-05-06 &   4.86 &  1.50 \\
  0833$+$585 &  \mbox{\em VLA} & 2000-11-05 &   1.42 &  2.44 \\
  0859$+$470 &  \mbox{\em VLA} & 2000-11-05 &   4.86 &  2.51 \\
  0953$+$254 &  \mbox{\em VLA} & 2000-11-05 &   1.42 &  1.93 \\
  1116$+$128 &  \mbox{\em VLA} & 2000-11-05 &   4.86 &  1.70 \\
  1303$-$827 & \mbox{\em ATCA} & 1993-07-13 &   8.64 &  4.17 \\
  1502$+$106 &  \mbox{\em VLA} & 2000-11-05 &   4.86 &  1.94 \\
  1622$-$297 & \mbox{\em ATCA} & 1994-07-29 &   8.64 &  9.06 \\
  1823$+$568 &  \mbox{\em VLA} & 2000-11-05 &   4.86 &  4.32 \\
  2201$+$315 &  \mbox{\em VLA} & 2000-11-05 &   1.42 &  2.64 \\
  2230$+$114 &  \mbox{\em VLA} & 2000-11-05 &   4.86 &  4.09 \\
\enddata
\end{deluxetable}

\begin{figure}[htp]
 {\plotone{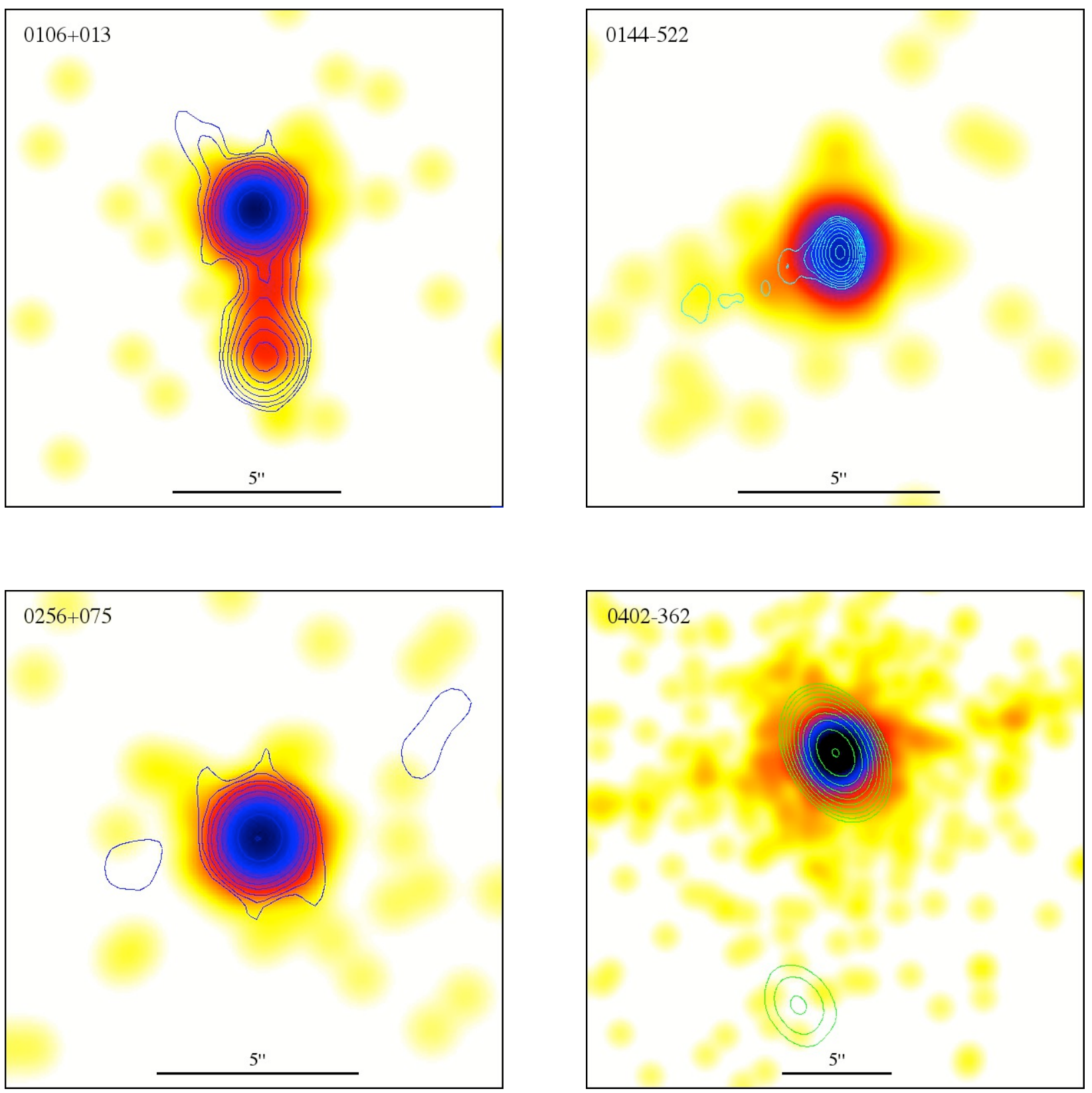}}
  \caption{X-ray images from {\em Chandra} observations,
  with contours from Australia
  Telescope Compact Array or Very Large Array (VLA) images.
  The radio surface brightness contours increase by a factor of 2 and start at 5
  times the rms noise, as given in Table~\ref{tab:radioCont}.
  The X-ray images were convolved with 1\arcsec\ Gaussians and then
  binned at 0.0492\arcsec, a tenth of a {\em Chandra} pixel.
  The color scale for all images is logarithmic,
  from 0.5 counts/beam (yellow) to 2500 counts/beam
  (black).  See the text for comments on individual objects.
  A readout streak is apparent
  in the X-ray map of 2201$+$315.
  } \label{fig:images}
\end{figure}

\addtocounter{figure}{-1}

\begin{figure}[htp]
  \plotone{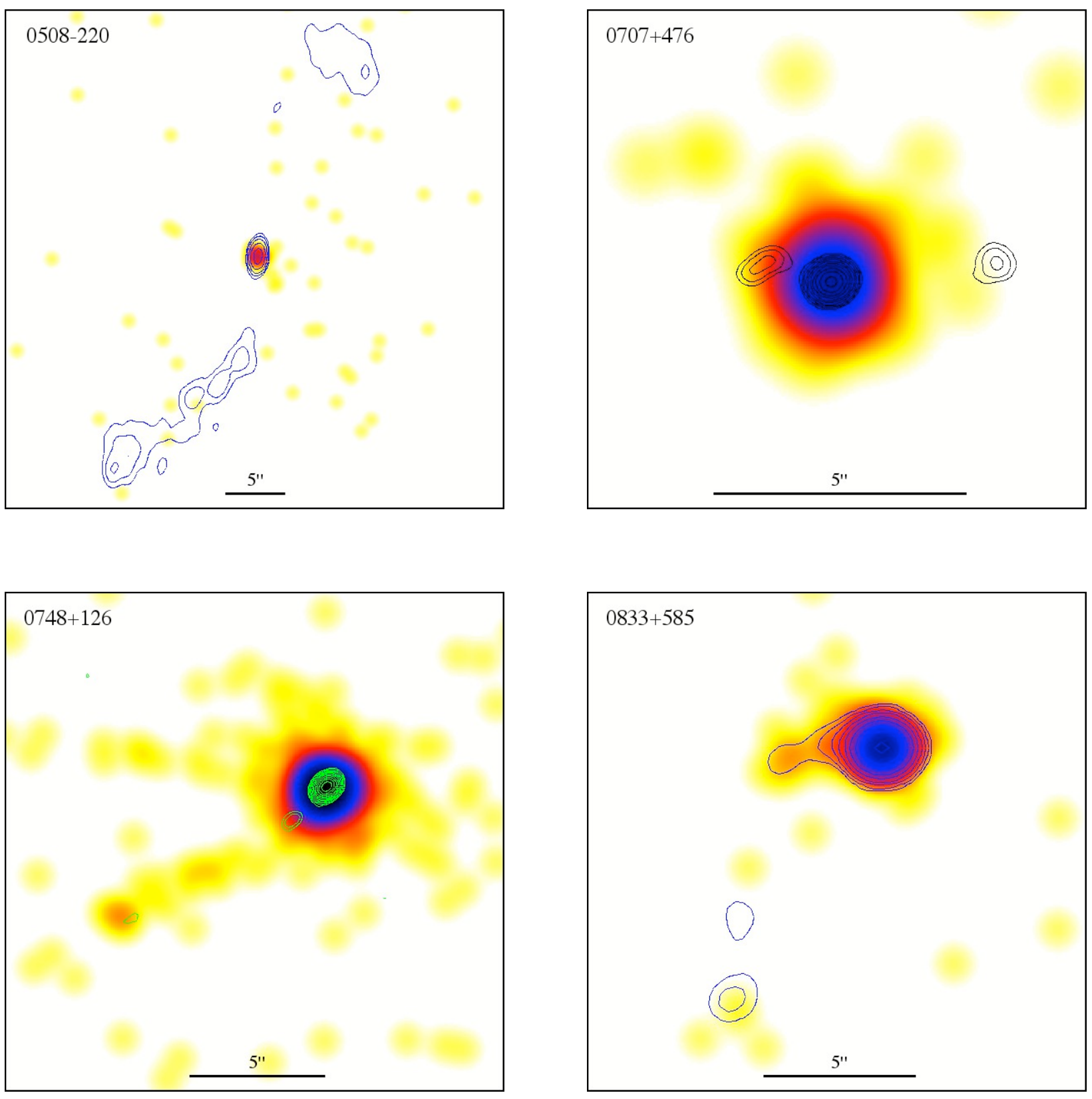}
  \caption{continued.}
\end{figure}

\addtocounter{figure}{-1}

\begin{figure}[htp]
  \plotone{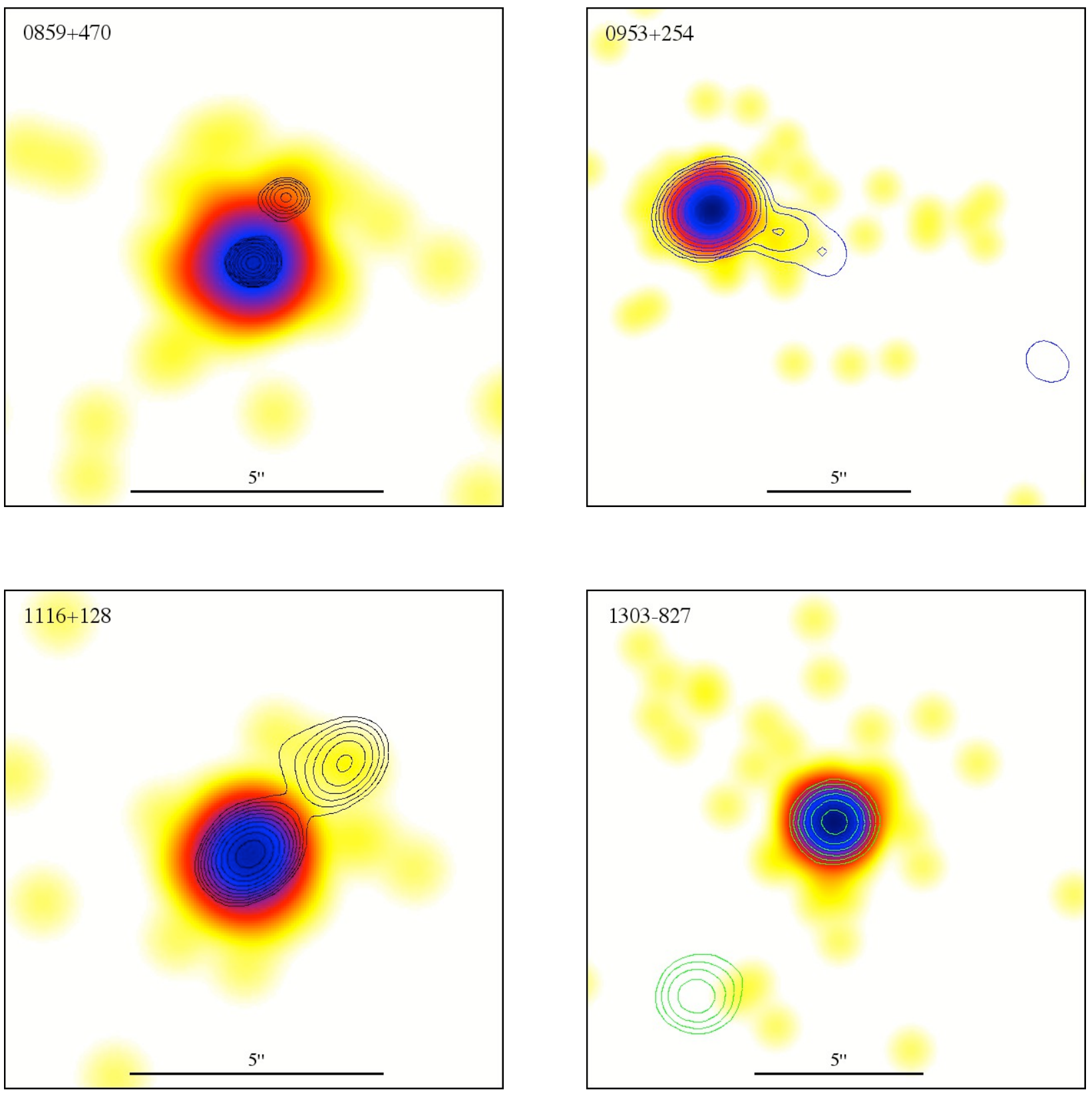}
  \caption{continued.}
\end{figure}

\addtocounter{figure}{-1}

\begin{figure}[htp]
 \plotone{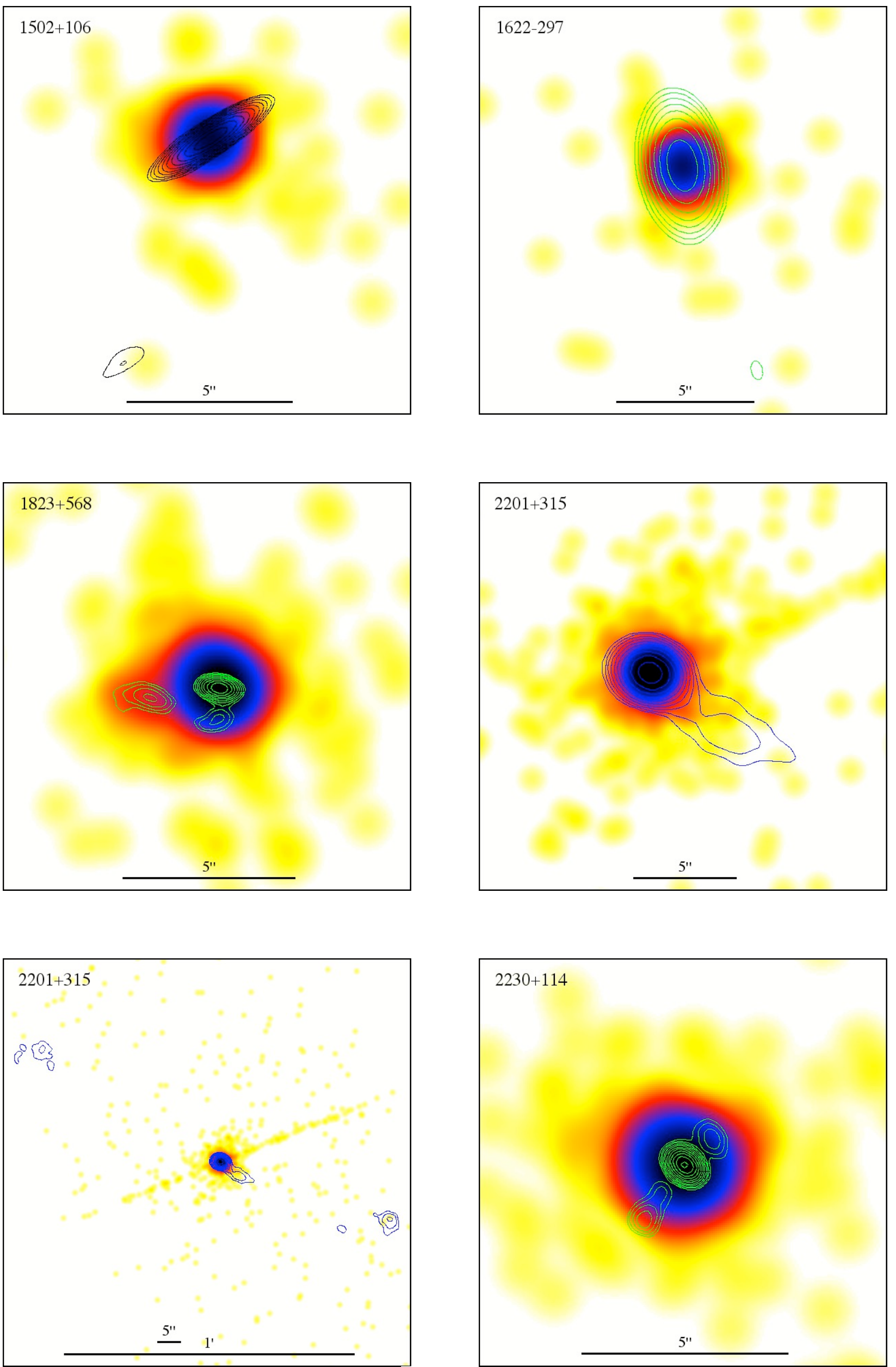}
  \caption{continued.}
\end{figure}

\subsection{Core Spectral Fits}

\label{sec:corefits}

 {The X-ray spectrum of the nucleus for each source was measured 
using the CIAO v4.7 software \citep{2006SPIE.6270E..1VF} and CALDB 4.6.5 calibration database.
On-source counts were extracted
  from a circle of radius $1.25''$ with local background sampled from
  a source-centered annulus, using a pie slice to exclude resolved
  X-ray jet
  emission where detected.  Spectral data between 0.4 and 7 keV were binned to
  a minimum of 25 counts per bin and were 
  fitted using the $\chi^2$ statistic in XSPEC \citep{arnaud}, initially to a power-law model of fixed Galactic absorption.
  If the fit was good (the majority of cases) no additional components were added.  If not,
  intrinsic absorption or a thermal component was added to
  the model to find an improved fit, and in some cases a pileup model was required.
The results are given in
Table~\ref{tab:cores}, where the notes to the table or an entry in the
$N_{\rm H_{\rm int}}$ column
identify cases where a model more complex than a power law with Galactic
absorption was used.}  The power-law slope, $\Gamma_{\rm x}$, is the
photon spectral index, and so is $\alpha + 1$ where $\alpha$ is the
energy spectral index more commonly used in radio astronomy ($S
\propto \nu^{-\alpha}$).  

\begin{figure}
\begin{center}
\plotone{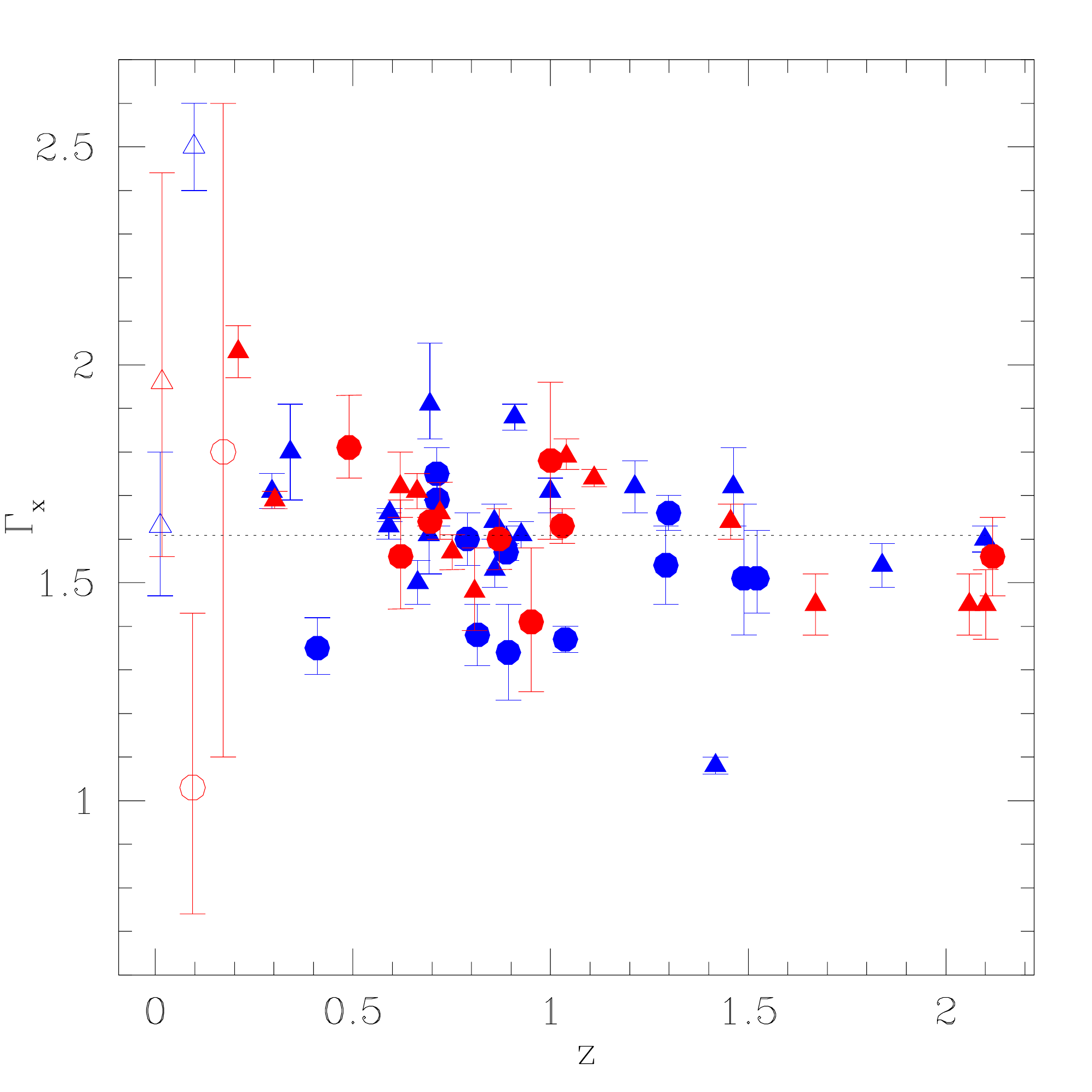}
\caption{X-ray spectral indices of the cores plotted against redshift.
Objects with X-ray jet emission (this paper) are
shown as triangles, and those without as circles.  Unfilled symbols
mark the five objects at $z < 0.2$ that are excluded from the calculation of
the central value of the distribution (dotted line).  {The 32 sources
  that appear as $\gamma$-ray detections in the {\it Fermi} LAT 3LAC
  catalog \citep{2015ApJ...810...14A} are shown in blue; the other 24 in red.}
\label{fig:corespec} }
\end{center}
\end{figure}

The X-ray spectral indices are plotted against redshift in
Figure~\ref{fig:corespec}.  We follow practice dating from the {\it
  Einstein\/} Observatory of assuming that the underlying
spectral-index distribution has a normal distribution, and maximize
the likelihood to find the best-fit underlying mean and dispersion
\citep{1988ApJ...326..680M, 1989ESASP.296..719W}.  For the 51 objects at $z > 0.2$, with
90\% joint-confidence uncertainties, we find $\bar\Gamma_{\rm x} =
1.61 \pm 0.05$ and $\sigma = 0.15^{+0.04}_{-0.03}$.  These
uncertainties are improved with respect to Paper I, and further
confirm the flatter spectral index found in radio-loud quasars as
compared with radio-quiet quasars for which $\bar\Gamma_{\rm x}
\approx 1.9$ \citep{2000MNRAS.316..234R}.  Our results are consistent with
\citet{2006MNRAS.366..339B} who, from studying the X-ray spectra of radio-loud
quasars and radio galaxies matched in extended radio power, conclude
that the X-ray emission of core-dominated quasars is dominated by a
beamed inverse-Compton jet component that is flatter in spectrum than
other emission.  The model of a radio-loud quasar's X-ray spectrum
being comprised of both isotropic and a beamed jet component was first
proposed based on {\it Einstein\/} data due to a larger X-ray to radio
flux ratio with increasing core dominance \citep{1987ApJ...313..596W}, and the
model is supported by more recent flux comparisons for larger samples
\citep{2011ApJ...726...20M}.

\begin{figure}
\begin{center}
\plotone{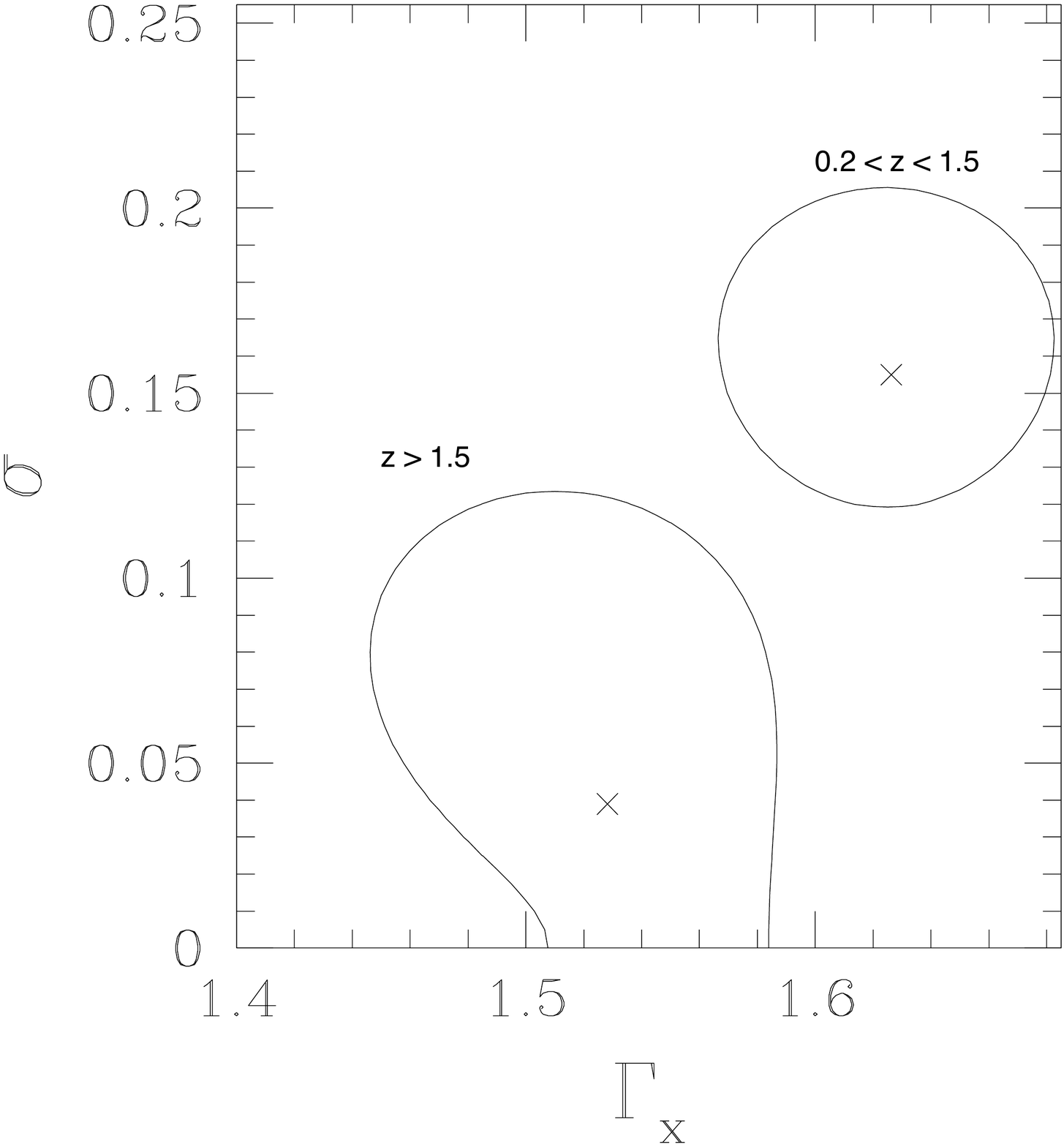}
\caption{90\% confidence contours (4.61 above the minimum value of
  $-2\ln$ likelihood) for the two interesting parameters of mean
  spectral index and intrinsic dispersion.  The comparison between the
  7 quasars at $z > 1.5$ and the 44 at $0.2 < z < 1.5$ finds less than
  1\% probability that they are drawn from the same parent
  distribution.
\label{fig:contit} }
\end{center}
\end{figure}

\begin{figure}
\begin{center}
\plotone{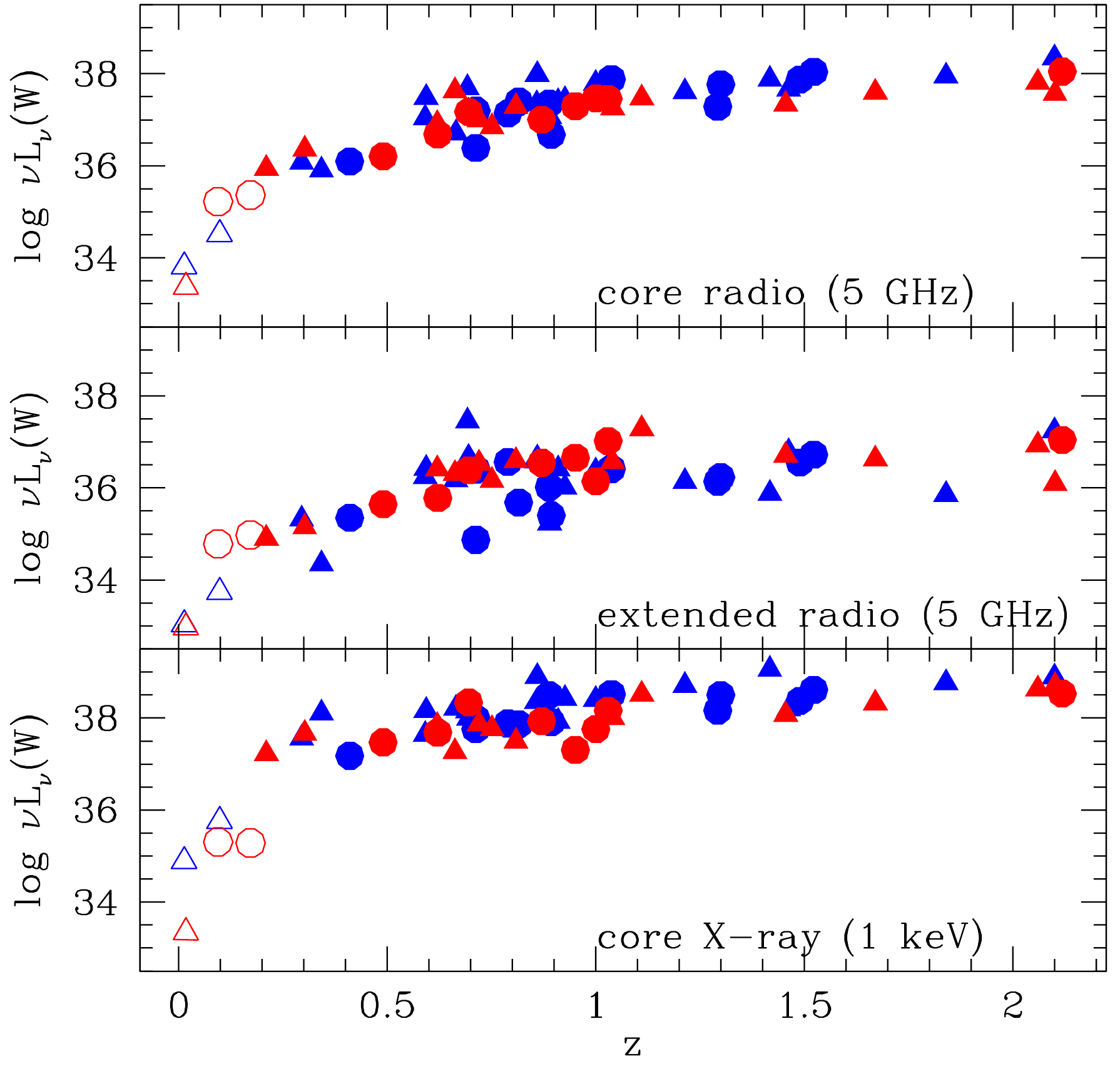}
\caption{{Rest-frame isotropic luminosities (disregarding beaming) against redshift for emissions in
  5-GHz core radio  (top), 5-GHz extended radio (middle) and 1-keV X-ray
  (bottom), showing the extent to which luminosity and redshift are
  related in our sample. Colors and symbols are as in Figure~\ref{fig:corespec}.}
\label{fig:seqfig} }
\end{center}
\end{figure}

Figure~\ref{fig:corespec} hints at decreased $\Gamma_{\rm x}$ and $\sigma$ at
high redshift: the seven objects above $z= 1.5$ (five of which have
X-ray jets) give $\bar\Gamma_{\rm x} = 1.53^{+0.06}_{-0.08}$ and $\sigma =
0.04^{+0.08}_{-0.04}$.  Figure~\ref{fig:contit} compares 90\% joint
confidence contours in spectral index and dispersion for the 7 quasars
at $z > 1.5$ and the 44 at $0.2 < z < 1.5$. The plot implies that the
probability that the two subsamples are drawn from the same
parent distribution is $< 1\%$.  The two caveats to this result are that the
choice of a dividing redshift of 1.5 is guided by the observations,
and the dependence of luminosity with redshift in the sample means
that any tendency towards flatter spectral index and a tighter
distribution may be more associated with higher luminosity than higher
redshift.  {Figure \ref{fig:seqfig} shows the dependence of
  luminosity with redshift in the radio (with core and
  extended jet emission shown separately) and the X-ray.  The trend is
  most obvious in the core radio, as expected from the flux-density
  thresholds applied during sample selection, but can also be seen in the X-ray and, with
  somewhat larger scatter, in the extended radio emission.  
  Figures~\ref{fig:corespec} and \ref{fig:seqfig} both differentiate by color
  the sources that show as {\it Fermi} $\gamma$-ray detections in the LAT
  3LAC catalog \citep{2015ApJ...810...14A}, and by symbol the sources with
  extended X-ray jet emission as found in this paper.  It is noticeable that
  the {\it Fermi\/} detections (57\%) are distributed across the redshift,
  luminosity, and X-ray spectral-index range of our sources rather
  than being clustered in any particular range.  There is also no obvious association between
  the detection of $\gamma$-rays and resolved X-ray jets.
  Radio-loud quasars detected in
  $\gamma$-rays have been found to participate with BL Lac objects in what has been termed
  the `blazar sequence', whereby the spectral energy distributions (SEDs)
  spanning radio and $\gamma$-ray are `bluer' as bolometric luminosity increases
  \citep{1998MNRAS.299..433F}. As with the luminosities shown in Figure
  \ref{fig:seqfig}, blazar-sequence bolometric luminosities
   are calculated
  assuming that the emission is isotropic, and the sequence
  is modelled as
  a growth of the inverse-Compton relative to lower-energy
  synchrotron hump in the SED, such that
  $\gamma$-ray emission dominates the
  luminosity for powers above about $10^{38}$ W.
  While a physical understanding of the blazar sequence and the extent to
  which selection effects contribute remain matters for debate
  \citep[see e.g.,][]{2012MNRAS.420.2899G},
  \citet{2017MNRAS.469..255G} have argued empirically from the average SEDs of radio-loud
  quasars in the {\it $\gamma$-ray-selected\/} 3LAC catalog that
  observed X-ray spectral index becomes flatter with
  increasing isotropic bolometric luminosity.  In time, more of the sources in our
  {\it radio-selected\/} sample may gather
  {\it Fermi\/} detections, and allow
  core X-ray spectral index to be looked
  at in a  statistically more meaningful way in the context of radio to
  $\gamma$-ray SED,
  luminosity, redshift, extended-jet characteristics, and beaming
  parameters.}

\clearpage

\startlongtable
\begin{deluxetable}{llrcrllcrr}
\tablewidth{0pc}
\tablecaption{Quasar X-ray Core Parameters \label{tab:cores} }
\tabletypesize{\scriptsize}
\tablehead{
\colhead{B1950}& \colhead{z} & \colhead{OBSID} & \colhead{Count Rate}& \colhead{Streak Rate} & \colhead{$N_{\rm H_{\rm Gal}}$\tablenotemark{a}}  
& \colhead{$\Gamma_{\rm x}$} & \colhead{$N_{\rm H_{\rm int}}$} & \colhead{$S_{\rm x}$\tablenotemark{b}} & \colhead{$\chi^2/$(dof)} \\
\colhead{} & & & \colhead{(cps)} & \colhead{(cps)} & \colhead{($10^{21}$ cm$^{-2}$)} & \colhead{} & \colhead{($10^{22}$ cm$^{-2}$)} & \colhead{(nJy)} & \colhead{} }
\startdata
$0106+013$ &2.099& 10380,10799 & $0.108 \pm 0.004$ & $0.09 \pm 0.06$  & 0.280 & $1.60 \pm 0.03$ & \nodata & $154\pm 4$ & 330.8/300 \\
$0144-522$\tablenotemark{e} &0.098& 10366 & $0.089 \pm 0.004$ & $0.07 \pm 0.07$  & 0.339 & $2.5 \pm 0.1$ & \nodata & $94 \pm 4$ & 7.2/16 \\
$0208-512$\tablenotemark{d,f} &0.999& 4813 & 0.305 $\pm$ 0.008& 0.54 $\pm$ 0.15& 0.294 & $1.71^{+0.03}_{-0.05}$ & \nodata & $245\pm 4$& 236.6/206 \\ 
$0229+131$\tablenotemark{d} &2.059& 3109 & 0.111 $\pm$ 0.005& 0.08 $\pm$ 0.10& 0.83  & $1.45\pm 0.07$ & \nodata & 106 $\pm$  6  & 19.3/24 \\ 
$0234+285$ &1.213& 4898 & $0.283 \pm 0.007$ & $0.30 \pm 0.09$ & 0.842 & $1.72 \pm 0.06$ &$0.5 \pm 0.1$ & $317_{-16}^{+17}$ & 60.6/73\\
$0256+075$ &0.893& 10375 & $0.155 \pm 0.005$ & $0.10 \pm 0.07$  & 1.147 & $1.34 \pm 0.11$& \nodata & $130_{-16}^{+18}$ & 30.6/28\\
$0402-362$\tablenotemark{f} &1.417& 10374 
& $1.337 \pm 0.016$ & $1.40 \pm 0.23$  
& 0.080 &  $1.08\pm 0.02$&
\nodata & $857\pm 14$ &196.9/196 \\
$0413-210$\tablenotemark{d}&0.808& 3110
&0.063 $\pm$ 0.004& 0.09 $\pm$ 0.11
& 0.239 & $1.48^{+0.10}_{-0.09}$ &
 \nodata & $57\pm 4$ & 6.3/11 \\ 
$0454-463$ &0.858& 4893 
& $0.361 \pm 0.008$ & $0.46 \pm 0.13$  
& 0.235 & $1.64\pm 0.04$ &
\nodata & $327 \pm 9$ & 85.5/64\\
$0508-220$ &0.172& 10367 
& $0.010 \pm 0.001$ & $-0.01 \pm 0.04$ 
& 0.257 & $1.8^{+0.8}_{-0.7}$&
\nodata  &$10^{+6}_{-4}$ & 0.7/1\\
$0707+476$ &1.292& 10368 
& $0.103 \pm 0.004$ & $0.11 \pm 0.07$  &
0.806 & $1.54 \pm 0.09$ &
\nodata & $90\pm6$ & 22.1/17\\
$0745+241$\tablenotemark{c}&0.410& 3111
&0.160 $\pm$ 0.006& 0.23 $\pm$ 0.16 
& 0.516 & $1.35^{+0.07}_{-0.06}$ &
 \nodata & $131^{+7}_{-6}$  & 23.5/22 \\ 
$0748+126$ &0.889& 10376 
& $0.449 \pm 0.009$ & $0.34 \pm 0.12$  &
0.360 & $1.60\pm 0.03$ &
\nodata & $385\pm9$ & 61.9/76\\
$0820+225$ &0.951& 4897 
& $0.034 \pm 0.003$ & $0.11 \pm 0.07$  
& 0.390 & $1.41^{+0.17}_{-0.16}$ &
\nodata & $27\pm 3$ & 4.8/4\\
$0833+585$ &2.101& 7870 
& $0.160 \pm 0.007$ & $0.01 \pm 0.05$  
& 0.443 & $1.45\pm0.08$ &
\nodata & $122\pm7$& 31.5/19\\
$0858-771$\tablenotemark{c}&0.490& 3112
&0.130 $\pm$ 0.005& 0.52 $\pm$ 0.21 
& 1.021 & $1.81^{+0.12}_{-0.07}$ &
 \nodata & $145^{+15}_{-7}$ & 18.1/17 \\ 
$0859+470$ &1.462& 10371 
& $0.075 \pm 0.004$ & $0.01 \pm 0.04$  &
0.192 & $1.72\pm0.09$ &
\nodata & $63\pm4$ & 12.1/14\\
$0903-573$\tablenotemark{c}&0.695& 3113
&0.123 $\pm$ 0.005& 0.00 $\pm$ 0.08& 3.212 & $1.91^{+0.14}_{-0.08}$ &
 \nodata & $199^{+26}_{-12}$ & 14.9/16 \\ 
$0920-397$\tablenotemark{d}&0.591& 5732,7220-1,7223
&0.105 $\pm$ 0.005& 0.11 $\pm$ 0.12& 2.147 & 1.63 $\pm$ 0.03        &
 \nodata & $149^{+3}_{-9}$   & 173.4/174 \\ 
$0923+392$\tablenotemark{f} &0.695 & 3048 
& $0.655 \pm 0.006$ & $0.84 \pm 0.10$  &
0.143 & $1.64\pm0.01$ &
\nodata & $514\pm5$ & 207.3/209\\
$0953+254$ &0.712& 10377 
& $0.134 \pm 0.005$ & $0.06 \pm 0.06$  &
0.267 & $1.75\pm 0.06$ &
\nodata & $121\pm5$ & 15.9/26\\
$0954+556$ &0.909& 4842 
& $0.109 \pm 0.002$ & $0.10 \pm 0.03$ &  
0.089 & $1.88\pm0.03$ & 
\nodata &  $91\pm2$ &  110.5/112\\
$1030-357$\tablenotemark{d}&1.455& 5730
&0.069 $\pm$ 0.004& 0.14 $\pm$ 0.14& 0.609 & $1.64\pm 0.04$ &
 \nodata & $52\pm 2$ & 57.8/52 \\ 
$1040+123$ &1.029& 2136 
& $0.178 \pm 0.004$ & $0.09 \pm 0.05$  &
0.287 & $1.63\pm0.04$ &
\nodata & $139\pm3$ & 58.0/62\\
$1046-409$\tablenotemark{c}&0.620& 3116
&0.196 $\pm$ 0.007&-0.01 $\pm$ 0.09& 0.829 & 
$1.72^{+0.08}_{-0.07}$ &
 \nodata & $202^{+11}_{-9}$ & 16.4/23 \\ 
$1055+018$ &0.888& 2137 
& $0.594 \pm 0.008$ & $0.58 \pm 0.13$ 
& 0.396 & $1.57 \pm 0.02$ &
\nodata & $442\pm6$ & 143.5/149\\
$1055+201$\tablenotemark{g} &1.110& 5733 
& $0.214 \pm 0.007$ & $0.12 \pm 0.08$  
& 0.182 & $1.74\pm 0.02$ &
\nodata & $247 \pm 3$ & 163.3/176\\
$1116+128$ &2.118& 10373 
& $0.083 \pm 0.004$ & $0.10 \pm 0.07$  
& 0.240 & $1.56 \pm 0.09$ &
\nodata &  $70\pm4$ & 15.0/15\\
$1116-462$ & 0.713 & 4891 
& $0.207 \pm 0.006$ & $0.32 \pm 0.11$  
& 1.040 & $1.69\pm 0.06$ &
\nodata & $210\pm8$ & 27.6/39\\
$1145-676$\tablenotemark{d,f}& 0.210 & 3117 
&0.292 $\pm$ 0.008& 0.49 $\pm$ 0.23
& 3.189 & 2.03 $\pm$ 0.06 &
 \nodata & $540\pm25$ & 119.3/111 \\ 
$1202-262$\tablenotemark{d}&0.789& 3118
&0.147 $\pm$ 0.005& 0.16 $\pm$ 0.13
& 0.708 & 1.60 $\pm$ 0.06 &
 \nodata & $137\pm 6$ & 19.5/26 \\ 
$1251-713$ &\nodata& 4892 
& $0.043 \pm 0.003$ & $-0.04\pm  0.03$ 
& 2.119 & $1.78\pm0.18$ &
\nodata & $53\pm6$ & 6.6/6\\
$1258-321$\tablenotemark{c,e} &0.01704& 3119
&0.011 $\pm$ 0.002& 0.09 $\pm$ 0.16
& 0.575 & $1.96^{+0.48}_{-0.40}$ &
 \nodata & $14^{+6}_{-3}$ & 6.5/2 \\ 
$1303-827$ & 0.870 & 10365 
& $0.132 \pm 0.005$ & $0.06 \pm 0.06$  &
0.720 & $1.60\pm0.07$ &
\nodata & $122\pm5$ & 16.0/24\\
$1343-601$\tablenotemark{c,e} &0.01292& 3120
&0.252 $\pm$ 0.007& 0.53 $\pm$ 0.27 
& 10.6  & $1.63^{+0.17}_{-0.16}$ & 
1.08 $\pm$ 0.24 & $870^{+240}_{-180}$ &35.1/35 \\
$1354+195$\tablenotemark{h,f} &0.720& 2140 
& $0.583 \pm 0.008$ & $0.93 \pm 0.15$  
& 0.223 & $1.66^{+0.07}_{-0.06}$ &
\nodata & $519\pm23$  &  165.1/142\\
$1421-490$ &0.662& 5729 
& $0.049 \pm 0.003$ & $0.07 \pm 0.06$  
& 1.457 & $1.71\pm0.04$ &
\nodata & $47\pm2$ & 72.0/62\\
$1424-418$\tablenotemark{c}&1.522& 3121
& 0.189 $\pm$ 0.007 & 0.08 $\pm$ 0.12
& 0.805 & $1.51^{+0.11}_{-0.08}$ &
 \nodata & $181^{+20}_{-11}$ & 13.8/23 \\ 
$1502+106$ &1.839 & 10378 
& $0.204 \pm 0.006$ & $0.19 \pm 0.09$  
& 0.237 & $1.54\pm0.05$ &
\nodata & $167\pm6$ & 39.2/39\\
$1622-297$ &0.815& 10370 
& $0.149 \pm 0.005$ & $0.09 \pm 0.07$  
& 1.528 & $1.38\pm0.07$ &
\nodata & $138\pm8$ & 22.7/27\\
$1641+399$\tablenotemark{f} &0.5928& 10379 
& $0.625 \pm 0.008$ & $1.14 \pm 0.16$  
& 0.104 & $1.66_{-0.02}^{+0.01}$ &
\nodata & $489\pm5$ &  238.6/245\\
$1642+690$ &0.751& 2142 
& $0.154 \pm 0.004$ & $0.12 \pm 0.06$  
& 0.434 & $1.57\pm 0.04$ &
\nodata & $119\pm4$ &  38.7/42\\
$1655+077$\tablenotemark{c}&0.621& 3122 
&0.165 $\pm$ 0.006& 0.09 $\pm$ 0.14
& 0.626 & $1.56^{+0.13}_{-0.12}$ &
 \nodata & $153^{+21}_{-16}$ & 38.3/22 \\ 
$1655-776$\tablenotemark{c,e} &0.0947& 3123 
&0.031 $\pm$ 0.003& 0.02 $\pm$ 0.11
& 0.830 & $1.03^{+0.40}_{-0.29}$ &
 \nodata & $41^{+23}_{-9}$ & 4.2/2 \\ 
$1823+568$ &0.664& 10369 
& $0.516 \pm 0.010$ & $0.64 \pm 0.16$  
& 0.411 & $1.50\pm 0.05$ &
$0.16\pm 0.05$ & $449\pm20$ & 81.6/84 \\
$1828+487$\tablenotemark{c,f} & 0.692 & 3124
&0.421 $\pm$ 0.009 & 0.73 $\pm$ 0.25 
& 0.66 & 1.61 $\pm$ 0.09 &
 \nodata & 344 $\pm$  10 & 74.6/61 \\ 
$1928+738$\tablenotemark{f} & 0.302 & 2145 
& $0.781 \pm 0.010$ & $1.08 \pm 0.16$  
& 0.800 & $1.69\pm 0.02$ &
\nodata & $705\pm10$ & 195.2/162\\
$2007+777$\tablenotemark{f} &0.342& 5709 
& $0.226 \pm 0.003$ & $0.29 \pm 0.11$  
& 0.879 & $1.80\pm 0.11$ &
$0.13\pm0.03$ & $1460\pm200$ &  177.1/189\\
$2052-474$\tablenotemark{c}&1.489& 3125
&0.113 $\pm$ 0.005&  0.48 $\pm$ 0.21
& 0.404 & $1.51^{+0.17}_{-0.13}$ &
 \nodata & $109^{+18}_{-10}$ & 10.3/15 \\ 
$2101-490$\tablenotemark{d}& 1.040 & 5731
&0.060 $\pm$ 0.003&   0.07 $\pm$ 0.11
& 0.341 & $1.79^{+0.04}_{-0.03}$ &
 \nodata & $84^{+3}_{-2}$ & 90.0/106 \\ 
$2123-463$ &1.67& 4890 
& $0.098 \pm 0.004$ & $0.07 \pm 0.06$  
& 0.330 & $1.45\pm0.07$ &
\nodata & $80\pm4$ &  19.9/21\\
$2201+315$\tablenotemark{f,i} &0.295 & 9283 
& $0.580 \pm 0.008$ & $0.75 \pm 0.13$  &
1.170 & $1.71\pm 0.04$ &
\nodata &  $586\pm16$ &  135.7/141\\
$2230+114$ & 1.037 & 10372 
& $0.473 \pm 0.009$ & $0.23 \pm 0.10$  
& 0.499 & $1.37 \pm 0.03$ &
\nodata & $374\pm76$  &  76.5/79 \\
$2251+158$\tablenotemark{d,f} &0.859& 3127  
& 0.656 $\pm$ 0.012 &2.89 $\pm$ 0.50
& 0.713 & $1.53^{+0.05}_{-0.04}$ &
 \nodata & $1225^{+270}_{-170}$  & 126.6/121 \\ 
$2255-282$ &0.926& 4894 
& $0.383 \pm 0.007$ & $0.18 \pm 0.08$  
& 0.228 & $1.61\pm 0.03$ &
\nodata & $328\pm7$ &  88.7/82\\
$2326-477$ &1.299& 4896 
& $0.199 \pm 0.005$ & $0.15 \pm 0.07$  
& 0.169 & $1.66\pm0.04$ &
\nodata  & $174\pm5$ & 46.3/54\\
\enddata
\tablecomments{Count rates and streak rates are for the OBSIDs used
  for the extended analysis.  Core spectra use deeper exposures where
  available. Errors in spectral parameters are $1\sigma$.}
\tablenotetext{a}{From the COLDEN program provided by the CXC, using
data from \citet{1990ARAA..28..215D}, except for 0229+131 \cite[from][]{1996ApJS..105..369M} and
1828+487 and 2251+158 \cite[from][]{1989AJ.....97..777E}.}
\tablenotetext{b}{$S_{\rm x}$ is the flux density at 1 keV from spectral fits.
One may roughly estimate $S_{\rm x}$ by scaling the count rate by 1000
nJy/(count/s).}
\tablenotetext{c}{Results from Paper I.}
\tablenotetext{d}{Results updated from Paper I, using either longer exposure
  and/or improved pileup or absorption model.}
\tablenotetext{e}{Not quasar -- low-redshift object.}
\tablenotetext{f}{Pileup model in {\sc xspec} has been applied.  Ratio of streak rate
to count rate is a guide to the relative importance of the pileup correction; it also depends on the window mode of the 
OBSID used.}
\tablenotetext{g}{Variability between this OBSID and 7795/7796 taken
  in a
less favourable window mode with regards pileup.}
\tablenotetext{h}{Other available OBSIDs available but in less favourable window
  modes with regards pileup.} 
\tablenotetext{i}{Structured residuals were improved significantly here with the inclusion of a themal (apec) component of
$kT \approx 0.3$ keV.}
\end{deluxetable}

\clearpage

\subsection{Imaging results}

We tested for the detection of X-rays from a jet using a simple Poisson test, as
in Paper I, for counts in a rectangular region of appropriate width extending over a specific
angular range ($\theta_i$, $\theta_o$) from the core at
a specific position angle.
The radio images were used to define the position angles and lengths
of possible jets.  Most jets are clearly defined as one-sided structures
but in a few ambiguous cases the pc-scale images were used to define
the jet direction, when available.
The parameters of the selection regions are given in Table~\ref{tab:jetresults}.
The width of the rectangle was
3\arcsec\ except for 0144$-$522, 0505$-$220,
0748$+$126, 0953$+$254, 1116$+$128, and 2201$+$315, where the
jets bend substantially, so the
rectangles were widened to 4-8\arcsec.
Profiles of the radio emission along the jets are shown in
Fig.~\ref{fig:profiles}.  In order to eliminate X-ray counts
from the wings of the quasar
core, a profile was computed at 90\arcdeg\ to the jet and subtracted
{giving the net counts, $C_{\rm net}$, in Table~\ref{tab:jetresults}}.
{Except for Q0106$+$013, there was insufficient signal to provide interesting
limits on the X-ray spectral indices of the jet without contamination by the much brighter
quasar core.
CIAO was used to extract the jet spectrum of Q0106$+$013, which was fit to
a power law (as used for quasar core fitting in \S\ref{sec:corefits}) using
isis,}\footnote{An excellent collection of {\tt isis} scripts is maintained at
{\tt http://www.sternwarte.uni-erlangen.de/isis/}.} {giving
$\Gamma_{\rm x} = 1.60^{+0.46}_{-0.27}$ and negligible $N_{\rm H}$. }
The X-ray counts in the same rectangular region defined by the radio data
were compared to a similar sized region on the opposite side of
the core for the Poisson test.
We set the critical probability for detection of an X-ray jet
to 0.0025, which yields a 5\% chance that there might be one
false detection in a set of 20 sources.
Histograms of the X-ray emission along the jets are shown in
Fig.~\ref{fig:profiles}.  The jet and counter-jet position angles
are compared, providing a qualitative view of the X-ray emission along
the jets.  No counter-jets are apparent in the X-ray images.

\begin{deluxetable}{rrrrrrrrrrrr}
\tablewidth{0pc}
\tabletypesize{\normalsize}
\tablecaption{Quasar Jet X-ray Measurements \label{tab:jetresults} }
\tablehead{
\colhead{Target} & \colhead{PA} 
	& \colhead{$\theta_i$} & \colhead{$\theta_o$} 
	& \colhead{$S_r$\tablenotemark{a}} & \colhead{$\nu_r$} & \colhead{$C_{\rm net}$} & \colhead{Count Rate} & \colhead{$S_x$} 
	& \colhead{$\alpha_{rx}$} & \colhead{$P_{jet}$\tablenotemark{a}} & \colhead{X?\tablenotemark{b}} \\
\colhead{ } & \colhead{(\arcdeg)} 
	& \colhead{(\arcsec)} & \colhead{(\arcsec)} 
	& \colhead{(mJy)} & \colhead{(GHz)} & \colhead{} & \colhead{(10$^{-3}$ cps)} & \colhead{(nJy)} & \colhead{} 
	& \colhead{} & \colhead{} }
\startdata
0106$+$013 & -175 &  1.5 &  6.0 &  487.8 $\pm$ 2.7 &  1.42 &  87 &   9.90 $\pm$  1.11 &      9.9 &     0.94 $\pm$ 0.01 & 1.00e-10 & Y \\
0144$-$522 &  100 &  1.5 & 20.0 &   15.7 $\pm$ 1.6 & 17.73 &  12 &   2.15 $\pm$  0.91 &      2.1 &     0.96 $\pm$ 0.03 & 1.30e-04 & Y \\
0256$+$075 &  100 &  1.5 &  6.0 &    5.8 $\pm$ 1.7 &  1.42 &   2 &   0.36 $\pm$  0.50 & $<$  1.9 & $>$ 0.79            & 1.85e-01 & N \\
0402$-$362 &  170 &  1.5 & 14.0 &   47.4 $\pm$ 1.6 &  8.64 &  22 &   4.12 $\pm$  2.23 &      4.1 &     0.95 $\pm$ 0.03 & 2.32e-03 & Y \\
0508$-$220 &  150 &  1.5 & 25.0 &  516.3 $\pm$15.5 &  1.42 &  -5 &  -0.94 $\pm$  0.78 & $<$  1.4 & $>$ 1.04            & 9.62e-01 & N \\
0707$+$476 &   85 &  1.5 &  6.0 &   12.7 $\pm$ 2.5 &  4.86 &  -2 &  -0.38 $\pm$  0.46 & $<$  1.0 & $>$ 0.92            & 9.08e-01 & N \\
0748$+$126 &  130 &  1.5 & 15.0 &   14.5 $\pm$ 2.1 &  4.86 &  22 &   3.92 $\pm$  1.55 &      3.9 &     0.85 $\pm$ 0.02 & 8.98e-05 & Y \\
0833$+$585 &   90 &  1.5 &  5.0 &    9.0 $\pm$ 1.4 &  1.42 &  17 &   4.51 $\pm$  1.27 &      4.5 &     0.77 $\pm$ 0.02 & 1.00e-10 & Y \\
0859$+$470 &  -20 &  1.0 &  4.0 &  114.0 $\pm$ 4.1 &  4.86 &  10 &   1.78 $\pm$  0.76 &      1.8 &     1.01 $\pm$ 0.02 & 7.63e-05 & Y \\
0953$+$254 & -115 &  1.5 & 15.0 &   30.3 $\pm$ 2.3 &  1.42 &   4 &   0.71 $\pm$  0.71 & $<$  2.9 & $>$ 0.85            & 8.39e-02 & N \\
1116$+$128 &  -45 &  1.5 & 10.0 &  101.6 $\pm$ 2.8 &  4.86 &   2 &   0.36 $\pm$  0.57 & $<$  2.1 & $>$ 1.00            & 2.15e-01 & N \\
1303$-$827 &  140 &  1.5 &  8.0 &   77.2 $\pm$ 3.2 &  8.64 &   2 &   0.36 $\pm$  0.56 & $<$  2.0 & $>$ 1.02            & 2.15e-01 & N \\
1502$+$106 &  160 &  1.5 &  9.0 &    6.4 $\pm$ 2.4 &  4.86 &   6 &   1.25 $\pm$  0.69 &      1.2 &     0.87 $\pm$ 0.04 & 2.84e-03 & Y \\
1622$-$297 & -160 &  1.5 & 10.0 &   69.9 $\pm$ 5.6 &  8.64 &   4 &   0.71 $\pm$  0.71 & $<$  2.9 & $>$ 0.99            & 8.39e-02 & N \\
1823$+$568 &   90 &  1.0 &  4.0 &  113.5 $\pm$ 6.2 &  4.86 &  60 &  10.75 $\pm$  2.20 &     10.8 &     0.91 $\pm$ 0.01 & 1.00e-10 & Y \\
2201$+$315 & -110 &  1.5 & 40.0 &  158.1 $\pm$ 5.1 &  1.42 &  27 &   2.95 $\pm$  1.35 &      2.9 &     0.94 $\pm$ 0.02 & 3.35e-04 & Y \\
2230$+$114 &  140 &  1.0 &  4.0 &   86.2 $\pm$ 6.4 &  4.86 &   6 &   1.28 $\pm$  1.63 & $<$  6.2 & $>$ 0.82            & 1.40e-01 & N \\
\enddata
\tablecomments{The jet radio flux density is measured at $\nu_r$ for the same region as for
the X-ray count rate, given by the PA, $\theta_i$, and $\theta_o$ values.  The X-ray flux
density is given at 1 keV assuming a conversion of 1 $\mu$Jy/(count/s), which is good
to $\sim$ 10\% for power law spectra with low column densities and X-ray spectral indices
near 0.5.}
\tablenotetext{a}{The quantity $P_{jet}$ is defined as the chance that there are more
counts than observed in the specified region under the null hypothesis that the counts
are background events.}
\tablenotetext{b}{The jet is defined to be detected if $P_{jet} < 0.0026$ (see text).}
\end{deluxetable}

\begin{figure}
 {\includegraphics[height=20cm]{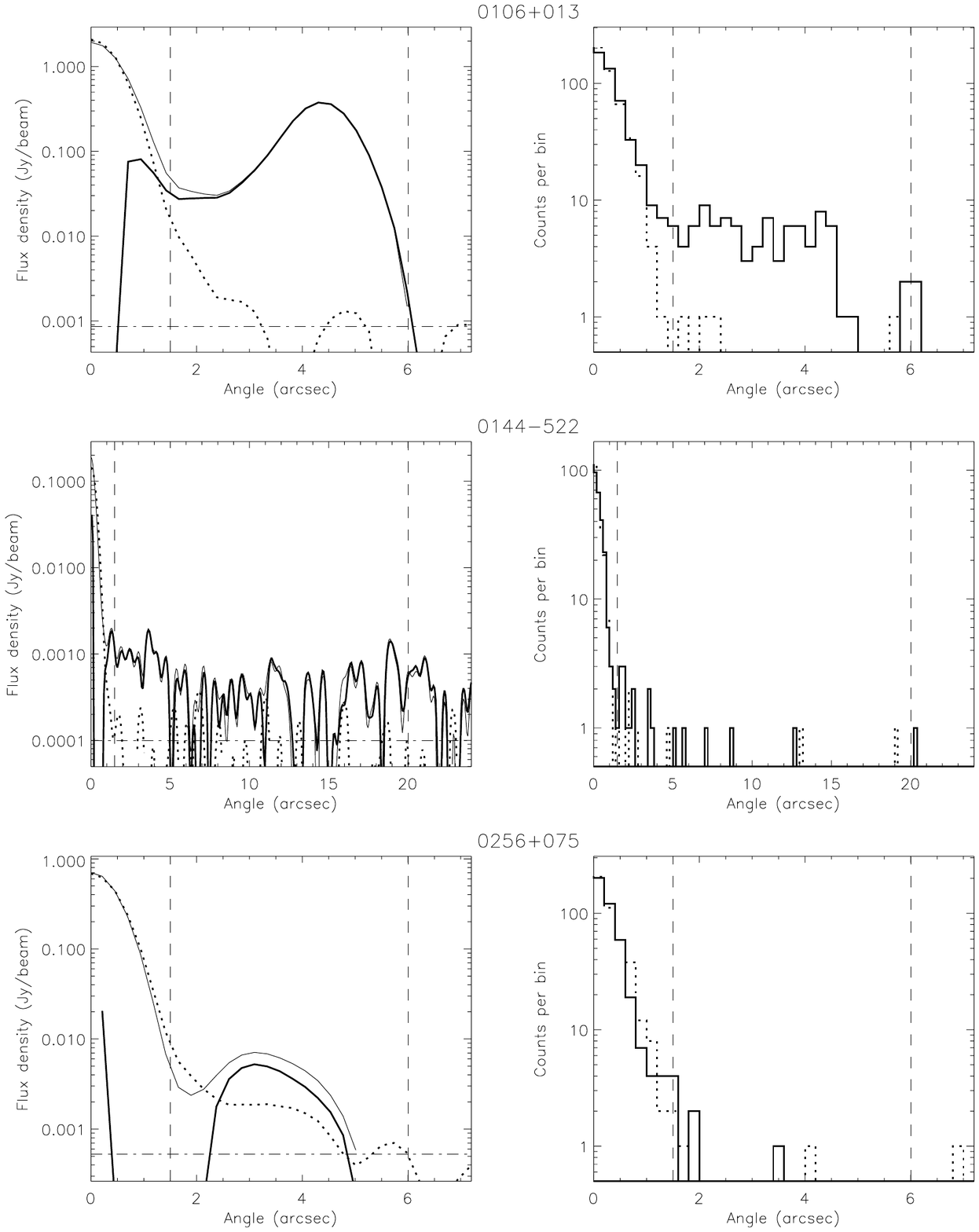}}
  \caption{Profiles of the radio emission {(left) and X-ray counts
  from Chandra (right)} along the jet.  The position
  angles of the jets are defined in Table~\ref{tab:jetresults}.
  The vertical dashed lines demarcate the jet regions.
  Dotted lines {in the radio panels} give the profiles at
  a position angle 90\arcdeg\ clockwise
  from the jet to avoid counter-jets.
  The bold, solid lines {(left)} give the differences between the profiles along the
  jet and perpendicular to it, nulling the core effectively.
  The horizontal dash-dot lines {(left)} are set
  to the average noise levels in each radio map.
  {Because there are no clearly detected
  counter-jets in the X-ray images, dotted lines (right) give the profiles at
  a position angle 180\arcdeg\ clockwise from the jet.}
  } \label{fig:profiles}
\end{figure}

\addtocounter{figure}{-1}

\begin{figure}
 {\includegraphics[height=20cm]{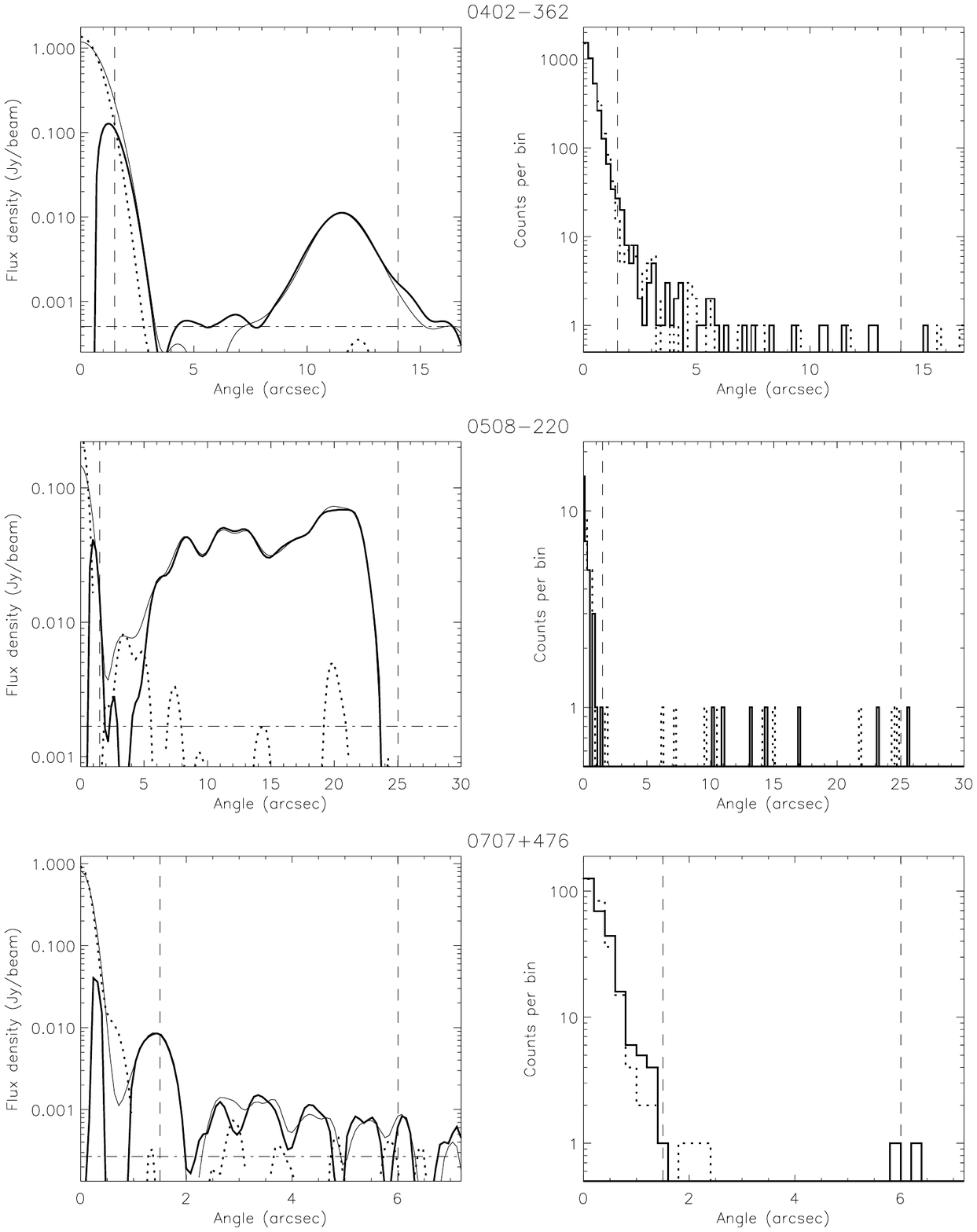}}
  \caption{continued.}
\end{figure}

\addtocounter{figure}{-1}

\begin{figure}
 {\includegraphics[height=20cm]{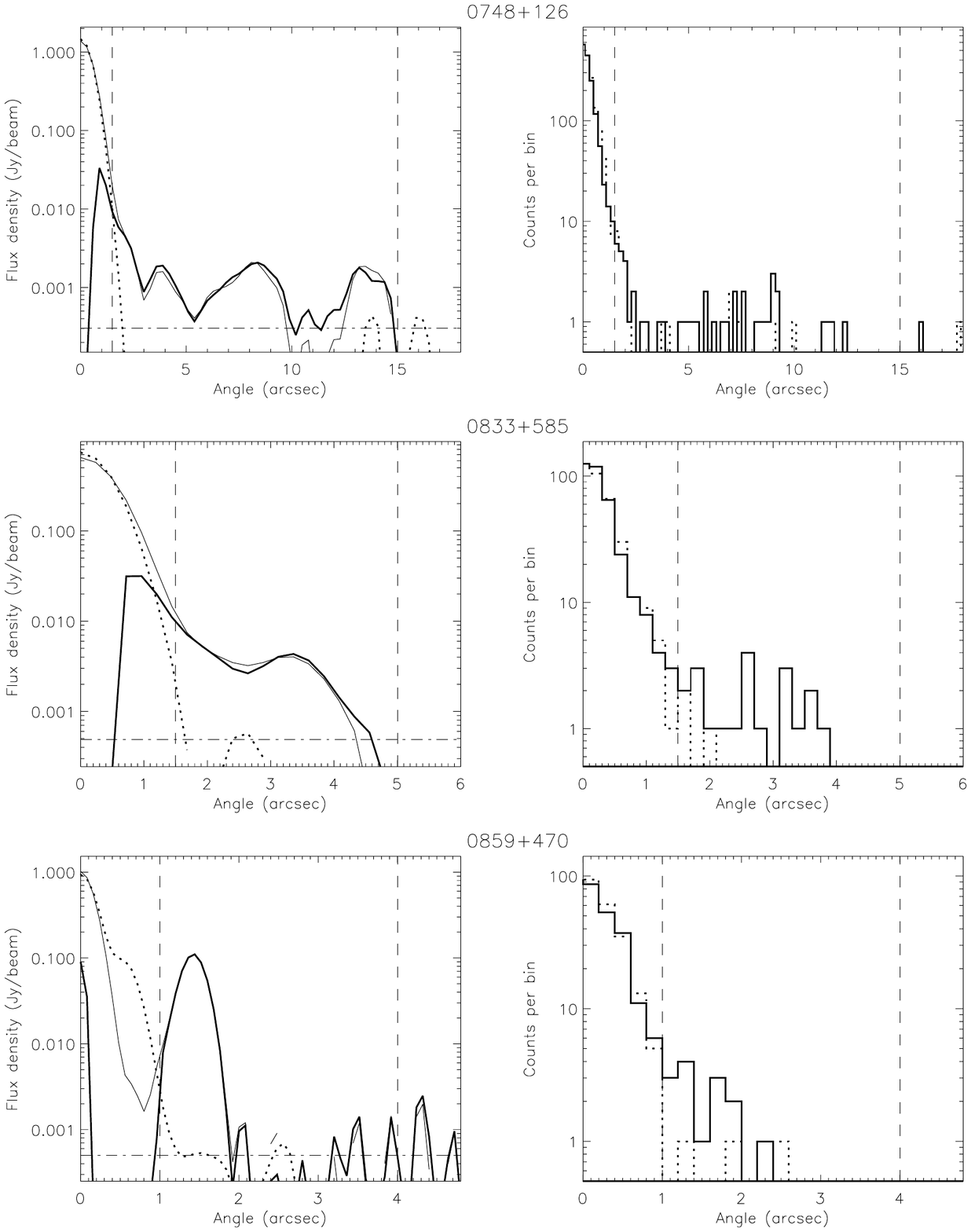}}
  \caption{continued.}
\end{figure}

\addtocounter{figure}{-1}

\begin{figure}
 {\includegraphics[height=20cm]{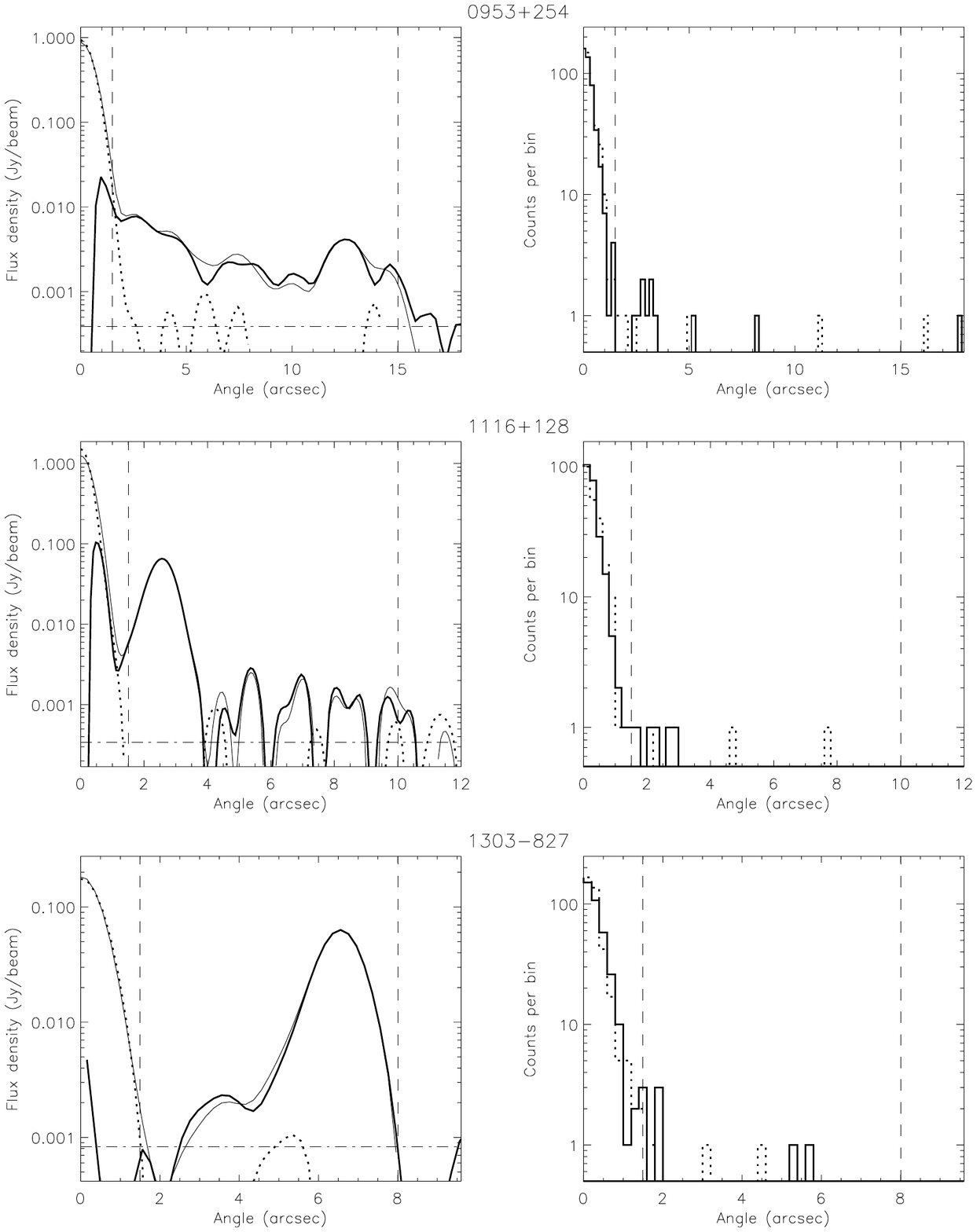}}
  \caption{continued.}
\end{figure}

\addtocounter{figure}{-1}

\begin{figure}
 {\includegraphics[height=20cm]{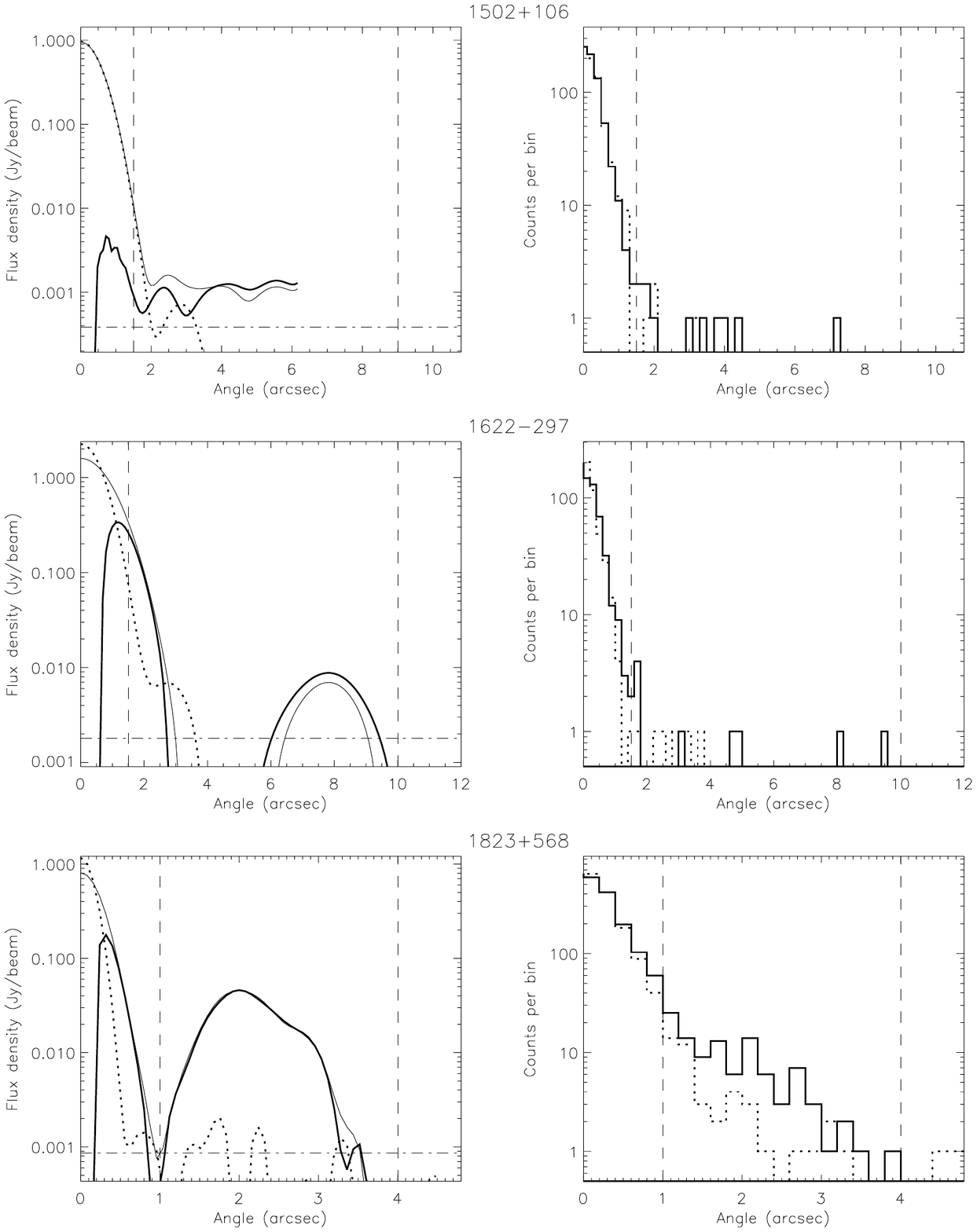}}
  \caption{continued.}
\end{figure}

\addtocounter{figure}{-1}

\begin{figure}
 {\includegraphics[height=20cm]{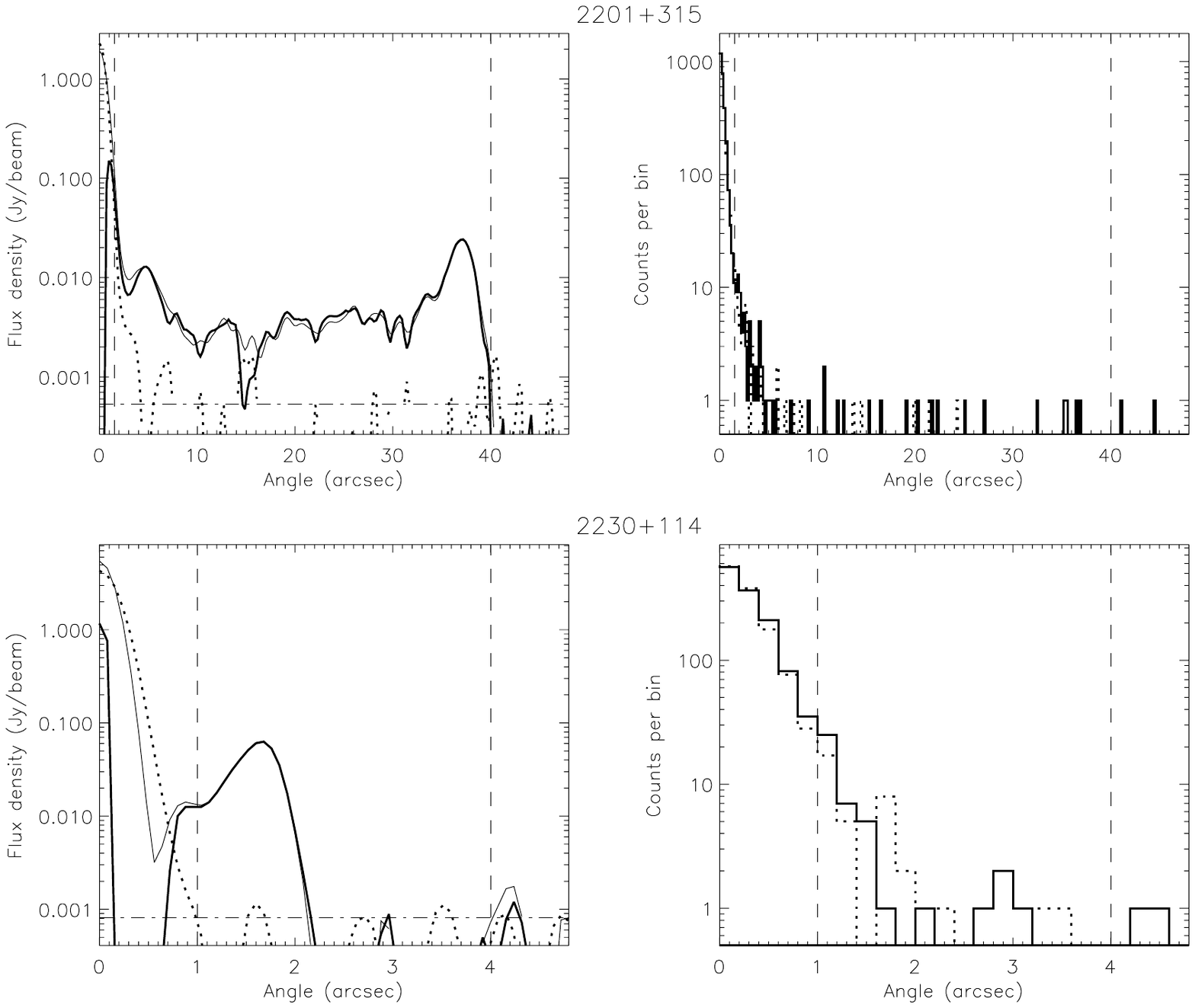}}
  \caption{continued.}
\end{figure}

Jet X-ray flux densities (Table~\ref{tab:jetresults})
were computed from count rates using the conversion factor
1 count/s $=$ 1 $\mu$Jy.  This
conversion is accurate to within 10\% for typical power law spectra.
The spectral index from radio to X-ray is computed using $\alpha_{rx} = 
-\log(S_x/S_r) / \log(\nu_x/\nu_r)$, where $\nu_x = 2.42 \times 10^{17}$ Hz
and $\nu_r$ depends on the map used.

\begin{deluxetable}{lccrc}
\tablecaption{Hubble Space Telescope Observations \label{tab:hstobs} }
\tablehead{
\colhead{Target} & \colhead{HST ID} & \colhead{Exposure (s)} & \colhead{Date}}
\startdata
  0144$-$522 & ib9402 &  2845.4 & 2010-04-12 \\
    0256$+$075 & ib9412 &  2545.4 & 2010-01-13 \\
  0402$-$362 & ib9411 &  2695.4 & 2010-07-16 \\
  0508$-$220 & ib9403 &  2695.4 & 2010-07-18 \\
    0707$+$476 & ib9405 &  2695.4 & 2009-10-04 \\
    0748$+$126 & ib9415 &  2545.4 & 2009-10-12 \\
    0859$+$470 & ib9408 &  2695.4 & 2009-10-01 \\
    0953$+$254 & ib9413 &  2695.4 & 2010-11-21 \\
    1116$+$128 & ib9410 &  2545.4 & 2010-02-21 \\
  1303$-$827 & ib9404 &  2995.4 & 2010-05-18 \\
    1502$+$106 & ib9414 &  2545.4 & 2010-03-14 \\
  1622$-$297 & ib9406 &  2695.4 & 2010-05-08 \\
    1823$+$568 & ib9407 &  2845.4 & 2009-10-12 \\
    2230$+$114 & ib9409 &  2545.4 & 2009-10-12 \\
\enddata
\end{deluxetable}

\begin{deluxetable}{lccr}
\tablecaption{IR Knot Fluxes \label{tab:hstfluxes} }
\tablehead{
\colhead{Target} & \colhead{$S_{\rm BG,noise}$} & \colhead{$S_{\rm knot}$} & \colhead{$\alpha_{ri}$} \\
\colhead{ } & \colhead{($\mu$Jy)} & \colhead{($\mu$Jy)} }
\startdata
0144$-$522 &  0.515 &  1.744 $\pm$  0.865 & $>$  0.66 \\
0256$+$075 &  0.022 & -0.053 $\pm$  0.032 & $>$  0.92 \\
0402$-$362 &  0.015 &  8.334 $\pm$  0.535 & $>$  0.72 \\
0508$-$220 &  0.014 & -0.166 $\pm$  0.025 & $>$  1.16 \\
0707$+$476 &  1.121 &  0.679 $\pm$  1.858 & $>$  0.67 \\
0748$+$126 &  0.338 & -1.260 $\pm$  0.463 & $>$  0.76 \\
0859$+$470 &  0.225 &  0.032 $\pm$  0.281 & $>$  1.10 \\
0953$+$254 &  0.046 & -0.066 $\pm$  0.081 & $>$  0.87 \\
1116$+$128 &  0.031 & -0.322 $\pm$  0.056 & $>$  1.21 \\
1303$-$827 &  0.013 &  0.017 $\pm$  0.028 & $>$  1.33 \\
1502$+$106 &  0.022 & -0.069 $\pm$  0.030 & $>$  0.92 \\
1622$-$297 &  0.027 &  0.072 $\pm$  0.048 & $>$  1.06 \\
1823$+$568 &  0.379 & -1.071 $\pm$  0.577 & $>$  0.96 \\
2230$+$114 &  0.167 & -0.626 $\pm$  0.249 & $>$  1.07 \\
\enddata
\tablecomments{
 Jet knot flux densities were measured in 0.5$\arcsec$
 radius circles
 centered at the peak of the radio emission, while the $1 \sigma$ background
 noise fluxes are from locations with comparable confusion.
 Limits to the spectral index between the radio and IR bands is
 given as $\alpha_{ri}$.
 Confusion involves quasar and stellar diffraction spikes and foreground
 galaxies (as in the case of 0402$-$362).
 These values are meant to be indicative, primarily; more accurate
 measurements for two sources are given in Tables \ref{tab:0508} and \ref{tab:0144}.
 See text for details.
}
\end{deluxetable}

\subsection{Optical Jet Measurements}

Images were obtained using the {\it Hubble} Space Telescope
Wide Field Camera 3 (WFC3) IR channel and the F160W filter
(Table~\ref{tab:hstobs}).  {
The drizzled images were examined using
{\tt SAOImage ds9 v7.6};
Fig.~\ref{fig:optical} shows the HST images overlaid with
contours from the radio maps after registering images to the cores.
Simple IR flux limits for point-like knots were determined using {\tt ds9}
by placing a 0.$\arcsec$5 radius aperture at the location of the peak
radio flux for each jet.  Often, these positions suffered from
some confusion with foreground sources, the host galaxy, or stellar diffraction spikes.
In order to take such confusion into account to first order, a background
aperture of the same size was placed in a comparable position relative
to the source of confusion -- e.g., on the opposite side of the galaxy or
diffraction spike.
The {\tt ds9 Analysis/Statistics} tool\footnote{{This tool is found in {\tt ds9 v7} as a  contextual
menu item after selecting ``{\tt Get Information...}'' from {\tt Regions} on the main menu
for each region.
A page with the features of {\tt v7} can be found at {\tt http://ds9.si.edu/doc/new.html}.}}
reports the total electron rate, $r$ (in e$^- {\rm s}^{-1}$), and the region's variance per
pixel, $V$, for $n$ pixels, where $n = 48$ for the drizzled image 
scale of 0.1283$\arcsec$ per pixel.  For $r_b$ in the background region, $r_k$ in the
knot, and associated variances $V_b$ and $V_k$, the background noise is
$ (n V_b)^{1/2}$, and the net source rate is $r_k - r_b$ $\pm$ $(n V_b + n V_k)^{1/2}$.
The result is scaled by $1.505 \times 10^{-7}$ Jy s/e$^-$, the inverse sensitivity
of the F160W filter (as found in the image header's {\tt photfnu} keyword).
The fluxes are given in Table~\ref{tab:hstfluxes}.
Without confusion, the typical flux noise for point-like sources is about 0.014 $\mu$Jy,
so a 3$\sigma$ flux limit would be about 0.05 $\mu$Jy at the so-called pivot
wavelength of 1.537 $\mu$m, or $\nu = 1.95 \times 10^{14}$ Hz.
} {
Fluxes were corrected for enclosed energy by dividing by 0.854, based
on the on-line WFC3 encircled energy table for 0.$\arcsec$5 radius apertures
at a wavelength of 1.5 $\mu$m.}\footnote{{The table is available
at {\tt http://www.stsci.edu/hst/wfc3/analysis/ir\_ee}.} }
{Limits to the spectral index between the radio and IR bands is
given as $\alpha_{ri}$, where $S_r = S_{\rm IR} \nu^{\alpha_{ri}}$.
Limits to $S_{\rm IR}$ were set to $3 \sigma_{\rm knot}$ plus the
measured flux, if positive.
In almost all cases, $\alpha_{ri}$ is greater than 0.8, a commonly observed
spectral index in the radio band for jet knots, indicating that a synchrotron model
would break between the radio and IR bands for these knots.
}

{
In two cases, 0144$-$522 and 0508$-$220, the host galaxies were sufficiently bright
that it is difficult to see jet-related emission on the images.
Purely elliptical models of each of these galaxies were created using the {\tt iraf ellipse}
tasks, after masking bright stars and galaxies that can distort the fits.
The results from fitting these two observations are shown in Figs.~\ref{fig:0144} and \ref{fig:0508}.
See the notes on individual sources for details.
}

\begin{figure}
 {\plotone{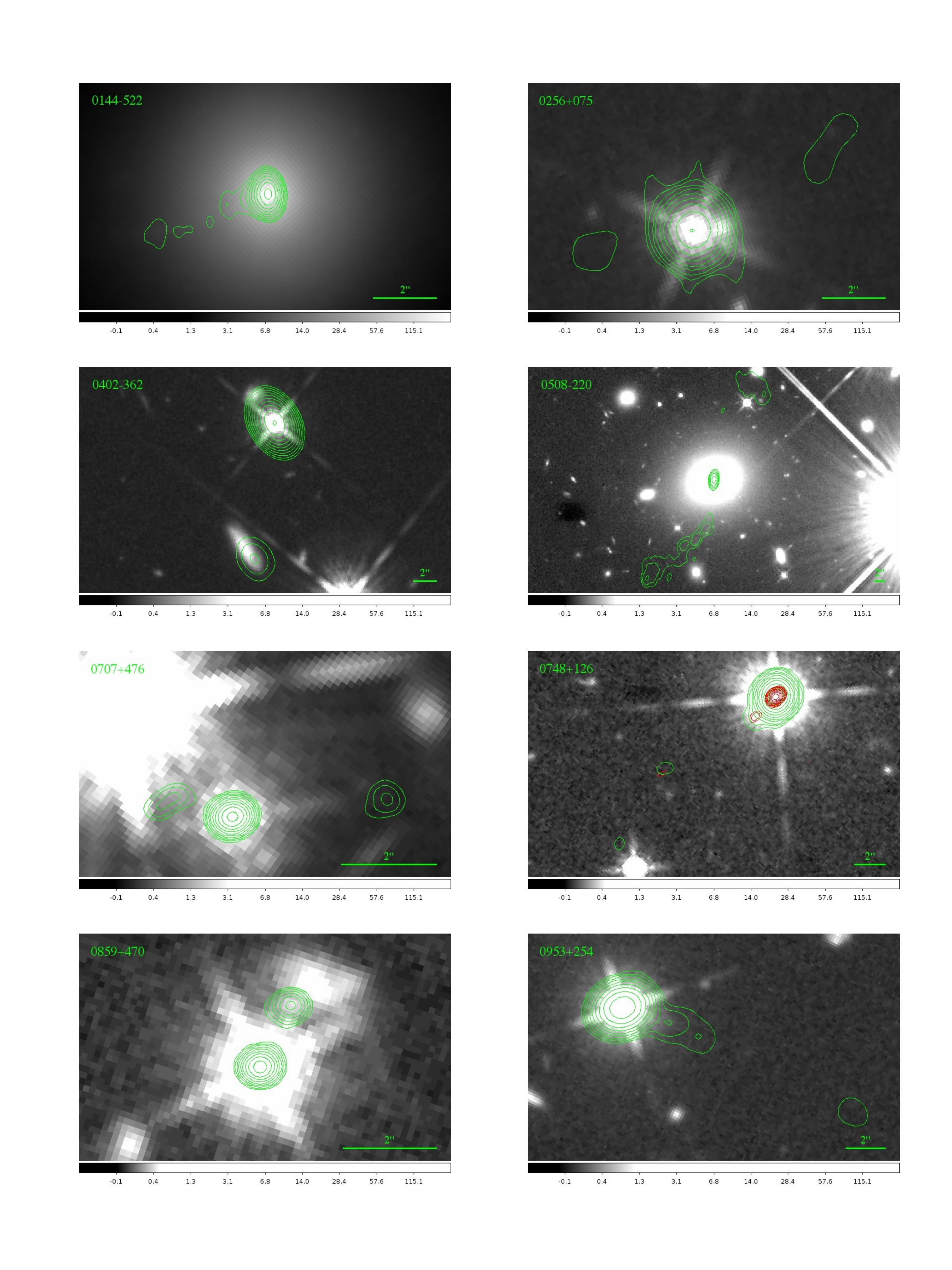}}
\caption{Overlays of radio contours on drizzled infrared images (grayscale)
in the F160W filter with the HST Wide Field Camera 3.
{Radio contours are the same as in Fig.~\ref{fig:images}.}
Even without subtracting galaxy models or point source functions
for the quasar cores, there are very few detections of kpc-scale
jet emission.  For 0144$-$522, the kpc-scale jet is observed
within the profile of the bright host galaxy and in 0859$+$470, there
is a possible feature associated with a knot about 1\arcsec\ from the
core to the northwest.
{Galaxy-subtracted images of 0144$-$522 and 0508$-$220 are shown
in Fig.~\ref{fig:0144} and Fig.~\ref{fig:0508}.}
\label{fig:optical} }
\end{figure}

\addtocounter{figure}{-1}

\begin{figure}[htp]
  \plotone{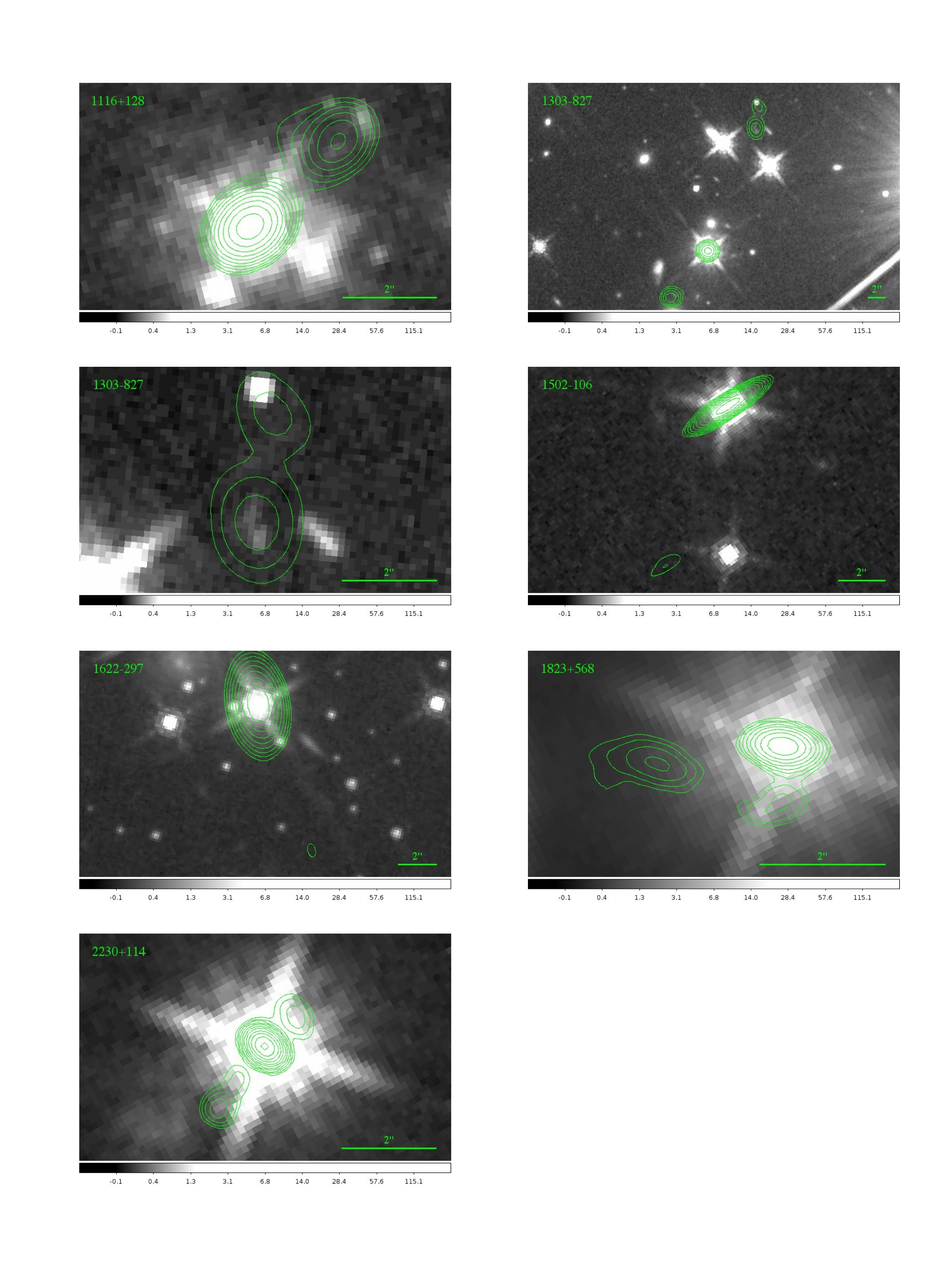}
  \caption{continued.}
\end{figure}

\subsection{Notes on Individual Sources}
\label{sec:sources}

In this section, we present qualitative descriptions of the X-ray and
radio morphologies shown in Figure~\ref{fig:images} and describe the
directions of any pc scale jets.  Profiles of the 
radio and X-ray emission along the jets are given in Figure~\ref{fig:profiles}.
All position angles (PAs) are
defined as positive when east of north with due north defining zero.
Unless noted,
VLBI imaging information comes from the MOJAVE project \cite{2009AJ....138.1874L}.

\subsubsection*{0106$+$013 (4C $+$01.02)}

The X-ray image was first published by \citet{2011ApJ...730...92H}.
The jet extends almost directly south for about 6\arcsec.  The X-ray
emission appears to end in the middle
of the terminal hotspot, which
peaks at about 4.5\arcsec\ from the core.
The quasar was not part of our observing program, so we do not
have an HST image of it.
VLBI mapping shows a jet at a PA
of -120\arcdeg\ with the greatest superluminal
motion of the quasars in our sample: 24.4$c$.

\subsubsection*{0144$-$522 (PKS B0144$-$522)}

\begin{figure}
 {\plotone{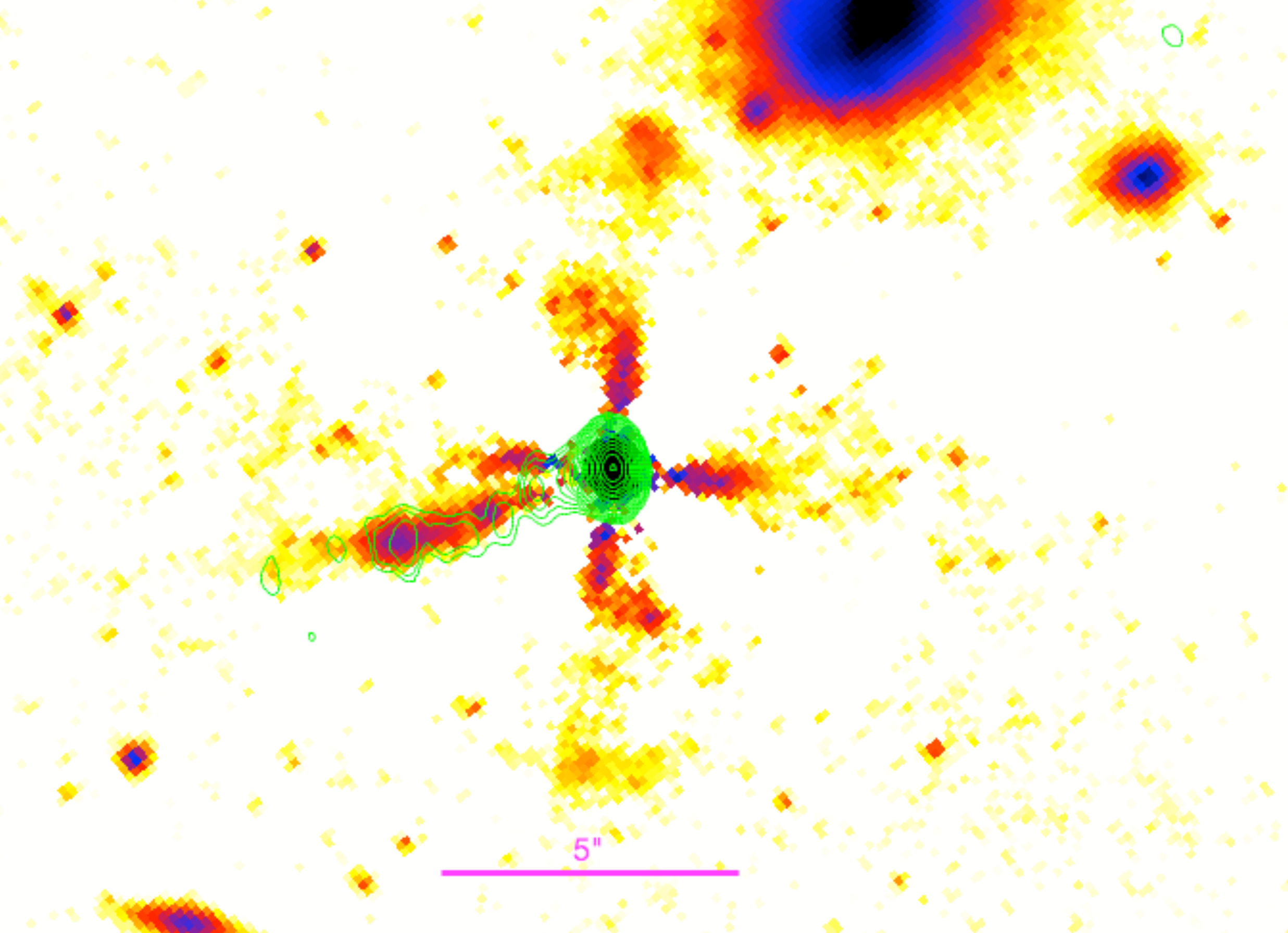}}
\caption{ {WFC3 image of 0144$-$522 after subtracting elliptical
 flux contours to eliminate the host galaxy light. No attempt was made to model
  the quasar core.  The radio contours start at 0.3 mJy/beam and increase by
  a factor of $2^{1/2}$ per contour.
 }
\label{fig:0144} }
\end{figure}

\begin{figure}[htp]
  \plotone{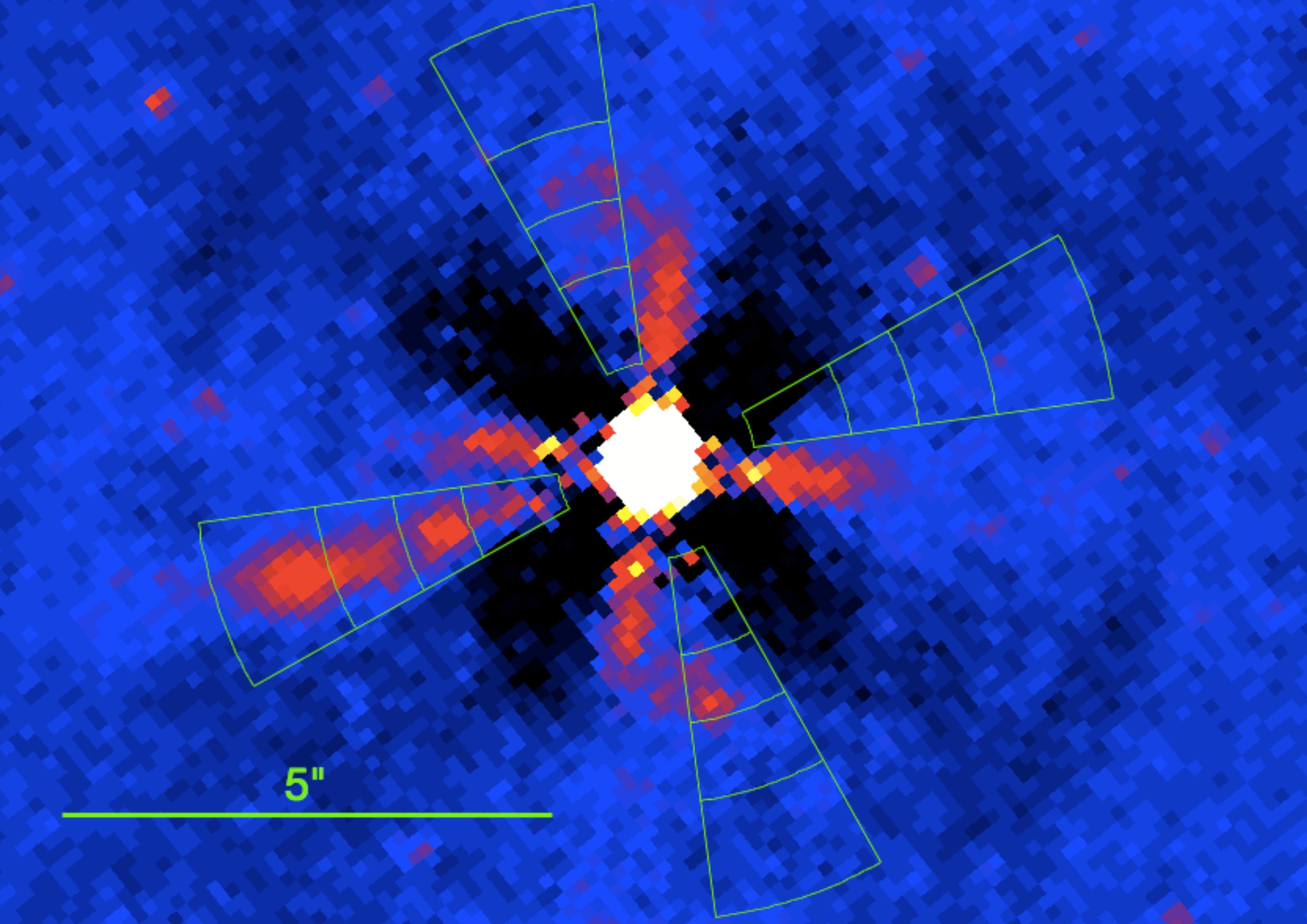}
  \caption{{Same as Fig.~\ref{fig:0144} but with
  regions used for jet analysis (to east) and for background
  (to north, west, and south).
  Fluxes in the different annular arcs of the jet are given
  in Table~\ref{tab:0144}.
  }
\label{fig:0144regions} }
\end{figure}

\begin{figure}[htp]
  \plotone{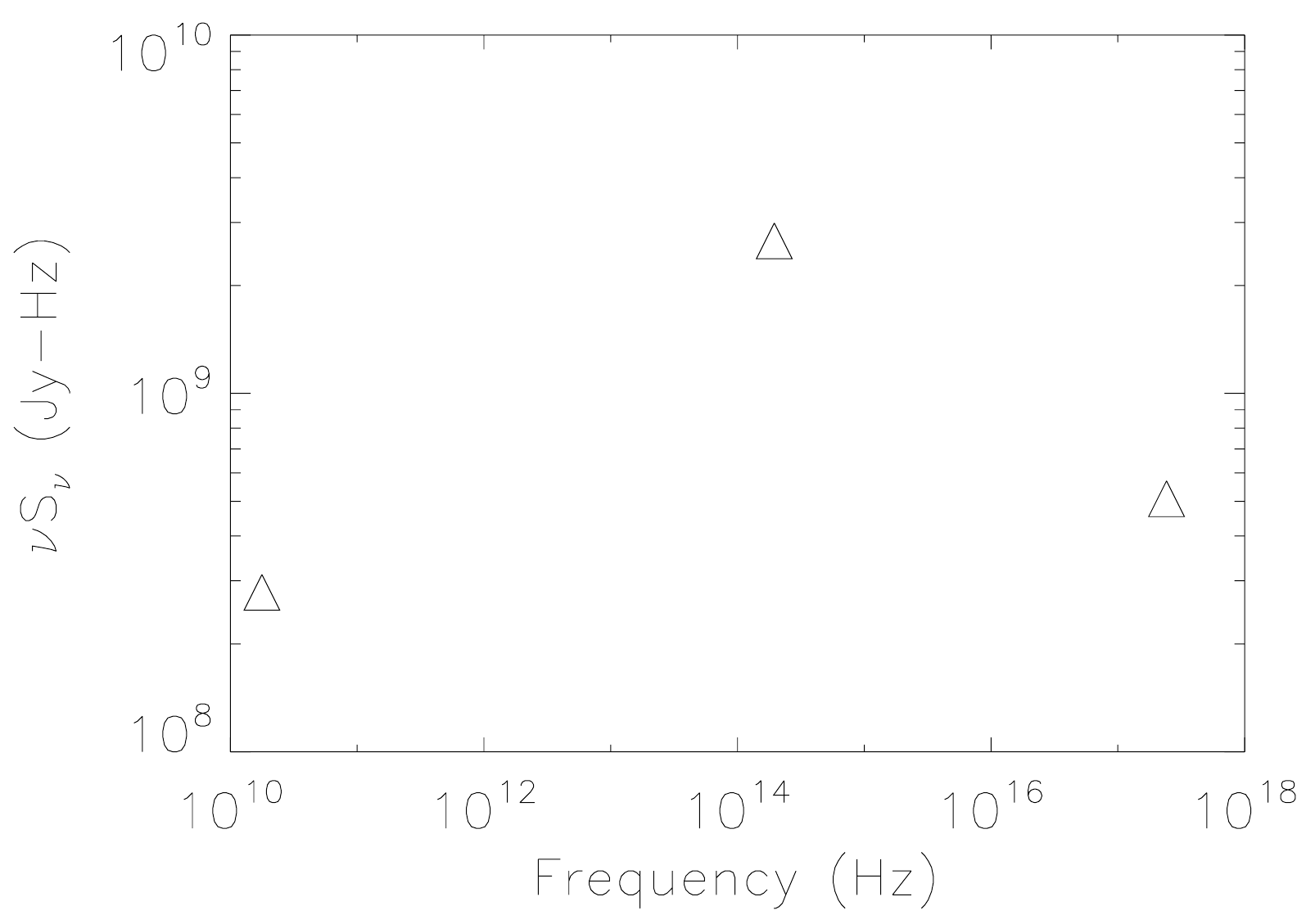}
  \caption{{Spectral energy distribution for the total
  jet in 0144$-$522.  A single-population synchrotron model would
  require a break between the IR and X-ray bands.
  }
\label{fig:0144sed} }
\end{figure}

The radio emission from the jet extends 20\arcsec\ to the east, curving northeast
while becoming significantly more diffuse and shows no hotspot,
reminiscent of an FR I type morphology but one-sided.
{Simple galaxy modeling and subtraction (Fig.~\ref{fig:0144}) was done
with ellipse fitting in IRAF}
because the host galaxy is so bright in the HST image.
There {is a clear} detection
of the inner 3\arcsec\ of the jet in both the X-ray and optical.
{The galaxy-subtracted HST data were analyzed separately using
regions defined in Fig.~\ref{fig:0144regions}, with background regions
comparably placed with regard to the core's diffraction spikes.  Results
are given in Table~\ref{tab:0144}.}
{
The regions are not circular, so we applied an estimated aperture correction
of 15\% to the measured fluxes.
Fig.~\ref{fig:0144sed} shows the spectral energy distribution (SED) that one
obtains using the radio and X-ray fluxes from Table~\ref{tab:jetresults} and totaling
the IR fluxes from Table~\ref{tab:0144}.
We find $\alpha_{ri} = 0.76$ using these data, indicating that a
single-population synchrotron model could explain the radio to IR SED.
The X-ray flux is well below the extrapolation of the radio-IR spectrum,
indicating that there must be a break in the spectrum, still consistent
with a single-population synchrotron model.
}

\begin{deluxetable}{lccr}
\tablecaption{IR Measurements of 0144$-$522 Jet Regions \label{tab:0144} }
\tablehead{
\colhead{Region} & \colhead{$\theta_i$} & \colhead{$\theta_o$} & \colhead{$S_{\rm IR}$} \\
\colhead{ } & \colhead{(\arcsec)} & \colhead{(\arcsec)}  & \colhead{($\mu$Jy)} }
\startdata
  WK1.5 &  1.0 & 2.0 &  2.15 $\pm$   0.38 \\
 WK2.35 &  2.0 & 2.7 &  2.03 $\pm$   0.34 \\
  WK3.1 &  2.7 & 3.5 &  3.26 $\pm$   0.34 \\
  WK4.1 &  3.5 & 4.7 &  6.22 $\pm$   0.67 \\
\enddata
\tablecomments{
 The jet optical flux densities were measured in the regions
 shown in Fig.~\ref{fig:0144regions},
 defined by $\theta_i$ and $\theta_o$, measured from the quasar core
 in an annular region between position angles 97$\arcdeg$ and 119$\arcdeg$ E of N.
}
\end{deluxetable}

\subsubsection*{0256$+$075 (PKS B0256$+$075)}

The VLA image shows faint, small lobes to the northeast and due west.
The latter is closer to the core and defines the region of interest for
X-ray analysis.  No X-ray emission was detected from the radio jet.
The HST image shows what may be a knot about 1\arcsec\ due west
of the core but the radio map shows no clear extension in this direction.

\subsubsection*{0402$-$362 (PKS B0402$-$362)}

The {\em Chandra} data show a marginally detected excess of flux in the box defined
to include the southern radio hotspot.
The HST image shows an edge-on spiral galaxy at the edge of the south hotspot, {so
the detection listed in Table~\ref{tab:hstfluxes} is likely to be a vast overestimate of any
IR flux from the hotspot.}

\subsubsection*{0508$-$220 (PKS B0508$-$220)}

\begin{figure}
 {\plotone{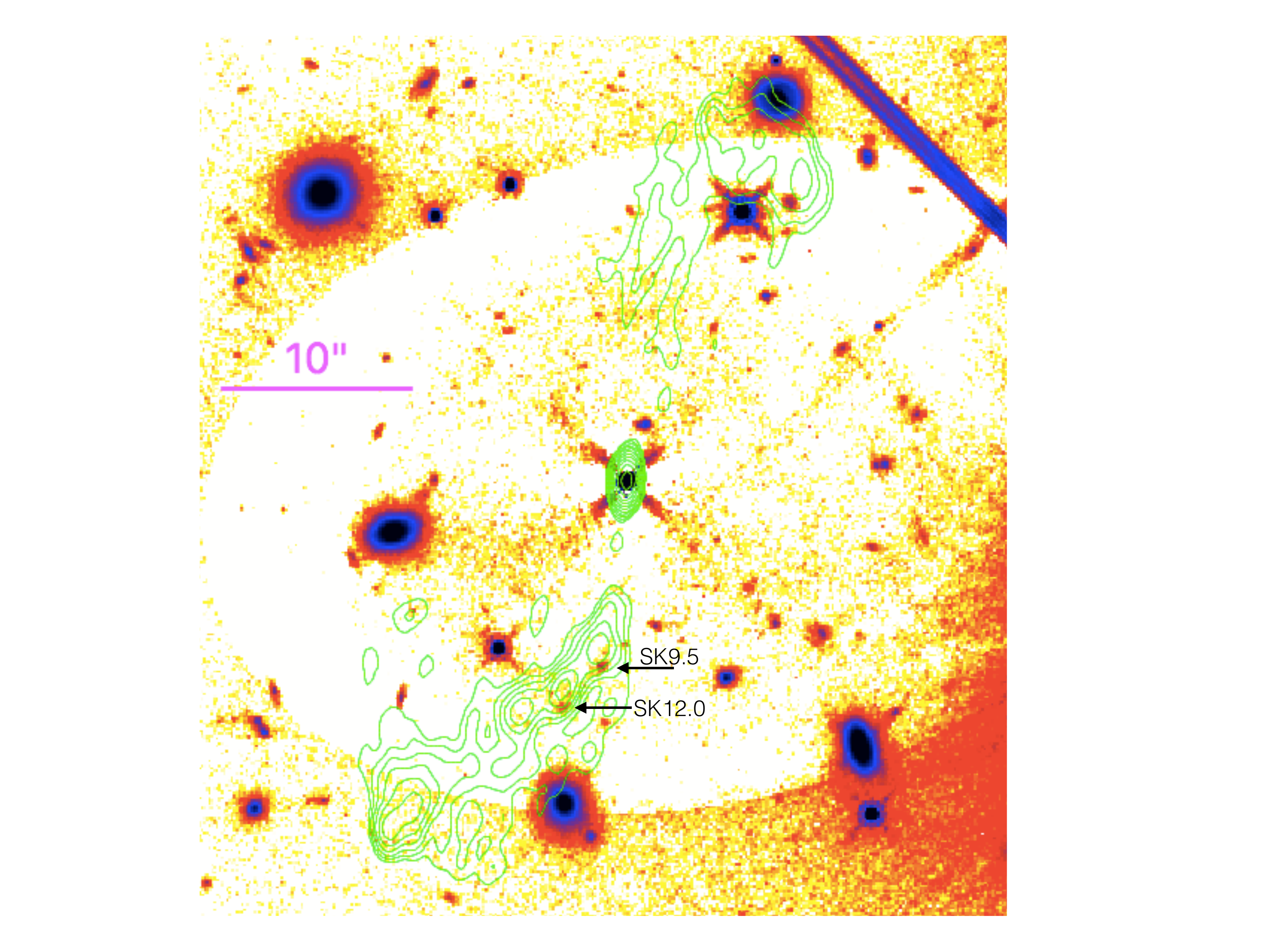}}
\caption{ {WFC3 image of 0508$-$220 after subtracting elliptical
 flux contours to eliminate the host galaxy light. No attempt was made to model
 the quasar core.  A bright star to the SW biases the contour fits, causing over-subtraction
 of the galaxy in part of the jet region.  The radio contours start at 5 mJy/beam and increase by
  a factor of $2^{1/2}$ per contour.
 }
\label{fig:0508} }
\end{figure}

The primary radio structure starts about 5\arcsec\ south of the core and curves to the
east, ending 25\arcsec\ from the core.  A lobe is found to the northwest extending
about as far as the southern lobe. The {\em Chandra} data do not
show a significant detection.  There is a rather bright elliptical galaxy at
the location of the core in the HST image { so elliptical contours were
fit in a manner similar to that for 0144$-$522.
The result of the galaxy subtraction is shown in Fig.~\ref{fig:0508}.  There
are two faint sources near radio knots about 10$\arcsec$ to the south of the
core.  The fluxes of these possible knots and their positions were measured from
the HST image using {\tt ds9} and results are given in Table~\ref{tab:0508}.  It is not clear
if these sources are related to the radio-emitting knots.}

\begin{deluxetable}{lccrr}
\tablecaption{IR Measurements of Possible 0508$-$220 Jet Knots \label{tab:0508} }
\tablehead{
\colhead{Knot} & \colhead{$\alpha$} & \colhead{$\delta$} & \colhead{$\theta$} & \colhead{$S_{\rm IR}$} \\
\colhead{ } & \colhead{(\arcdeg)}  & \colhead{(\arcdeg)} & \colhead{(\arcsec)}  & \colhead{($\mu$Jy)} }
\startdata
   SK9.5 &   77.75249 & -22.03465 &  9.49 & 1.052 $\pm$ 0.135 \\
  SK11.9 &   77.75310 & -22.03522 & 11.93 & 0.949 $\pm$ 0.069 \\
\enddata
\tablecomments{
 The jet optical flux densities were measured in 0.5$\arcsec$ radius circles
 centered at the given coordinates ($\alpha$, $\delta$),
 at distances $\theta$ from the core.
}
\end{deluxetable}

\subsubsection*{0707$+$476 (B3 0707$+$476)}

There are weak knots to the east and west of the core, with the west
one being slightly farther at about 3\arcsec\ from the core.  The jet continues
beyond the east knot about 6\arcsec\ (apparent in the radio profile),
so we somewhat arbitrarily
define the jet direction to be toward the east.  VLBI imaging shows
a pc scale jet to the northeast \citep{2004ApJ...609..539K} but
the apparent velocities given by \citet{2004ApJ...609..539K} are negative
(but marginally significant)
so we quote the absolute value and assume that the core location is not
precisely known.  The core lies quite close to a bright
star in the HST image, avoiding its southern diffraction spike but
the bright eastern knot lies practically on top of the spike so no jet
features are apparent in the HST image.

\subsubsection*{0748$+$126 (PKS B0748$+$126)}

The source has a straight radio jet to the southeast ending at a hotspot
15\arcsec\ from the core.  The PA of the VLBI jet is only 15\arcdeg\
from that of the kpc scale jet and superluminal motion
at just under 20$c$ was found.  The jet and hotspot are both clearly
detected in the 5.6 ks observation.  No features are apparent in
the HST image at any of the radio jet knots.  

\subsubsection*{0833$+$585 (SBS 0833$+$585)}

The radio jet starts out to the east but goes through an
apparent bend at a right angle about 4\arcsec\ from the
core.  The section before the bend is clearly detected
in X-rays and there is a marginal detection
also of the hotspot at the end, about 10\arcsec\ to
the southwest.

\subsubsection*{0859$+$470 (4C $+$47.29)}

The extended radio emission consists of a single knot at
3\arcsec\ northwest of the core, detected by {\em Chandra}.
There is an extended feature slightly farther out from the
core in the HST image and there appears to be
optical emission associated with the knot.

\subsubsection*{0953$+$254 (B2 0954$+$25A)}

The radio jet starts out at a PA of -115\arcdeg, as
in VLBI images, and wiggles
a few times before ending at a hotspot 15\arcsec\ from
the core.  No corresponding features are detected in the X-ray band.
Similarly, there are no clear knot associations in
the HST image.

\subsubsection*{1116$+$128 (4C $+$12.39)}

The extended radio emission consists of a single knot at
2.5\arcsec\ northwest of the core.  The knot is not
detected in the {\em Chandra} data.
No features are apparent in
the HST image at the radio knot.

\subsubsection*{1303$-$827 (PKS 1302$-$82)}

The extended radio emission consists of a single knot at
6.5\arcsec\ southeast of the core and a knot pair
about 16\arcsec\ north-northwest of the core.
The X-ray jet searched was set to the SE to include
the closest knot.
No knots are detected by {\em Chandra}.
In the HST image, there is faint, resolved flux from the
closer of the NNW knot pair, shown in a inset in
Fig.~\ref{fig:optical}.

\subsubsection*{1502$+$106 (B2 0954$+$25A)}

The VLBI jet is oriented at a PA of 120\arcdeg, while the kpc scale
jet is at a PA of 160\arcdeg\ (see Table~\ref{tab:bends}) and is marginally
detected.  The VLA image of the jet published by \citet{2007ApJS..171..376C} shows a
continuous jet all the way to the termination, about 9\arcsec\ from
the core.
\citet{2013AJ....146..120L} found a maximum apparent speed of VLBI
knots of 17.5$c$.
The HST image shows no clear detection of any part of the kpc scale jet.

\subsubsection*{1622$-$297 (PKS B1622$-$297)}

The kpc radio jet is weak and not detected with {\em Chandra}.
There is a pc-scale jet at a PA of -70\arcdeg
with maximum apparent speed of 18.6$c$ \citep{2013AJ....146..120L}.
There are no apparent features in the HST image that
are associated with radio or X-ray emission.

\subsubsection*{1823$+$568 (4C $+$56.27)}

There is a pc-scale jet at a PA of -160\arcdeg\ with
maximum apparent speed of 26.17 $c$ \citep{2013AJ....146..120L}.
The kpc scale jet is first oriented almost due south but bends
about 1\arcsec\ from the core to
the east, where it is clearly detected in the {\em Chandra} image
at 2.5\arcsec\ from the core.
There are no apparent features in the HST image that
are associated with radio or X-ray emission, although the southern
portion lies along a diffraction spike.

\subsubsection*{2201$+$315 (4C $+$31.63)}

The X-ray image was first published by \citet{2011ApJ...730...92H}.
This quasar has a VLBI components with apparent
velocities up to 8.3$c$ along a PA of -145\arcdeg\ \citep{2013AJ....146..120L}.  In the VLA image,
the jet is -37\arcsec long at a PA of -110\arcdeg,
terminating in a hotspot \citep{2007ApJS..171..376C}.
To the northeast, there is a lobe
about 45\arcsec from the core.  
The radio jet is detected in the {\em Chandra} data
only out to about 4\arcsec from the core.
The quasar was not part of our observing program, so we do not
have an HST image of it.

\subsubsection*{2230$+$114 (CTA 102, 4C $+$11.69)}

There is a pc-scale jet at a PA of 160\arcdeg\ with
maximum apparent speed of 8.6$c$ \citep{2013AJ....146..120L}.
There are radio-emitting knots to the NW and SE sides of the core
but no clear association with X-ray emission.
There are no apparent features in the HST image that
are associated with the radio knots.

\section{Discussion}

\subsection{Detection Statistics}

We detected 9 X-ray emitting jets among the 17 sources that
complete our sample.  For the remainder of the paper, we will
combine the results for the full sample of sources, as
described in Papers I and II and the present paper.
A total of 33 jets were found
with X-ray emission out of 56 sources, for a 59\% detection rate,
nearly identical to rates found in Papers I and II.
The detection fraction is unchanged if only sources
with $z > 0.1$ are considered.

Of the full sample, 30 were in the A subsample, selected on extended flux,
and 26 were in the B-only list, selected based on morphology but with
extended flux too faint for the A list limit. 
Jets were detected in 21 of the 30 sources in the A list
for a detection rate of 70 $\pm$ 8\%.
This detection rate is similar
to that obtained by \citet{2004ApJ...608..698S} and in Paper I.
The jet detection rate for the B-only subsample
is not as high: 12 of 26 jets are detected (46 $\pm$ 10\%).
However, at the 90\% confidence level, we cannot rule out the
possibility that the A and B subsample detection probabilities
are the same.

\subsection{Modeling the X-ray Jet Emission}

A hypothesis that bears testing with these data is that the
X-ray emission results from the inverse Compton scattering of CMB
photons by relativistic electrons and that the bulk motion
of the jet is highly relativistic and aligned close to the line of sight.

\subsubsection{Distribution of $\alpha_{rx}$ and Redshift Dependence}

\begin{figure}
{\includegraphics{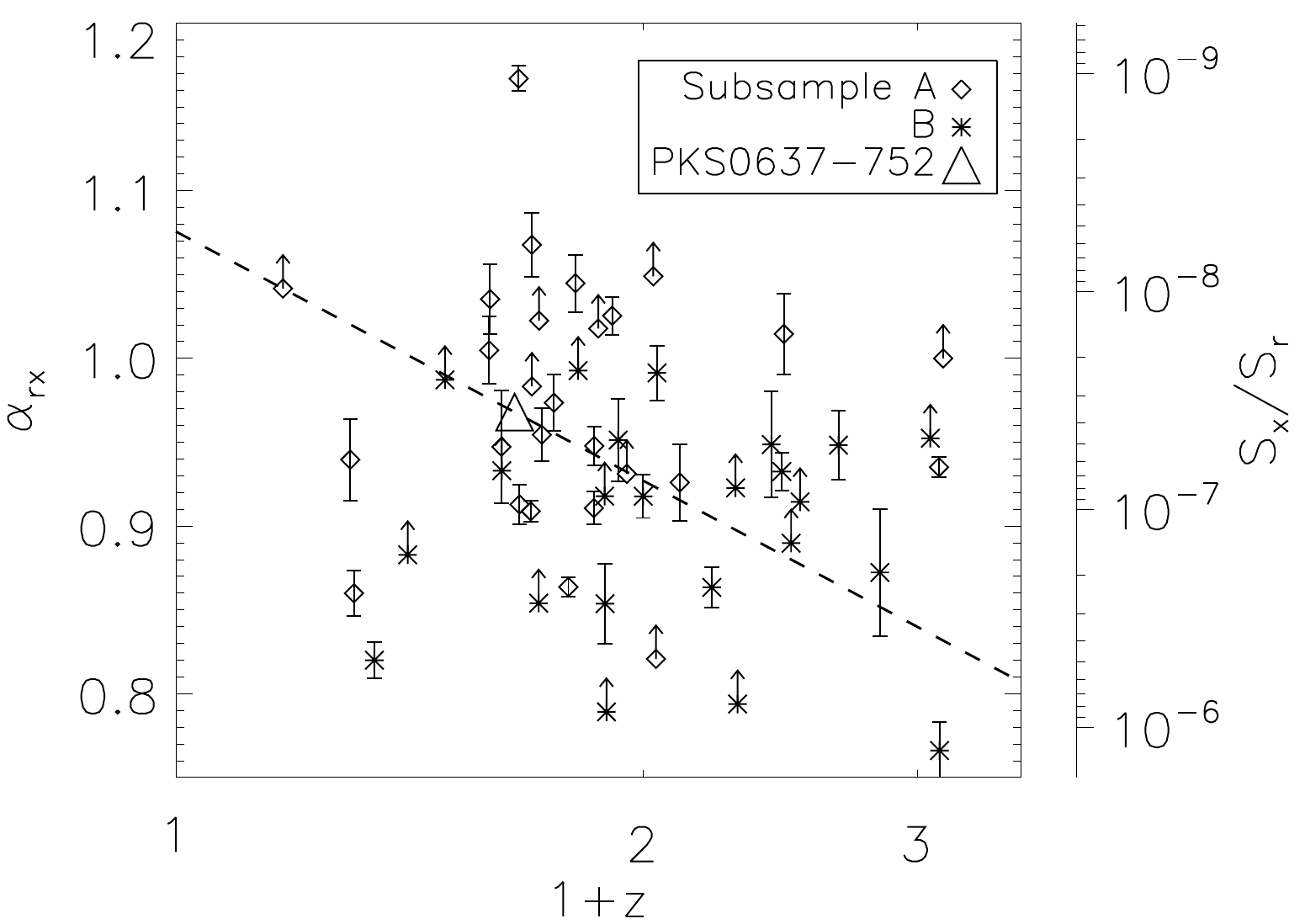}}
\caption{Plot of $\alpha_{rx}$ against redshift.
A value of $\alpha_{rx}$ of 1.0 indicates that there is
equal power per logarithmic frequency interval in both the X-ray
and radio bands.  The right-hand axis gives the ratio of the
X-ray and radio flux densities ($S_x$ and $S_r$).
As a reference, the result for PKS 0637-752 is indicated.
The dashed line gives the dependence of $\alpha_{rx}$ on $z$ under
the assumptions that the X-ray emission results only from inverse Compton
scattering off of the cosmic microwave background and that the beaming parameters
for all jets are the same as those of PKS~0637$-$752.  In this model,
the X-ray to radio flux density ratio would increase as
$(1+z)^{3+\alpha}$ (where we assume $\alpha = 0.5$)
but such a dependence is not apparent.
\label{fig:alpharx-z} }
\end{figure}

Values of $\alpha_{rx}$, defined by $S_x = S_r (\nu_r/\nu_x)^{\alpha_{rx}}$, are given in
Table~\ref{tab:beaming} and shown in fig.~\ref{fig:alpharx-z} as a function of $z$.
We use values of $\nu_r$ from Table~\ref{tab:radioCont}
and $\nu_x = 2.4 \times 10^{17}$ Hz.
A change of about 0.13 in $\alpha_{rx}$ results
from a $\times 10$ change in the X-ray flux relative to the radio flux.

\begin{figure}
\includegraphics{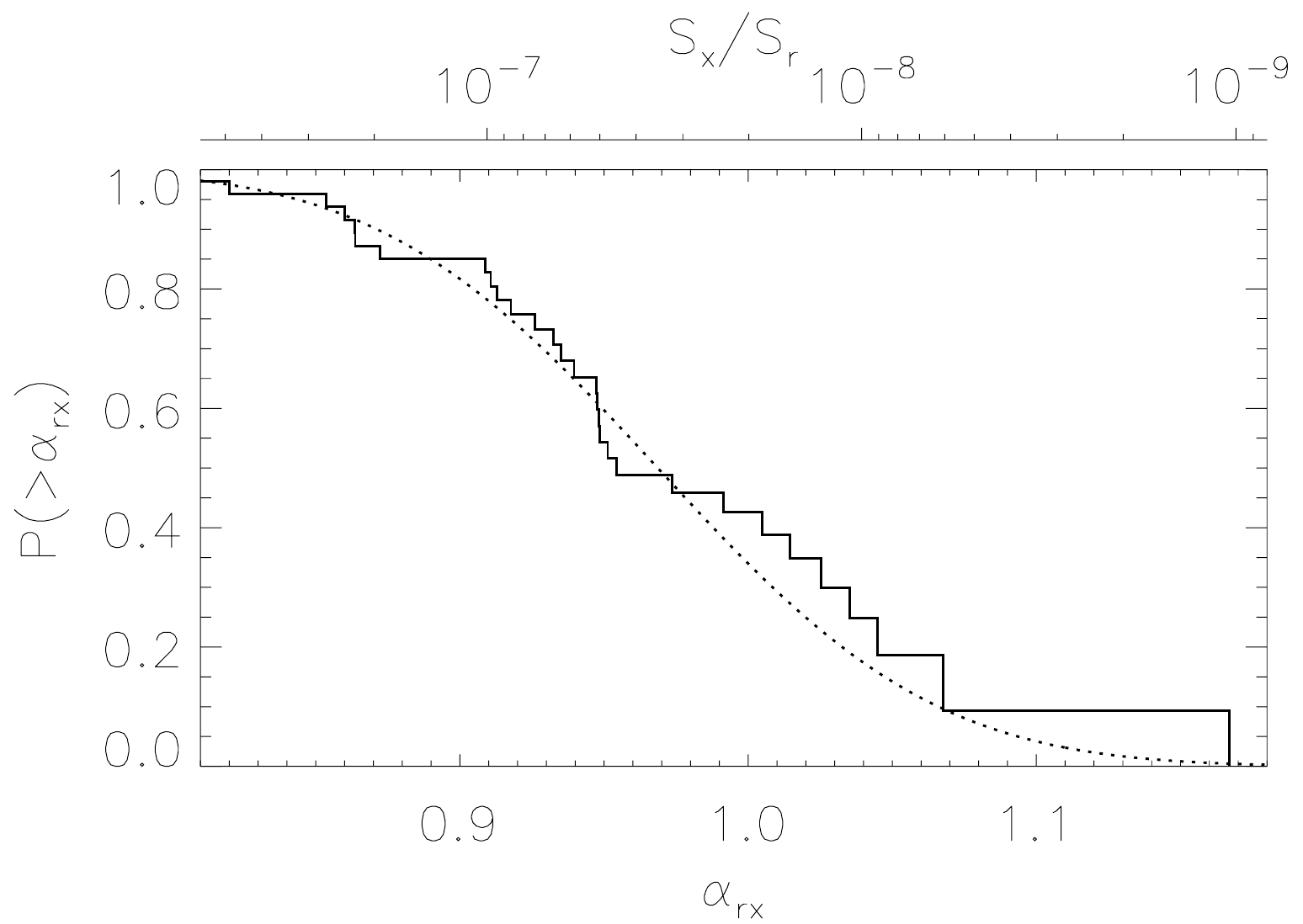}
\caption{Distribution of $\alpha_{rx}$ for
our sample.  Upper limits are handled by using the Kaplan-Meier
method.  {\it Dotted Line:} Model obtained from integrating
a normal distribution of $\alpha_{rx}$ with mean 0.974 and $\sigma$
of 0.077.
\label{fig:arxdistribution} }
\end{figure}

\begin{figure}
\epsscale{0.8}
\includegraphics[scale=0.8]{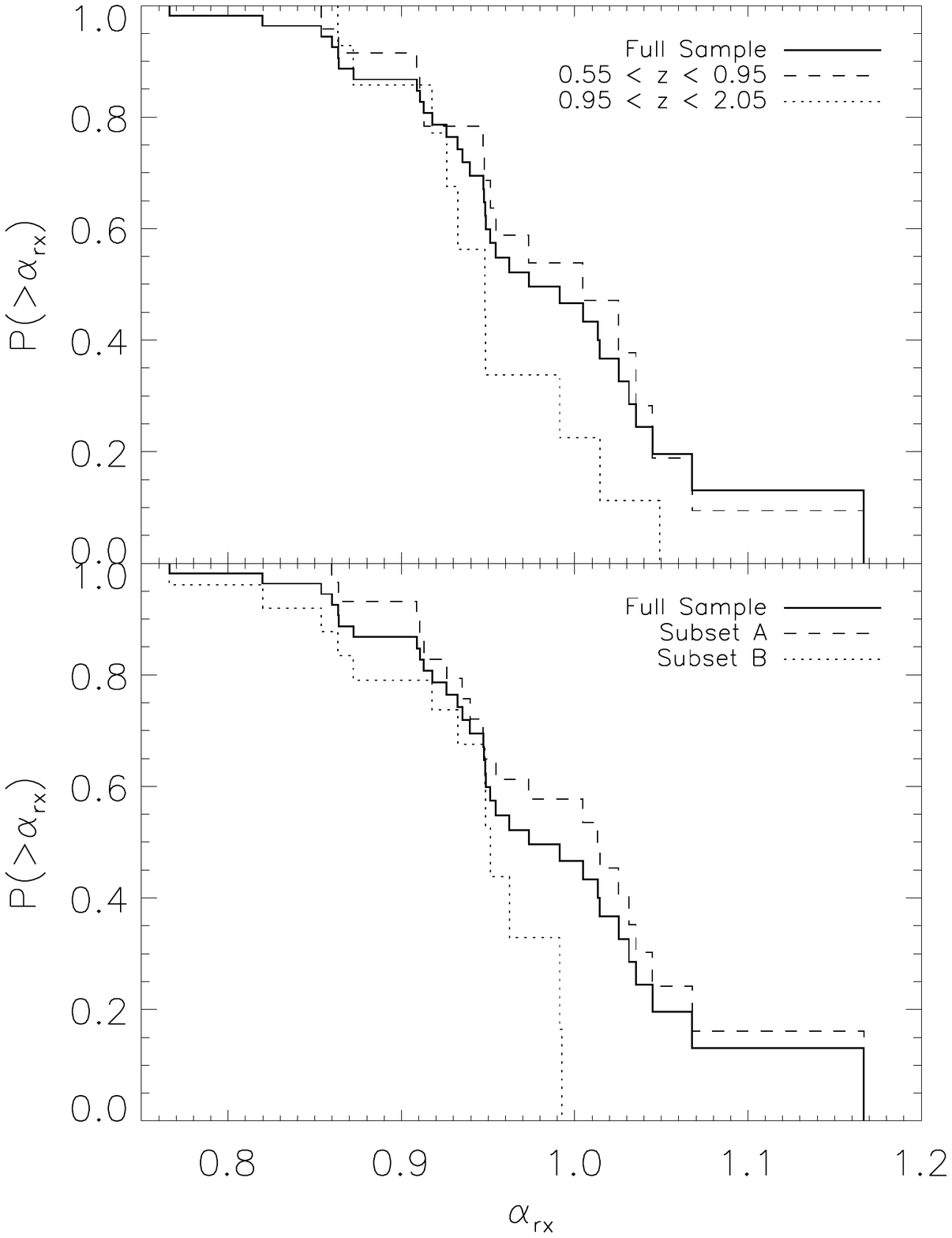}
\caption{Distribution of $\alpha_{rx}$ for
our sample, split into subsets by two different
criteria.  Upper limits are handled by using the Kaplan-Meier
method.  {\it Top:} The sample is divided by
redshift (excluding one with an unknown redshift).
The high $z$ subsample has marginally smaller values of
$\alpha_{rx}$; i.e., the jets' X-ray flux densities are slightly larger relative to their
radio flux densities than for the low $z$ sample.
{\it Bottom:} The sample is divided according to the A or B category
(see \S\ref{sec:sample}).  The B subset shows slightly smaller values
of $\alpha_{rx}$, than the A subset but we cannot rule out
that A and B targets were detected at the same rate at the
90\% confidence level.
\label{fig:arxdistribution2} }
\end{figure}

The observed distribution of $\alpha_{rx}$, using the likelihood method that includes
detections and limits (from Paper II), is shown in Fig.~\ref{fig:arxdistribution}.
The unbinned distribution was fit with a likelihood method to a Gaussian; the
best fit mean was $ 0.974 \pm 0.012$ and the dispersion was $\sigma = 0.077 \pm 0.008$.
The excellent fit indicates that $S_x/S_r$ follows a log-normal distribution well
but with large dispersion -- a FWHM of a factor of 24.
The dependence of the $\alpha_{rx}$ distribution on $z$ or selection
criterion (A or B) is weak, as found in Paper II
and shown in fig.~\ref{fig:arxdistribution2}.

\begin{figure}
\includegraphics[angle=180,scale=0.75]{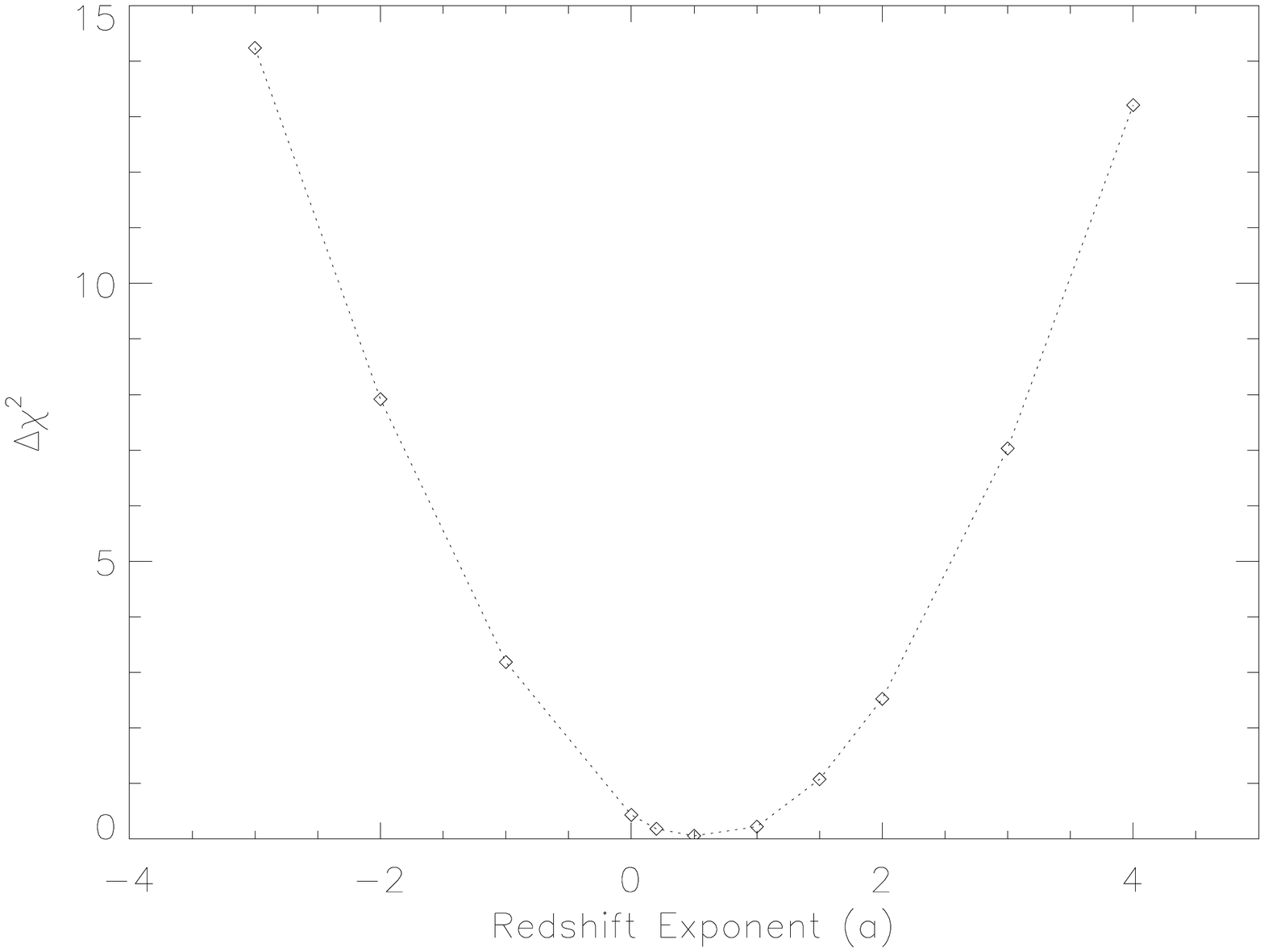}
\caption{Dependence of $\Delta \chi^2$ on $a$, where $S_x/S_r \propto (1+z)^a$,
assuming that the distribution of intrinsic magnetic
fields is not redshift-dependent.
In the IC-CMB model, $a = 3+\alpha$;
this dependence is ruled out at better
than 99.5\% confidence for $\alpha > 0.5$. 
\label{fig:arxtest} }
\end{figure}

As in Paper II, we
define a quantity that is derived from the observed data for each source,
$Q \equiv R B_1^{1+\alpha}$, where
{$R =
  S_{\rm x} \nu_{\rm x}^\alpha / (S_{\rm r} \nu_{\rm r}^\alpha)$, $\alpha$ is the spectral
index in the radio band, and} $B_1$ is the minimum energy magnetic
field strength in the rest frame of the jet under the assumption that relativistic beaming
is unimportant (i.e., the bulk Doppler factor $\delta = 1$).
{As in Paper I, $B_1$ \citep[defined originally by][]{2002ApJ...565..244H} 
is computed using observables such as the
luminosity distance to the source $d_L(z)$,
the observed radio flux density, and the angular size of the emission region (as given
in Table~\ref{tab:jetresults}), and is mildly dependent on assumed or estimated
quantities such as $\alpha$, the frequency limits of the synchrotron spectrum,
the filling factor, and baryon energy fraction.  For
this paper, we assume that all quantities except
$d_L$ that are required to compute $B_1$ are independent of redshift.
Under this assumption, $Q \propto (1+z)^{3+\alpha}$
in the IC-CMB model, as shown in Paper II.}
However, our fit to $Q \propto (1+z)^{a}$
gives $a = 0.88 \pm 0.90$ (at 90\% confidence, Fig.~\ref{fig:arxtest}).
We find $a = 3+\alpha$ is rejected at better than 99.5\% confidence
for $\alpha > 0.5$.
{Furthermore, we also reject $a=2$ at better than 90\% confidence;
this value of $a$ would result from an implicit dependence of $B_1$ on $z$ if
path lengths through jets are independent of $z$, as shown in Paper II.}
Thus, if the IC-CMB mechanism
is responsible for most of the X-ray emission from quasar jets, then other jet
parameters such as the magnetic field or Lorentz factor must depend on $z$ or $d_L$
in a compensatory fashion.

\subsubsection{Angles to the Line of Sight}

As in Papers I and II, we computed the angles to
the line of sight for these kpc scale jets
under the assumptions that 1) X-rays arise from the
IC-CMB mechanism, and 2) all jets have a common
Lorentz factor, $\Gamma$.
For $\Gamma = 15$, we find that $\theta$ ranges from
6\arcdeg\ to 13\arcdeg\ for the quasars in our sample
(see Table~\ref{tab:beaming}).
We can also determine limits to $\Gamma$ and $\theta$ under
the assumption of the IC-CMB model.
From the IC-CMB solution in Paper I, \citet{2011ApJ...730...92H}
determined that there is a minimum $\Gamma$ for a detected
source that is associated with $\theta = 0$ (or $\mu \equiv \cos \theta = 1$):
$\Gamma_{\rm min} = K/[2(K-1)^{1/2}]$, where $K$ is a combination
of observables (also listed in Table~\ref{tab:beaming}; see Paper I).
Similarly, there is a solution to the IC-CMB equation
for $\beta$ for an assumed value of $\theta$:

\begin{equation}
\label{eq:beta}
\beta = \frac{-1-\mu +2 K \mu \pm (1+2 \mu  - 4 K \mu + \mu^2 
	+ 4 K \mu^3)^{1/2}}{2 K \mu^2}
\end{equation}

\noindent
\citep{2006ESASP.604..643M} and there is a maximum value of $\theta$,
$\theta_{\rm max}$, which is obtained by finding $\mu$ for which the term in parentheses
in Eq.~\ref{eq:beta} is zero; for large $K$, $\theta_{\rm max} \approxeq (K)^{-1/2}$.
These limiting values of $\Gamma$ and $\theta$
are given in Table~\ref{tab:beaming} for cases where the
kpc-scale jet was detected.  The uncertainties on $K$ are typically 10-20\%,
giving uncertainties in $\theta_{\rm max}$ and $\Gamma_{\rm min}$ of
5-10\%, so $\Gamma > 1.3$ at 95\% confidence for all X-ray detections.

We then compare these angles to the range of angles that would be
inferred using information from pc-scale jets observed in VLBI studies.
This method is described in appendix C of Paper II.
Briefly, the method assumes that one may use the core's maximal apparent superluminal (SL)
motion, $\beta_{app}$, to estimate the angle of the pc-scale jet to the line
of sight via $\sin \theta_{\rm pc} \approx (2 \beta_{\rm app})^{-1}$.
Then, using
the observed position angles of the pc- and kpc-scale jets, $\phi$, we determine
the 10\% probability limits on the intrinsic angle to the line of sight for
the kpc-scale jet, $\theta_{\rm kpc}$.  These values are given in
Table~\ref{tab:bends}, along with the most probable value.
{We note that eq. C2 of Paper II was given incorrectly, and should have read}

\begin{equation}
{ \tan \zeta = \frac{\tan \eta \sin \theta}{ \sin \phi  - \tan \eta \cos \phi \cos \theta}  }
\label{eq:tanzeta}
\end{equation}

\noindent
{where $\zeta$ is the magnitude of the jet bend in the frame
of the quasar host galaxy, $\phi$ is a phase angle
giving the rotation of the bent jet about the axis defined by the jet before the
bend, and $\eta$ is the apparent bend, as projected on the sky (see Paper II, Appendix C.)
However, the change has no effect,
because the values in the tables here and in Paper II were actually computed using
this correct expression (i.e., Eq. \ref{eq:tanzeta}).
}

\begin{figure}
\includegraphics{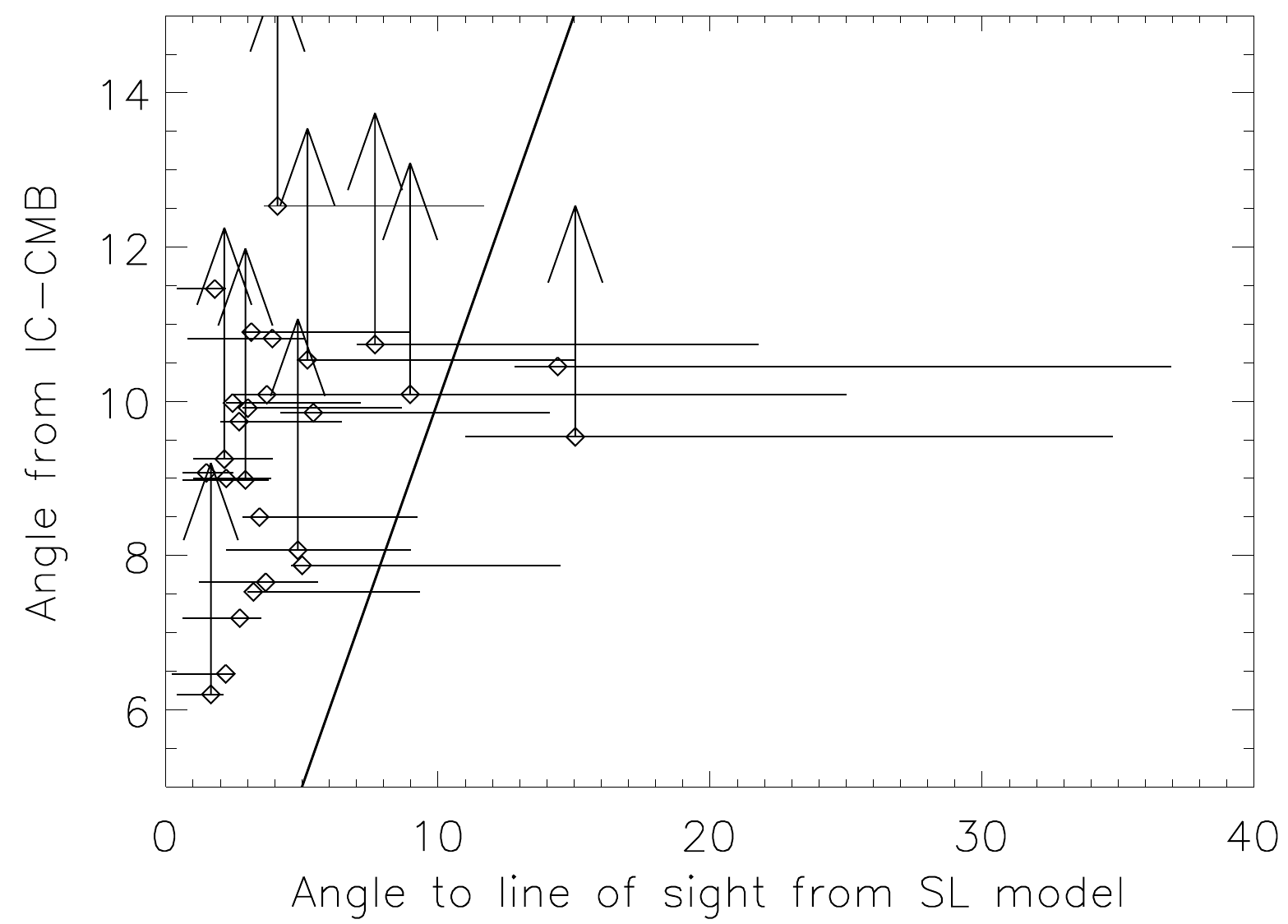}
\caption{Comparison of angles of kpc-scale jets to the line of sight for two computation
methods -- one using VLBI observations of superluminal (SL) motion ($\theta_{\rm kpc}$
in Table~\ref{tab:bends}) and the
other assuming that the kpc-scale X-rays result from the IC-CMB model with $\Gamma = 15$
($\theta$ in Tables~\ref{tab:beaming} and \ref{tab:bends}).
The abscissa is determined from geometric constraints using the
difference between the position angles of the
pc-scale and kpc-scale jets by the method described in Paper II.
The IC-CMB model is used to derive an angle to the
line of sight.  The solid line indicates
where these two angles are equal.  Angles from
the IC-CMB calculation are generally $\times$2 larger than those based on geometry and
superluminal motion of the pc-scale jet.  Thus, one may infer that the
jets decelerate substantially from pc scales to kpc scales, given that it is highly
unlikely that that jets predominantly bend away
from the line of sight between pc and kpc scales or that $\Gamma$ reaches values
of order 50 on kpc scales.
\label{fig:angles} }
\end{figure}

Values of the PA of the VLBI component and its maximum
apparent transverse velocity, $\beta_{\rm app} c$, are taken
from \citet{2013AJ....146..120L}, and \citet{2016AJ....152...12L}.\footnote{Values
in Paper II were obtained from the
the MOJAVE web site, {\tt http://www.physics.purdue.edu/astro/MOJAVE/index.html},
which have been updated.}
For 1354$+$195, we  estimated a maximum apparent speed
of $243 \pm 1~ \mu$arcsec yr$^{-1}$
based on 4 temporally-spaced epochs from the
MOJAVE 15 GHz VLBA archive \citep{2009AJ....137.3718L}.
The range of $\theta_{\rm kpc}$ is plotted against $\theta$ from the IC-CMB
calculation in Fig.~\ref{fig:angles}.  This figure shows that the method
based on SL motion of the pc-scale jets and their bends give significantly
smaller angles to the line of sight than the IC-CMB method.

We can bring $\theta$ into closer agreement with $\theta_{\rm kpc}$ by
increasing the value of $\Gamma$.  Our choice of $\Gamma = 15$ was informed
by population modeling \citep{2007ApJ...658..232C} of
superluminal sources; on pc-scales, $\Gamma$ appears
to have a broad distribution between 0 and 30.
To reduce the IC-CMB angles requires increasing $\Gamma$ because $\delta$ is
approximately fixed by the IC-CMB model but $1-\beta \cos[\theta]$ approaches 0 faster
than $\Gamma^{-1}$ can compensate.
We find that $\Gamma > 100$ is needed to achieve $\theta < \theta_{\rm kpc}$ for half of the sample.
If instead we require that at least half of the IC-CMB angles
be below the {\bf maximum} allowed $\theta_{\rm kpc}$ (at 10\% probability), then $\Gamma = 23$ suffices.
This value of $\Gamma$ is still somewhat higher than found from the MOJAVE population,
{whose distribution of $\beta_{\rm app}$ is consistent with} our parent sample (Paper II).

As previously noted by \citet{2011ApJ...730...92H} and
in paper II, jet bends are insufficient to explain the large values of $\theta$
but jets could decelerate substantially from pc to kpc scales.
If the IC-CMB model holds for all kpc-scale jets detected in X-rays,
then $\Gamma_{\rm min} > 2$ for over half of them, so the jets would still be
relativistic on kpc scales, regardless of bending between pc and kpc scales.
\citet{2011ApJ...730...92H} also pointed out that jet bending by a few degrees
is {\it required} for many cases, as we also find (see minimum values of $\zeta$
in Table~\ref{tab:bends}).
However, the solution for $\theta$ is a steep function of $\theta$ for small $\Gamma$ and $\theta$
\cite[see Fig. 4 of ][]{2011ApJ...730...92H}, so $\Gamma$ must be very close to $\Gamma_{\rm min}$
in order to bring $\theta$ into better agreement with $\theta_{\rm kpc}$.  To achieve
$\theta < \theta_{\rm kpc}$ for half of the sample requires that $\Gamma$ be just 1.5\% larger
than $\Gamma_{\rm min}$; essentially $\Gamma = \Gamma_{\rm min}$ to within the
statistical uncertainties under the deceleration hypothesis.

\clearpage

\startlongtable
\begin{deluxetable}{rcrrrrrrrrr}
\tablewidth{0pc}
\tablecaption{Jet Beaming Model Parameters \label{tab:beaming} }
\tablehead{
\colhead{Target} & \colhead{z} &  \colhead{A/B} & \colhead{$\alpha_{rx}$} & \colhead{$R_1$\tablenotemark{a}}
	& \colhead{$V$\tablenotemark{b}} & \colhead{$B_1$\tablenotemark{c}} &
	\colhead{$K$\tablenotemark{d}} & \colhead{$\theta$\tablenotemark{e}}  &
	\colhead{$\Gamma_{\rm min}$\tablenotemark{f}} & \colhead{$\theta_{\rm max}$\tablenotemark{f}}\\
\colhead{ } & \colhead{}  & \colhead{}  & \colhead{}  & \colhead{($10^{-3}$)} & \colhead{(pc$^3$)} &
	\colhead{($\mu$G)} & \colhead{} & \colhead{(\arcdeg)}  & \colhead{} & \colhead{(\arcdeg)}}
\startdata
0106$+$013 &  2.099 & A &     0.94 &        77.3 &  1.2e+12 &  160. &      14.7 &       10 &        2.0 &       15 \\
0208$-$512 &  0.999 & B &     0.92 &       132.8 &  1.0e+12 &   75. &      23.6 &        9 &        2.5 &       12 \\
0229$+$131 &  2.059 & B & $>$ 0.95 &  $<$   55.8 &  1.2e+12 &   82. & $<$   6.5 &  $>$  13 &   \nodata & \nodata \\
0234$+$285 &  1.213 & B &     0.86 &       300.5 &  2.2e+12 &   51. &      20.4 &        9 &        2.3 &       13 \\
0256$+$075 &  0.893 & B & $>$ 0.79 &  $<$ 1227.3 &  9.4e+11 &   26. & $<$  31.7 &  $>$   8 &   \nodata & \nodata \\
0402$-$362 &  1.417 & B &     0.95 &        78.2 &  3.3e+12 &   69. &      10.9 &       11 &        1.7 &       18 \\
0413$-$210 &  0.808 & A &     1.04 &        13.0 &  5.7e+11 &  127. &      13.6 &       10 &        1.9 &       16 \\
0454$-$463 &  0.858 & A &     0.91 &       149.8 &  1.3e+12 &   69. &      27.0 &        8 &        2.6 &       11 \\
0508$-$220 &  0.172 & A & $>$ 1.04 &  $<$   10.3 &  2.6e+11 &   45. & $<$  10.5 &  $>$  11 &   \nodata & \nodata \\
0707$+$476 &  1.292 & B & $>$ 0.92 &  $<$  113.5 &  1.2e+12 &   53. & $<$  11.4 &  $>$  11 &   \nodata & \nodata \\
0745$+$241 &  0.410 & B & $>$ 0.88 &  $<$  230.3 &  4.7e+11 &   34. & $<$  30.2 &  $>$   8 &  \nodata & \nodata \\
0748$+$126 &  0.889 & B &     0.85 &       385.9 &  2.8e+12 &   33. &      21.0 &        9 &        2.3 &       13 \\
0820$+$225 &  0.951 & A & $>$ 0.93 &  $<$   83.3 &  7.7e+11 &   54. & $<$  13.7 &  $>$  10 &  \nodata & \nodata \\
0833$+$585 &  2.101 & B &     0.77 &      1899.7 &  9.0e+11 &   55. &      30.0 &        8 &        2.8 &       11 \\
0858$-$771 &  0.490 & B & $>$ 0.99 &  $<$   40.3 &  5.4e+11 &   52. & $<$  15.7 &  $>$  10 &   \nodata & \nodata \\
0859$+$470 &  1.462 & A &     1.01 &        22.3 &  8.0e+11 &  120. &       9.0 &       11 &        1.6 &       19 \\
0903$-$573 &  0.695 & A &     1.07 &        10.1 &  6.4e+11 &  123. &      13.1 &       10 &        1.9 &       16 \\
0920$-$397 &  0.591 & A &     1.00 &        29.8 &  1.4e+12 &   64. &      14.3 &       10 &        2.0 &       15 \\
0923$+$392 &  0.695 & A & $>$ 0.98 &  $<$   38.8 &  4.0e+11 &   77. & $<$  17.2 &  $>$  10 &  \nodata & \nodata \\
0953$+$254 &  0.712 & B & $>$ 0.85 &  $<$  359.4 &  2.2e+12 &   28. & $<$  21.1 &  $>$   9 &   \nodata & \nodata \\
0954$+$556 &  0.909 & A &     1.03 &        18.4 &  7.4e+11 &   87. &      10.0 &       11 &        1.7 &       18 \\
1030$-$357 &  1.455 & B &     0.93 &       103.0 &  3.9e+12 &   78. &      13.8 &       10 &        1.9 &       16 \\
1040$+$123 &  1.029 & A & $>$ 1.05 &  $<$   12.1 &  1.0e+12 &  109. & $<$   8.8 &  $>$  12 &   \nodata & \nodata \\
1046$-$409 &  0.620 & A &     0.95 &        80.0 &  6.2e+11 &   51. &      18.9 &        9 &        2.2 &       13 \\
1055$+$018 &  0.888 & B & $>$ 0.92 &  $<$  123.9 &  4.9e+12 &   42. & $<$  14.2 &  $>$  10 &   \nodata & \nodata \\
1055$+$201 &  1.110 & A &     0.93 &        92.2 &  4.5e+12 &   49. &      11.1 &       11 &        1.7 &       17 \\
1116$+$128 &  2.118 & A & $>$ 1.00 &  $<$   28.9 &  2.2e+12 &  114. & $<$   6.0 &  $>$  13 &   \nodata & \nodata \\
1116$-$462 &  0.713 & A & $>$ 1.02 &  $<$   22.0 &  4.1e+11 &  103. & $<$  16.5 &  $>$  10 &   \nodata & \nodata \\
1145$-$676 &  0.210 & B & $>$ 0.95 &  $<$   77.7 &  6.8e+11 &   32. & $<$  21.3 &  $>$   9 &  \nodata & \nodata \\
1202$-$262 &  0.789 & A &     0.86 &       335.5 &  1.2e+12 &   66. &      43.7 &        7 &        3.3 &        9 \\
1303$-$827 &  0.870 & A & $>$ 1.02 &  $<$   23.9 &  1.3e+12 &   74. & $<$  10.3 &  $>$  11 &   \nodata & \nodata \\
1354$+$195 &  0.720 & A &     0.95 &        64.8 &  3.1e+12 &   49. &      14.2 &       10 &        2.0 &       15 \\
1421$-$490 &  0.662 & A &     1.17 &         2.4 &  9.8e+11 &  216. &      10.8 &       11 &        1.7 &       18 \\
1424$-$418 &  1.522 & B & $>$ 0.91 &  $<$  140.6 &  8.1e+11 &  105. & $<$  20.8 &  $>$   9 &   \nodata & \nodata \\
1502$+$106 &  1.839 & B &     0.87 &       277.4 &  2.0e+12 &   48. &      10.8 &       11 &        1.7 &       18 \\
1622$-$297 &  0.815 & B & $>$ 0.99 &  $<$   36.7 &  1.6e+12 &   64. & $<$  12.1 &  $>$  11 &   \nodata & \nodata \\
1641$+$399 &  0.593 & A &     1.04 &        15.4 &  4.6e+11 &   98. &      15.1 &       10 &        2.0 &       15 \\
1642$+$690 &  0.751 & A &     0.97 &        46.2 &  7.9e+11 &   69. &      16.0 &       10 &        2.1 &       14 \\
1655$+$077 &  0.621 & B & $>$ 0.93 &  $<$   94.7 &  4.9e+11 &   47. & $<$  19.1 &  $>$   9 &   \nodata & \nodata \\
1823$+$568 &  0.664 & A &     0.91 &       135.3 &  4.6e+11 &   81. &      37.8 &        8 &        3.1 &        9 \\
1828$+$487 &  0.692 & A &     0.91 &       145.3 &  2.4e+11 &  129. &      60.3 &        6 &        3.9 &        7 \\
1928$+$738 &  0.302 & A &     0.86 &       321.3 &  7.4e+11 &   28. &      35.9 &        8 &        3.0 &       10 \\
2007$+$777 &  0.342 & B &     0.82 &       685.0 &  1.0e+12 &   18. &      32.7 &        8 &        2.9 &       10 \\
2052$-$474 &  1.489 & B & $>$ 0.89 &  $<$  214.4 &  6.7e+11 &   74. & $<$  18.9 &  $>$   9 &   \nodata & \nodata \\
2101$-$490 &  1.040 & B &     0.99 &        37.6 &  3.4e+12 &   63. &       9.4 &       11 &        1.6 &       19 \\
2123$-$463 &  1.670 & B &     0.95 &        87.6 &  1.5e+12 &   82. &      11.1 &       11 &        1.7 &       17 \\
2201$+$315 &  0.295 & A &     0.94 &        70.9 &  1.5e+12 &   28. &      15.4 &       10 &        2.0 &       15 \\
2230$+$114 &  1.037 & A & $>$ 0.82 &  $<$  691.9 &  7.0e+11 &   91. & $<$  68.3 &  $>$   6 &   \nodata & \nodata \\
2251$+$158 &  0.859 & A &     0.95 &        72.9 &  1.1e+12 &   97. &      25.5 &        8 &        2.6 &       11 \\
2255$-$282 &  0.926 & B &     0.95 &        68.5 &  1.6e+12 &   53. &      12.5 &       10 &        1.8 &       16 \\
2326$-$477 &  1.299 & B & $>$ 0.79 &  $<$ 1111.1 &  1.4e+12 &   32. & $<$  24.4 &  $>$   9 &   \nodata & \nodata \\
\enddata
\tablenotetext{a}{The ratio of the inverse Compton to synchrotron
   luminosities; see Paper I.}
\tablenotetext{b}{$V$ is the volume of the synchrotron emission region.}
\tablenotetext{c}{$B_1$ is the minimum energy magnetic field; see Paper I.}
\tablenotetext{d}{$K$ is a function of observable and assumed quantities;
  large values indicate stronger beaming in the IC-CMB model.  See Paper I
  for details.}
\tablenotetext{e}{The bulk Lorentz factor is assumed to  be 15.}
\tablenotetext{f}{{}Limits to $\Gamma$ and $\theta$ are calculated only when the jet is detected
  in X-rays.}
\end{deluxetable}

\clearpage

\begin{table}
\begin{center}
\tablewidth{0pc}
\caption{Quasar Jet Orientations\tablenotemark{a} \label{tab:bends} }
\begin{tabular}{rrrrcrrrrrrrr}
\tableline\tableline
Name & {PA$_{\rm kpc}$} & {PA$_{\rm pc}$} & {$\beta_{\rm app}$~~~~~} & Ref.\tablenotemark{b}
	& \multicolumn{3}{c}{{$\theta_{\rm kpc}$}} & \multicolumn{3}{c}{{$\zeta$\tablenotemark{c}}} 
	& {$\alpha_{rx}$\tablenotemark{d}} & {$\theta$\tablenotemark{d}} \\
{ } 	&	{ } 	&	{ } 	&	{ } 	&	{ } 	&	{min}	&	{mid}	&	{max}
	&	{min}	&	{mid}	&	{max}	& { } 	&  { } \\
\tableline
0106$+$013 & -175 &   -97 &  25.80 $\pm$ 2.80 &  1 &    2.2 &    2.4 &    7.2 &    1.1 &    1.5 &    6.9 &             0.94 &  10.0 \\
0229$+$131 &   20 &    80 &  14.00 $\pm$ 1.20 &  1 &    3.6 &    4.1 &   11.7 &    1.8 &    2.5 &   11.2 &    $>$  0.95 &  $>$ 12.5 \\
0234$+$285 &  -20 &    -9 &  20.72 $\pm$ 0.97 &  1 &    0.5 &    1.5 &    2.5 &    0.3 &    0.4 &    1.7 &             0.86 &   9.1 \\
0707$+$476 &  -90 &     2 &   8.33 $\pm$ 0.92 &  1 &    6.9 &    7.7 &   21.8 &    3.5 &    4.9 &   21.0 &    $>$  0.92 &  $>$ 10.7 \\
0745$+$241 &  -45 &   -59 &   6.60 $\pm$ 0.72 &  1 &    2.2 &    4.8 &    9.0 &    1.1 &    1.5 &    6.9 &    $>$  0.88 &  $>$  8.1 \\
0748$+$126 &  130 &   117 &  14.09 $\pm$ 0.92 &  1 &    0.9 &    2.2 &    3.9 &    0.4 &    0.6 &    2.8 &             0.85 &   9.0 \\
0859$+$470 &  -20 &   -16 &  16.10 $\pm$ 1.30 &  1 &    0.3 &    1.8 &    2.2 &    0.1 &    0.2 &    0.8 &             1.01 &  11.5 \\
0923$+$392 &   75 &   106 &   2.75 $\pm$ 0.54 &  1 &   10.9 &   15.0 &   34.7 &    5.5 &    7.7 &   31.5 &    $>$  0.98 &  $>$  9.5 \\
0953$+$254 & -115 &  -120 &   9.96 $\pm$ 0.26 &  1 &    0.5 &    2.9 &    3.8 &    0.3 &    0.4 &    1.6 &    $>$  0.85 &  $>$  9.0 \\
1055$+$018 &  180 &   -76 &   6.98 $\pm$ 0.68 &  1 &    8.0 &    9.0 &   25.0 &    4.0 &    5.6 &   24.1 &    $>$  0.92 &  $>$ 10.1 \\
1055$+$201 &  -10 &   -15 &   7.45 $\pm$ 0.98 &  1 &    0.7 &    3.9 &    5.1 &    0.4 &    0.5 &    2.3 &             0.93 &  10.8 \\
1202$-$262 &  -15 &   -19 &  10.70 $\pm$ 3.30 &  1 &    0.5 &    2.7 &    3.5 &    0.2 &    0.3 &    1.5 &             0.86 &   7.2 \\
1354$+$195 &  165 &   141 &   9.84 $\pm$ 0.70 &  2 &    2.3 &    3.7 &    8.4 &    1.2 &    1.6 &    7.3 &             0.95 &  10.1 \\
1502$+$106 &  160 &   101 &  18.20 $\pm$ 1.10 &  1 &    2.7 &    3.1 &    8.9 &    1.3 &    1.9 &    8.6 &             0.87 &  10.9 \\
1622$-$297 & -160 &   -84 &  12.00 $\pm$ 1.40 &  1 &    4.6 &    5.2 &   15.0 &    2.3 &    3.3 &   14.5 &    $>$  0.99 &  $>$ 10.5 \\
1641$+$399 &  -25 &   -86 &  19.29 $\pm$ 0.51 &  3 &    2.6 &    3.0 &    8.7 &    1.3 &    1.9 &    8.3 &             1.04 &   9.9 \\
1642$+$690 &  170 &  -162 &  14.56 $\pm$ 0.40 &  1 &    1.8 &    2.7 &    6.5 &    0.9 &    1.3 &    5.8 &             0.97 &   9.7 \\
1655$+$077 &  -50 &   -36 &  14.80 $\pm$ 1.10 &  1 &    0.9 &    2.1 &    3.9 &    0.5 &    0.7 &    2.9 &    $>$  0.93 &  $>$  9.3 \\
1823$+$568 &   90 &  -159 &  18.91 $\pm$ 0.37 &  1 &    2.9 &    3.2 &    9.3 &    1.4 &    2.0 &    9.0 &             0.91 &   7.5 \\
1828$+$487 &  -40 &   -37 &  13.07 $\pm$ 0.14 &  3 &    0.2 &    2.2 &    2.5 &    0.1 &    0.1 &    0.6 &             0.91 &   6.5 \\
1928$+$738 & -170 &  -178 &   8.16 $\pm$ 0.21 &  3 &    1.0 &    3.7 &    5.6 &    0.5 &    0.7 &    3.4 &             0.86 &   7.7 \\
2007$+$777 & -105 &    -4 &  12.60 $\pm$ 2.20 &  1 &    4.5 &    5.0 &   14.5 &    2.2 &    3.2 &   14.0 &             0.82 &   7.9 \\
2201$+$315 & -110 &  -147 &   8.28 $\pm$ 0.10 &  3 &    4.2 &    5.4 &   14.1 &    2.1 &    2.9 &   13.1 &             0.94 &   9.9 \\
2230$+$114 &  135 &   139 &  17.73 $\pm$ 0.87 &  1 &    0.3 &    1.6 &    2.1 &    0.1 &    0.2 &    0.8 &    $>$  0.82 &  $>$  6.2 \\
2251$+$158 &  -50 &   -91 &  13.80 $\pm$ 0.49 &  3 &    2.8 &    3.4 &    9.2 &    1.4 &    1.9 &    8.7 &             0.95 &   8.5 \\
2255$-$282 &  -70 &  -134 &   4.10 $\pm$ 0.37 &  1 &   12.7 &   14.4 &   37.0 &    6.3 &    8.9 &   35.3 &             0.95 &  10.5 \\
\tableline
\end{tabular}
\end{center}
\tablenotetext{a}{All angles are in degrees.  Position angles (PA) are defined relative to north, positive to the east.
  The min, mid, and max values give the minimum, 50\%, and 10\% probability points for the given angle.}
\tablenotetext{b}{References for values of $\beta_{\rm app}$: (1) \cite{2016AJ....152...12L}; (2) based
on 4 temporally-spaced epochs from the MOJAVE 15 GHz VLBA archive \citep{2009AJ....138.1874L},
yielding a maximum proper motion rate of $243 \pm 17~\mu$arcsec yr$^{-1}$; (3) \citet{2013AJ....146..120L}.}
\tablenotetext{c}{The quantity $\zeta$ is the angle between the pc-scale and kpc-scale jets in the frame of the quasar.
See Paper II.}
\tablenotetext{d}{From Table~\ref{tab:beaming}.}
\end{table}

\section{Conclusions}

We have reported new imaging results using the {\em Chandra} X-ray Observatory
for quasar jets selected from the radio sample originally defined by Paper I.
For the larger sample, we confirm many results in Papers I and II: 1) quasar jets can be readily detected
in X-rays using short {\em Chandra} observations, 2) no X-ray counterjets are detected,
3) the distribution of core photon indices is consistent
with a normal distribution with mean $1.61^{+0.04}_{-0.05}$
and dispersion $0.15^{+0.04}_{-0.03}$,
4) the IC-CMB model's prediction that $\alpha_{rx}$ should evolve strongly with $z$ is
not observed,
and 5) the line-of-sight angles of the kpc-scale jets are larger in the IC-CMB model
than inferred on pc scales, even if jet bending is allowed, possibly explained by
significant jet deceleration in the IC-CMB model.  {For the last point, we find it
important to note that inverse Compton scattering of CMB photons
by relativistic electrons in the jet must take place at some level.  The issue at stake
is whether the jet bulk Lorentz factors are still large on kpc scales, because jet
bending is insufficient to explain the observations.}

Our results add to the growing evidence of discrepancies between expectations of the
IC-CMB model and observations:
\begin{enumerate}
\item{Morphologies in the radio and X-ray band show significant
differences, perhaps
indicative of clumping \citep{2003A&A...403...83T}.}
\item{The X-ray and radio spectral indices do not agree for individual
  knots in 3C 273 \citep{2006ApJ...648..900J}}.
\item{Optical polarization indicates that the spectral component dominating
  the X-ray band is most likely synchrotron in origin in PKS 1136$-$135 and not from the
  same population that produces radio emission \citep{2013ApJ...773..186C}.}
\item{The proper motions of 3C 273 jet knots imply $\Gamma < 2.9$
  \citep{2016ApJ...818..195M}.}
\item{{Limits to $\gamma$-ray fluxes based on {\em Fermi} observations show severe
  tension with a simple extrapolation of the IC-CMB model \cite[e.g.,][]{2014ApJ...780L..27M,2017ApJ...835L..35M}.} }
\end{enumerate}

\noindent
These results demonstrate the need for ancillary data such as high resolution
optical and sub-mm imaging, X-ray spectral measurements, and polarimetry for testing either
synchrotron or IC-CMB models of the X-ray emission from kpc-scale
jets.

On the other hand, observations of high redshift quasars are consistent with
and perhaps best explained by the IC-CMB
model \citep{2016ApJ...816L..15S,2016ApJ...833..123M}.
There were only 11 sources in the study by \citet{2016ApJ...833..123M}, who found
marginally higher values of $\alpha_{rx}$ than for low $z$ quasars.
A larger study such as we have done on quasar jets with $z < 2.2$ should
also be carried out for a more definitive test of the IC-CMB model.

\acknowledgments

Support for this work was provided in part by the National Aeronautics and
Space Administration (NASA) through the Smithsonian Astrophysical Observatory (SAO)
contract SV3-73016 to MIT for support of the {\em Chandra} X-Ray Center (CXC),
which is operated by SAO for and on behalf of NASA under contract NAS8-03060.
Support was also provided by NASA under contract NAS 8-39073 to SAO.
This research has made use of data from the MOJAVE
database that is maintained by the MOJAVE team \citep{2009AJ....138.1874L}.
We thank M. Lister for providing MOJAVE results ahead of publication.
This research has made
use of the United States Naval Observatory (USNO) Radio Reference
Frame Image Database (RRFID).  The Australia Telescope Compact
Array is part of the Australia Telescope which is funded by the
Commonwealth of Australia for operation as a National Facility managed
by CSIRO.  This research has made use of the NASA/IPAC Extragalactic
Database (NED) which is operated by the Jet Propulsion Laboratory,
California Institute of Technology, under contract with the
National Aeronautics and Space Administration.

\vspace{5mm}
\facilities{CXO(ACIS), ATCA, VLA, HST(WFC3)}

\software{ciao \citep{2006SPIE.6270E..1VF}, IDL, SAOImage DS9, IRAF \citep{1993ASPC...52..173T}, ISIS \cite{2000ASPC..216..591H}}

\bibliographystyle{aasjournal} 

\begin{thebibliography}{}
\expandafter\ifx\csname natexlab\endcsname\relax\def\natexlab#1{#1}\fi
\providecommand{\url}[1]{\href{#1}{#1}}

\bibitem[{{Ackermann} {et~al.}(2015){Ackermann}, {Ajello}, {Atwood}, {Baldini},
  {Ballet}, {Barbiellini}, {Bastieri}, {Becerra Gonzalez}, {Bellazzini},
  {Bissaldi}, {Blandford}, {Bloom}, {Bonino}, {Bottacini}, {Brandt}, {Bregeon},
  {Britto}, {Bruel}, {Buehler}, {Buson}, {Caliandro}, {Cameron}, {Caragiulo},
  {Caraveo}, {Carpenter}, {Casandjian}, {Cavazzuti}, {Cecchi}, {Charles},
  {Chekhtman}, {Cheung}, {Chiang}, {Chiaro}, {Ciprini}, {Claus},
  {Cohen-Tanugi}, {Cominsky}, {Conrad}, {Cutini}, {D'Abrusco}, {D'Ammando}, {de
  Angelis}, {Desiante}, {Digel}, {Di Venere}, {Drell}, {Favuzzi}, {Fegan},
  {Ferrara}, {Finke}, {Focke}, {Franckowiak}, {Fuhrmann}, {Fukazawa},
  {Furniss}, {Fusco}, {Gargano}, {Gasparrini}, {Giglietto}, {Giommi},
  {Giordano}, {Giroletti}, {Glanzman}, {Godfrey}, {Grenier}, {Grove},
  {Guiriec}, {Hewitt}, {Hill}, {Horan}, {Itoh}, {J{\'o}hannesson}, {Johnson},
  {Johnson}, {Kataoka}, {Kawano}, {Krauss}, {Kuss}, {La Mura}, {Larsson},
  {Latronico}, {Leto}, {Li}, {Li}, {Longo}, {Loparco}, {Lott}, {Lovellette},
  {Lubrano}, {Madejski}, {Mayer}, {Mazziotta}, {McEnery}, {Michelson},
  {Mizuno}, {Moiseev}, {Monzani}, {Morselli}, {Moskalenko}, {Murgia}, {Nuss},
  {Ohno}, {Ohsugi}, {Ojha}, {Omodei}, {Orienti}, {Orlando}, {Paggi}, {Paneque},
  {Perkins}, {Pesce-Rollins}, {Piron}, {Pivato}, {Porter}, {Rain{\`o}},
  {Rando}, {Razzano}, {Razzaque}, {Reimer}, {Reimer}, {Romani}, {Salvetti},
  {Schaal}, {Schinzel}, {Schulz}, {Sgr{\`o}}, {Siskind}, {Sokolovsky}, {Spada},
  {Spandre}, {Spinelli}, {Stawarz}, {Suson}, {Takahashi}, {Takahashi},
  {Tanaka}, {Thayer}, {Thayer}, {Tibaldo}, {Torres}, {Torresi}, {Tosti},
  {Troja}, {Uchiyama}, {Vianello}, {Winer}, {Wood}, \&
  {Zimmer}}]{2015ApJ...810...14A}
{Ackermann}, M., {Ajello}, M., {Atwood}, W.~B., {et~al.} 2015, \apj, 810, 14

\bibitem[{{Arnaud}(1996)}]{arnaud}
{Arnaud}, K.~A. 1996, in Astronomical Society of the Pacific Conference Series,
  Vol. 101, Astronomical Data Analysis Software and Systems V, ed. G.~H.
  {Jacoby} \& J.~{Barnes}, 17

\bibitem[{{Belsole} {et~al.}(2006){Belsole}, {Worrall}, \&
  {Hardcastle}}]{2006MNRAS.366..339B}
{Belsole}, E., {Worrall}, D.~M., \& {Hardcastle}, M.~J. 2006, \mnras, 366, 339

\bibitem[{{Breiding} {et~al.}(2017){Breiding}, {Meyer}, {Georganopoulos},
  {Keenan}, {DeNigris}, \& {Hewitt}}]{2017ApJ...849...95B}
{Breiding}, P., {Meyer}, E.~T., {Georganopoulos}, M., {et~al.} 2017, \apj, 849,
  95

\bibitem[{{Burgess} \& {Hunstead}(2006)}]{2006AJ....131..114B}
{Burgess}, A.~M., \& {Hunstead}, R.~W. 2006, \aj, 131, 114

\bibitem[{{Cara} {et~al.}(2013){Cara}, {Perlman}, {Uchiyama}, {Cheung},
  {Coppi}, {Georganopoulos}, {Worrall}, {Birkinshaw}, {Sparks}, {Marshall},
  {Stawarz}, {Begelman}, {O'Dea}, \& {Baum}}]{2013ApJ...773..186C}
{Cara}, M., {Perlman}, E.~S., {Uchiyama}, Y., {et~al.} 2013, \apj, 773, 186

\bibitem[{{Celotti} {et~al.}(2001){Celotti}, {Ghisellini}, \&
  {Chiaberge}}]{2001MNRAS.321L...1C}
{Celotti}, A., {Ghisellini}, G., \& {Chiaberge}, M. 2001, \mnras, 321, L1

\bibitem[{{Cohen} {et~al.}(2007){Cohen}, {Lister}, {Homan}, {Kadler},
  {Kellermann}, {Kovalev}, \& {Vermeulen}}]{2007ApJ...658..232C}
{Cohen}, M.~H., {Lister}, M.~L., {Homan}, D.~C., {et~al.} 2007, \apj, 658, 232

\bibitem[{{Cooper} {et~al.}(2007){Cooper}, {Lister}, \&
  {Kochanczyk}}]{2007ApJS..171..376C}
{Cooper}, N.~J., {Lister}, M.~L., \& {Kochanczyk}, M.~D. 2007, \apjs, 171, 376

\bibitem[{{Dickey} \& {Lockman}(1990)}]{1990ARAA..28..215D}
{Dickey}, J.~M., \& {Lockman}, F.~J. 1990, \araa, 28, 215

\bibitem[{{Elvis} {et~al.}(1989){Elvis}, {Wilkes}, \&
  {Lockman}}]{1989AJ.....97..777E}
{Elvis}, M., {Wilkes}, B.~J., \& {Lockman}, F.~J. 1989, \aj, 97, 777

\bibitem[{{Fossati} {et~al.}(1998){Fossati}, {Maraschi}, {Celotti}, {Comastri},
  \& {Ghisellini}}]{1998MNRAS.299..433F}
{Fossati}, G., {Maraschi}, L., {Celotti}, A., {Comastri}, A., \& {Ghisellini},
  G. 1998, \mnras, 299, 433

\bibitem[{{Fruscione} {et~al.}(2006){Fruscione}, {McDowell}, {Allen},
  {Brickhouse}, {Burke}, {Davis}, {Durham}, {Elvis}, {Galle}, {Harris},
  {Huenemoerder}, {Houck}, {Ishibashi}, {Karovska}, {Nicastro}, {Noble},
  {Nowak}, {Primini}, {Siemiginowska}, {Smith}, \&
  {Wise}}]{2006SPIE.6270E..1VF}
{Fruscione}, A., {McDowell}, J.~C., {Allen}, G.~E., {et~al.} 2006, in
  \procspie, Vol. 6270, Society of Photo-Optical Instrumentation Engineers
  (SPIE) Conference Series, 62701V

\bibitem[{{Georganopoulos} \& {Kazanas}(2004)}]{2004ApJ...604L..81G}
{Georganopoulos}, M., \& {Kazanas}, D. 2004, \apjl, 604, L81

\bibitem[{{Ghisellini} {et~al.}(2017){Ghisellini}, {Righi}, {Costamante}, \&
  {Tavecchio}}]{2017MNRAS.469..255G}
{Ghisellini}, G., {Righi}, C., {Costamante}, L., \& {Tavecchio}, F. 2017,
  \mnras, 469, 255

\bibitem[{{Giommi} {et~al.}(2012){Giommi}, {Padovani}, {Polenta}, {Turriziani},
  {D'Elia}, \& {Piranomonte}}]{2012MNRAS.420.2899G}
{Giommi}, P., {Padovani}, P., {Polenta}, G., {et~al.} 2012, \mnras, 420, 2899

\bibitem[{{Hardcastle}(2006)}]{2006MNRAS.366.1465H}
{Hardcastle}, M.~J. 2006, \mnras, 366, 1465

\bibitem[{{Harris} \& {Krawczynski}(2002)}]{2002ApJ...565..244H}
{Harris}, D.~E., \& {Krawczynski}, H. 2002, \apj, 565, 244

\bibitem[{{Harris} \& {Krawczynski}(2006)}]{2006ARA&A..44..463H}
---. 2006, \araa, 44, 463

\bibitem[{{Hogan} {et~al.}(2011){Hogan}, {Lister}, {Kharb}, {Marshall}, \&
  {Cooper}}]{2011ApJ...730...92H}
{Hogan}, B.~S., {Lister}, M.~L., {Kharb}, P., {Marshall}, H.~L., \& {Cooper},
  N.~J. 2011, \apj, 730, 92

\bibitem[{{Houck} \& {Denicola}(2000)}]{2000ASPC..216..591H}
{Houck}, J.~C., \& {Denicola}, L.~A. 2000, in Astronomical Society of the
  Pacific Conference Series, Vol. 216, Astronomical Data Analysis Software and
  Systems IX, ed. N.~{Manset}, C.~{Veillet}, \& D.~{Crabtree}, 591

\bibitem[{{Jester} {et~al.}(2006){Jester}, {Harris}, {Marshall}, \&
  {Meisenheimer}}]{2006ApJ...648..900J}
{Jester}, S., {Harris}, D.~E., {Marshall}, H.~L., \& {Meisenheimer}, K. 2006,
  \apj, 648, 900

\bibitem[{{Jorstad} \& {Marscher}(2006)}]{2006AN....327..227J}
{Jorstad}, S.~G., \& {Marscher}, A.~P. 2006, Astronomische Nachrichten, 327,
  227

\bibitem[{{Kataoka} \& {Stawarz}(2005)}]{2005ApJ...622..797K}
{Kataoka}, J., \& {Stawarz}, {\L}. 2005, \apj, 622, 797

\bibitem[{{Kellermann} {et~al.}(2004){Kellermann}, {Lister}, {Homan},
  {Vermeulen}, {Cohen}, {Ros}, {Kadler}, {Zensus}, \&
  {Kovalev}}]{2004ApJ...609..539K}
{Kellermann}, K.~I., {Lister}, M.~L., {Homan}, D.~C., {et~al.} 2004, \apj, 609,
  539

\bibitem[{{Lister} {et~al.}(2009{\natexlab{a}}){Lister}, {Cohen}, {Homan},
  {Kadler}, {Kellermann}, {Kovalev}, {Ros}, {Savolainen}, \&
  {Zensus}}]{2009AJ....138.1874L}
{Lister}, M.~L., {Cohen}, M.~H., {Homan}, D.~C., {et~al.} 2009{\natexlab{a}},
  \aj, 138, 1874

\bibitem[{{Lister} {et~al.}(2009{\natexlab{b}}){Lister}, {Aller}, {Aller},
  {Cohen}, {Homan}, {Kadler}, {Kellermann}, {Kovalev}, {Ros}, {Savolainen},
  {Zensus}, \& {Vermeulen}}]{2009AJ....137.3718L}
{Lister}, M.~L., {Aller}, H.~D., {Aller}, M.~F., {et~al.} 2009{\natexlab{b}},
  \aj, 137, 3718

\bibitem[{{Lister} {et~al.}(2013){Lister}, {Aller}, {Aller}, {Homan},
  {Kellermann}, {Kovalev}, {Pushkarev}, {Richards}, {Ros}, \&
  {Savolainen}}]{2013AJ....146..120L}
{Lister}, M.~L., {Aller}, M.~F., {Aller}, H.~D., {et~al.} 2013, \aj, 146, 120

\bibitem[{{Lister} {et~al.}(2016){Lister}, {Aller}, {Aller}, {Homan},
  {Kellermann}, {Kovalev}, {Pushkarev}, {Richards}, {Ros}, \&
  {Savolainen}}]{2016AJ....152...12L}
---. 2016, \aj, 152, 12

\bibitem[{{Lovell}(1997)}]{lovell}
{Lovell}, J.~E.~J. 1997, {Ph.D.} thesis, University of Tasmania

\bibitem[{{Maccacaro} {et~al.}(1988){Maccacaro}, {Gioia}, {Wolter}, {Zamorani},
  \& {Stocke}}]{1988ApJ...326..680M}
{Maccacaro}, T., {Gioia}, I.~M., {Wolter}, A., {Zamorani}, G., \& {Stocke},
  J.~T. 1988, \apj, 326, 680

\bibitem[{{Marshall} {et~al.}(2006){Marshall}, {Jester}, {Harris}, \&
  {Meisenheimer}}]{2006ESASP.604..643M}
{Marshall}, H.~L., {Jester}, S., {Harris}, D.~E., \& {Meisenheimer}, K. 2006,
  in ESA Special Publication, Vol. 604, The X-ray Universe 2005, ed.
  A.~{Wilson}, 643

\bibitem[{{Marshall} {et~al.}(2005){Marshall}, {Schwartz}, {Lovell}, {Murphy},
  {Worrall}, {Birkinshaw}, {Gelbord}, {Perlman}, \&
  {Jauncey}}]{2005ApJS..156...13M}
{Marshall}, H.~L., {Schwartz}, D.~A., {Lovell}, J.~E.~J., {et~al.} 2005, \apjs,
  156, 13 (Paper I)

\bibitem[{{Marshall} {et~al.}(2011){Marshall}, {Gelbord}, {Schwartz}, {Murphy},
  {Lovell}, {Worrall}, {Birkinshaw}, {Perlman}, {Godfrey}, \&
  {Jauncey}}]{2011ApJS..193...15M}
{Marshall}, H.~L., {Gelbord}, J.~M., {Schwartz}, D.~A., {et~al.} 2011, \apjs,
  193, 15 (Paper II)

\bibitem[{{McKeough} {et~al.}(2016){McKeough}, {Siemiginowska}, {Cheung},
  {Stawarz}, {Kashyap}, {Stein}, {Stampoulis}, {van Dyk}, {Wardle}, {Lee},
  {Harris}, {Schwartz}, {Donato}, {Maraschi}, \&
  {Tavecchio}}]{2016ApJ...833..123M}
{McKeough}, K., {Siemiginowska}, A., {Cheung}, C.~C., {et~al.} 2016, \apj, 833,
  123

\bibitem[{{Meyer} {et~al.}(2017){Meyer}, {Breiding}, {Georganopoulos}, {Oteo},
  {Zwaan}, {Laing}, {Godfrey}, \& {Ivison}}]{2017ApJ...835L..35M}
{Meyer}, E.~T., {Breiding}, P., {Georganopoulos}, M., {et~al.} 2017, \apjl,
  835, L35

\bibitem[{{Meyer} \& {Georganopoulos}(2014)}]{2014ApJ...780L..27M}
{Meyer}, E.~T., \& {Georganopoulos}, M. 2014, \apjl, 780, L27

\bibitem[{{Meyer} {et~al.}(2015){Meyer}, {Georganopoulos}, {Sparks}, {Godfrey},
  {Lovell}, \& {Perlman}}]{2015ApJ...805..154M}
{Meyer}, E.~T., {Georganopoulos}, M., {Sparks}, W.~B., {et~al.} 2015, \apj,
  805, 154

\bibitem[{{Meyer} {et~al.}(2016){Meyer}, {Sparks}, {Georganopoulos},
  {Anderson}, {van der Marel}, {Biretta}, {Sohn}, {Chiaberge}, {Perlman}, \&
  {Norman}}]{2016ApJ...818..195M}
{Meyer}, E.~T., {Sparks}, W.~B., {Georganopoulos}, M., {et~al.} 2016, \apj,
  818, 195

\bibitem[{{Miller} {et~al.}(2011){Miller}, {Brandt}, {Schneider}, {Gibson},
  {Steffen}, \& {Wu}}]{2011ApJ...726...20M}
{Miller}, B.~P., {Brandt}, W.~N., {Schneider}, D.~P., {et~al.} 2011, \apj, 726,
  20

\bibitem[{{Murphy} {et~al.}(1993){Murphy}, {Browne}, \&
  {Perley}}]{1993MNRAS.264..298M}
{Murphy}, D.~W., {Browne}, I.~W.~A., \& {Perley}, R.~A. 1993, \mnras, 264, 298

\bibitem[{{Murphy} {et~al.}(1996){Murphy}, {Lockman}, {Laor}, \&
  {Elvis}}]{1996ApJS..105..369M}
{Murphy}, E.~M., {Lockman}, F.~J., {Laor}, A., \& {Elvis}, M. 1996, \apjs, 105,
  369

\bibitem[{{Reeves} \& {Turner}(2000)}]{2000MNRAS.316..234R}
{Reeves}, J.~N., \& {Turner}, M.~J.~L. 2000, \mnras, 316, 234

\bibitem[{{Sambruna} {et~al.}(2004){Sambruna}, {Gambill}, {Maraschi},
  {Tavecchio}, {Cerutti}, {Cheung}, {Urry}, \& {Chartas}}]{2004ApJ...608..698S}
{Sambruna}, R.~M., {Gambill}, J.~K., {Maraschi}, L., {et~al.} 2004, \apj, 608,
  698

\bibitem[{{Sbarufatti} {et~al.}(2009){Sbarufatti}, {Ciprini}, {Kotilainen},
  {Decarli}, {Treves}, {Veronesi}, \& {Falomo}}]{2009AJ....137..337S}
{Sbarufatti}, B., {Ciprini}, S., {Kotilainen}, J., {et~al.} 2009, \aj, 137, 337

\bibitem[{{Schechter} \& {Dressler}(1987)}]{1987AJ.....94..563S}
{Schechter}, P.~L., \& {Dressler}, A. 1987, \aj, 94, 563

\bibitem[{{Schwartz}(2002)}]{2002ApJ...569L..23S}
{Schwartz}, D.~A. 2002, \apjl, 569, L23

\bibitem[{{Schwartz} {et~al.}(2000){Schwartz}, {Marshall}, {Lovell}, {Piner},
  {Tingay}, {Birkinshaw}, {Chartas}, {Elvis}, {Feigelson}, {Ghosh}, {Harris},
  {Hirabayashi}, {Hooper}, {Jauncey}, {Lanzetta}, {Mathur}, {Preston},
  {Tucker}, {Virani}, {Wilkes}, \& {Worrall}}]{2000ApJ...540L..69S}
{Schwartz}, D.~A., {Marshall}, H.~L., {Lovell}, J.~E.~J., {et~al.} 2000, \apjl,
  540, L69

\bibitem[{{Schwartz} {et~al.}(2006){Schwartz}, {Marshall}, {Lovell}, {Murphy},
  {Bicknell}, {Birkinshaw}, {Gelbord}, {Georganopoulos}, {Godfrey}, {Jauncey},
  {Perlman}, \& {Worrall}}]{2006ApJ...640..592S}
---. 2006, \apj, 640, 592

\bibitem[{{Simionescu} {et~al.}(2016){Simionescu}, {Stawarz}, {Ichinohe},
  {Cheung}, {Jamrozy}, {Siemiginowska}, {Hagino}, {Gandhi}, \&
  {Werner}}]{2016ApJ...816L..15S}
{Simionescu}, A., {Stawarz}, {\L}., {Ichinohe}, Y., {et~al.} 2016, \apjl, 816,
  L15

\bibitem[{{Stawarz} {et~al.}(2004){Stawarz}, {Sikora}, {Ostrowski}, \&
  {Begelman}}]{2004ApJ...608...95S}
{Stawarz}, {\L}., {Sikora}, M., {Ostrowski}, M., \& {Begelman}, M.~C. 2004,
  \apj, 608, 95

\bibitem[{{Tavecchio} {et~al.}(2003){Tavecchio}, {Ghisellini}, \&
  {Celotti}}]{2003A&A...403...83T}
{Tavecchio}, F., {Ghisellini}, G., \& {Celotti}, A. 2003, \aap, 403, 83

\bibitem[{{Tavecchio} {et~al.}(2006){Tavecchio}, {Maraschi}, {Sambruna},
  {Gliozzi}, {Cheung}, {Wardle}, \& {Urry}}]{2006ApJ...641..732T}
{Tavecchio}, F., {Maraschi}, L., {Sambruna}, R.~M., {et~al.} 2006, \apj, 641,
  732

\bibitem[{{Tavecchio} {et~al.}(2000){Tavecchio}, {Maraschi}, {Sambruna}, \&
  {Urry}}]{2000ApJ...544L..23T}
{Tavecchio}, F., {Maraschi}, L., {Sambruna}, R.~M., \& {Urry}, C.~M. 2000,
  \apjl, 544, L23

\bibitem[{{Tody}(1993)}]{1993ASPC...52..173T}
{Tody}, D. 1993, in Astronomical Society of the Pacific Conference Series,
  Vol.~52, Astronomical Data Analysis Software and Systems II, ed. R.~J.
  {Hanisch}, R.~J.~V. {Brissenden}, \& J.~{Barnes}, 173

\bibitem[{{Worrall}(1989)}]{1989ESASP.296..719W}
{Worrall}, D.~M. 1989, in ESA Special Publication, Vol. 296, Two Topics in
  X-Ray Astronomy, Volume 1: X Ray Binaries. Volume 2: AGN and the X Ray
  Background, ed. J.~{Hunt} \& B.~{Battrick}

\bibitem[{{Worrall}(2009)}]{2009A&ARv..17....1W}
{Worrall}, D.~M. 2009, \aapr, 17, 1

\bibitem[{{Worrall} {et~al.}(1987){Worrall}, {Tananbaum}, {Giommi}, \&
  {Zamorani}}]{1987ApJ...313..596W}
{Worrall}, D.~M., {Tananbaum}, H., {Giommi}, P., \& {Zamorani}, G. 1987, \apj,
  313, 596

\end{thebibliography}

\end{document}